\documentclass[aps,prd,11pt,superscriptaddress,notitlepage,longbibliography,nofootinbib,tightenlines,preprintnumbers]{revtex4-2}
\usepackage{amsmath,amssymb,amsfonts}
\usepackage{graphicx}
\usepackage{color}

\usepackage{tikz}
\usetikzlibrary{decorations.text}
\usetikzlibrary{decorations.pathreplacing}
\usetikzlibrary{calc}

\usepackage{booktabs}
\usepackage{dsfont}
\usepackage[pdftex]{hyperref}
\hypersetup{colorlinks=true, linkcolor=darkred, citecolor=blue, linktoc=page}
\definecolor{darkred}{rgb}{0.8,0.1,0.1}
\usepackage{slashed}

\usepackage{physics}

\usepackage{subfigure}

\def\cC{{\cal C}}
\def\cF{{\cal F}}

\def\RR{\ensuremath{\mathbb R}}
\def\ZZ{\ensuremath{\mathbb Z}}

\def\cN{{\cal N}}

\def\cL{{\cal L}}
\def\cW{{\cal W}}
\def\cZ{{\cal Z}}

\def\cL{{\cal L}}

\DeclareMathOperator{\vol}{vol}

\DeclareMathOperator{\Li}{Li}

\makeatletter\def\l@subsubsection#1#2{}%
\makeatother

\def\Im{\mathop{\rm Im}}
\def\Re{\mathop{\rm Re}}

\begin{document}

\title{On the planar limit of 3d \texorpdfstring{$T_\rho^\sigma[SU(N)]$}{Trhosigma[SU(N)]}}

\author{Lorenzo Coccia}
\email{l.coccia@campus.unimib.it}

\affiliation{Dipartimento di Fisica, Universit\`a di Milano-Bicocca, I-20126 Milano, Italy}
\affiliation{INFN, Sezione di Milano-Bicocca, I-20126 Milano, Italy\\[2mm]}

\author{Christoph F.~Uhlemann} 

\email{uhlemann@umich.edu}
  
\affiliation{Leinweber Center for Theoretical Physics, Department of Physics
	\\
	University of Michigan, Ann Arbor, MI 48109-1040, USA}

\preprint{LCTP-20-28}

\begin{abstract}
We discuss a limit of 3d $T_\rho^\sigma[SU(N)]$ quiver gauge theories in which the number of nodes is large and the ranks scale quadratically with the length of the quiver. The sphere free energies and topologically twisted indices are obtained using supersymmetric localization. Both scale quartically with the length of the quiver and quadratically with $N$, with trilogarithm functions depending on the quiver data as coefficients. The IR SCFTs have well-behaved supergravity duals in Type IIB, and the free energies match precisely with holographic results. Previously discussed theories with $N^2\ln N$ scaling arise as limiting cases. Each balanced 3d quiver theory is linked to a 5d parent, whose matrix model is related and dominated by the same saddle point, leading to close relations between BPS observables.
\end{abstract}

\maketitle
\tableofcontents

\clearpage

\setlength{\parskip}{1.5 pt}

\section{Introduction}

There are many interesting superconformal field theories (SCFTs) in 3d, which can be loosely categorized according to the scaling of their free energies in the planar limit.
Famously, the free energy of the $\cN=6$ ABJM theories with holographic duals in M-theory scales like $N^{3/2}$ \cite{Aharony:2008ug,Aharony:2008gk,Drukker:2010nc}. 
For $\cN=2$  Chern-Simons-matter theories with duals in massive type IIA, the scaling is modified to $N^{5/3}$ \cite{Schwarz:2004yj,Gaiotto:2009mv,Guarino:2015jca}.
A large class of 3d SCFTs, denoted as $T_\rho^\sigma[G]$, can be obtained by considering 4d $\cN=4$ SYM with gauge group $G$ on an interval, with boundary conditions specified by two Young tableaux $\rho$ and $\sigma$ and separated by an S-duality wall \cite{Gaiotto:2008ak}.
For $G=SU(N)$ these theories can also be described as IR fixed points of 3d Yang-Mills-type quiver gauge theories.
Holographic duals for these theories in Type IIB supergravity were constructed in \cite{Assel:2011xz}, building on earlier work in \cite{DHoker:2007zhm,DHoker:2007hhe},
and the free energies for certain theories in this class, including $T[SU(N)]$, were matched to holographic results in \cite{Assel:2012cp}.
Remarkably, the free energies were found to scale like $N^2\ln N$.

In this paper we revisit the 3d $T_\rho^\sigma[SU(N)]$ theories, with the motivation to better understand their planar limit and the scaling of the free energies.
From the perspective of the supergravity duals, the $N^2\ln N$ scaling found in \cite{Assel:2012cp} is the result of a somewhat peculiar limit in which certain brane sources run off and stretch out the internal space to produce the logarithmic scaling. 
The  $\mathcal O(N^2)$ part is sensitive to higher-curvature corrections.
From the field theory perspective the scaling can be understood in a similar way:
The 3d SCFTs considered in \cite{Assel:2012cp} are the IR fixed points of quiver gauge theories with a large number of nodes, with the ranks of the gauge groups of the same order as the length of the quivers.
The matrix models resulting from supersymmetric localization can be reformulated in a way which is adapted to such long quiver gauge theories, following \cite{Uhlemann:2019ypp}, which was discussed for the $T[SU(N)]$ theory in \cite{Coccia:2020cku}.
The $N^2\ln N$ scaling emerges in this formulation from singular behavior of the localized partition function when evaluated on the large-$N$ saddle point, while the $\mathcal O(N^2)$ part is sensitive to corrections.

In this work we discuss a limit of 3d long quiver gauge theories in which the supergravity duals are free from runaway sources and the field theory computations do not lead to singularities.
We take a large number of nodes, $L$, and a large number of flavors at isolated interior nodes. Unlike the limit considered in \cite{Assel:2012cp}, we take the ranks of the gauge groups to scale quadratically with $L$. These theories can be understood as $T_\rho^\sigma[SU(N)]$ with $N=\mathcal O(L^2)$ where $\rho$ and $\sigma$ have $\mathcal O(L)$ rows with $\mathcal O(L)$ boxes each.
We will show that the free energies scale like $L^4$, or $N^2$, with coefficients given by trilogarithm functions
whose arguments depend on the data characterizing the quiver gauge theories, $\rho$ and $\sigma$.
These results will be produced both from field theory and supergravity, and we show that they match precisely.
We also discuss the topologically twisted index, and show that, in the spirit of the ``index theorem" of \cite{Hosseini:2016tor}, this index agrees up to a universal overall factor with the free energy on $S^3$.
From the perspective of the 4d $\cN=4$ SYM construction of the 3d $T_\rho^\sigma[SU(N)]$ theories, the $L^4$ scaling corresponds to the familiar $N^2$ scaling in 4d.
The theories considered in \cite{Assel:2012cp} will be recovered as special limiting cases, in which the trilogarithm functions appearing as coefficients of the leading $N^2$ terms reduce to logarithms.

The quartic scaling of the free energy with the length of the quiver is a feature also exhibited by a class of 5d SCFTs which arise as UV fixed points of long quiver gauge theories \cite{Gutperle:2017tjo,Fluder:2018chf,Uhlemann:2019ypp}.
The way the SCFTs relate to gauge theories in 3d and 5d differ: The 3d SCFTs arise as IR fixed points of UV-free gauge theories, while the 5d SCFTs arise as UV fixed points of IR-free gauge theories. The scaling of the ranks of the gauge groups is quadratic with the length of the quiver in 3d, but only linear in the 5d theories.
Moreover, the constraints that the gauge theories have to satisfy in order to obtain well-defined SCFTs in 3d and 5d are inequalities constraining the numbers of flavors in opposite directions.
Nevertheless, the constraints overlap for balanced theories.
For each 3d quiver gauge theory with all nodes balanced, we discuss a 5d parent theory for which the matrix model resulting from supersymmetric localization is related in a simple way and is dominated by the same saddle point, leading to simple relations between BPS observables in the planar limit.

The paper is organized as follows: In sec.~\ref{sec:quivers} we introduce the 3d quiver gauge theories whose IR fixed points we will study.
We review their brane realization in type IIB string theory and their supergravity duals.
In sec.~\ref{sec:F-loc} we derive general formulae for the free energies. 
We establish an ``index theorem", relating the topologically twisted index to the free energy in sec.~\ref{sec:index}.
In sec.~\ref{sec:sample} we present case studies, and derive explicit results for a sample of concrete theories.
In sec.~\ref{sec:5d} we discuss the relation to 5d long quiver SCFTs.
We conclude in sec.~\ref{sec:disc}.

\section{3d long quiver SCFTs}\label{sec:quivers}

We start with a characterization of the SCFTs to be discussed in the following in terms of their UV gauge theory descriptions.
Brane constructions and supergravity duals will be discussed afterwards.
The theories of interest are 3d Yang-Mills-type $\cN=4$ supersymmetric $U(\cdot)$ linear quiver gauge theories with $L$ nodes labeled by $t=1,\ldots, L$.
The general form is
\begin{align}\label{eq:quiver}
	U&(N_1)-U(N_2)-\ldots -U(N_{L-1})-U(N_L)
	\nonumber\\
	&\hskip 2mm |\hskip 14mm |\hskip 26mm |\hskip 17mm |
	\\
	&[k_1] \hskip 9mm [k_2] \hskip 19mm [k_{L-1}] \hskip 10mm [k_L]
	\nonumber
\end{align}
The dashes between the gauge nodes  denote hypermultiplets in the bifundamental representation,
and $[k_t]$ denotes $k_t$ fundamental flavors. 
The theories in (\ref{eq:quiver}) are $T_{\rho}^\sigma [SU(N)]$ theories,\footnote{We will use $T_\rho^\sigma[SU(N)]$ to refer to the IR SCFT and to the UV gauge theory, hoping that the distinction will be clear from context.}
 which were classified into good, bad and ugly in \cite{Gaiotto:2008ak}. The general quivers were spelled out previously e.g.\ in \cite{Cremonesi:2014uva}.
We will focus on the good theories, for which the number of flavors at each node is at least twice the number of colors. 

We will be interested in the limit where the gauge theories have a large number of nodes, $L\gg 1$.
In that limit the nodes can be labeled by an effectively continuous coordinate $z\in[0,1]$ along the quiver,
and the data $\lbrace N_t,k_t\rbrace$ is encoded in functions $N(z)$, $k(z)$ defined by
\begin{align}\label{eq:cont-rep}
	z&=\frac{t}{L}~,
	&
	N(z)&= N_{z L}~, 
	& k(z)&= k_{zL}~.
\end{align}
In the limit discussed in \cite{Assel:2012cp} the ranks of the gauge groups are $\mathcal O(L)$.
A prime example is the $T[SU(N)]$ theory, $(1)-(2)-\ldots -(N-1)-[N]$, which was cast in the above language in \cite{Coccia:2020cku}.
This is also the scaling considered for 5d quiver theories in \cite{Uhlemann:2019ypp}.
Here we will consider a different scaling, in which the majority of nodes has rank $N_t$ of $\mathcal O(L^2)$.
Concretely, we will take $N(z)$ to be a continuous, piece-wise linear function of $\mathcal O(L^2)$,
and we will also assume that the leading-order part of $N(z)$ vanishes at the boundaries of the quiver.
These assumptions will be relaxed and discussed in more detail later. For now we assume
\begin{align}\label{eq:lim}
	N(z)&=\mathcal O(L^2)~, & \lim_{z\rightarrow \lbrace 0,1\rbrace }L^{-2} N(z)&=0~.
\end{align}
Fundamental hypermultiplets will be attached to isolated nodes, such that their total number is $\mathcal O(L)$.
The nodes where $N(z)$ is linear, with no additional fundamentals attached, are balanced.
The kinks of $N(z)$ may be convex (curving away from the real axis) or concave (curving towards the real axis). The nodes at convex kinks have a flavor excess.
Concave kinks need additional fundamental hypermultiplets to bring the number of flavors at least to twice the number of colors.

\subsection{Brane construction}\label{sec:3d-branes}

\begin{figure}
	\begin{tikzpicture}[scale=1.5]
		\foreach \i in {-0.15,0,0.15} \draw[thick] (0,\i) -- (1,\i);
		\foreach \i in {-0.2,-0.066,0.066,0.2} \draw[thick] (1,\i) -- (2,\i);
		\foreach \i in {-0.2,-0.1,0,0.1,0.2} \draw[thick] (2,\i) -- (2.75,\i);
		
		\node at (3.25,0) {\ldots};
		\node[anchor=north] at (3.25,-1) {\ldots};	
		
		\foreach \i in {-0.2,-0.1,0,0.1,0.2} \draw[thick] (3.75,\i) -- (4.5,\i);
		\foreach \i in {-0.2,-0.066,0.066,0.2} \draw[thick] (4.5,\i) -- (5.5,\i);
		
		\foreach \i in {0,1,2,4.5,5.5}{ \draw[fill=gray] (\i,0) ellipse (2pt and 7pt);}
		
		\foreach \i in {-0.05,0.05} \draw (0.5+\i,-0.8) -- +(0,1.6);
		\foreach \i in {0} \draw (1.5+\i,-0.8) -- +(0,1.6);
		
		\foreach \i in {-0.075,0,0.075} \draw (5+\i,-0.8) -- +(0,1.6);
		
		\foreach \i in {-0.05,0.05} \draw (2.375+\i,-0.8) -- +(0,1.6);
		\foreach \i in {-0.05,0.05} \draw (4.125+\i,-0.8) -- +(0,1.6);
		
		\foreach \i in {1,2}{ \node[anchor=north] at ({-0.5+\i},-0.8) {\footnotesize $k_{\i}$};}
		\node[anchor=north] at (5,-0.8) {\footnotesize $k_{L}$};
		
		\node at (0.75,-0.35) {\footnotesize $N_1$};
		\node at (1.75,-0.35) {\footnotesize $N_2$};
		\node at (5.27,-0.35) {\footnotesize $N_{L}$};        
	\end{tikzpicture}
	\caption{Brane construction  for the quiver gauge theories in (\ref{eq:quiver}). The vertical lines represent D5-branes  oriented along the (012456) directions, the ellipses are NS5-branes oriented along (012789) and the horizontal lines are D3-branes extending along (0123).
		\label{fig:brane}}
\end{figure}
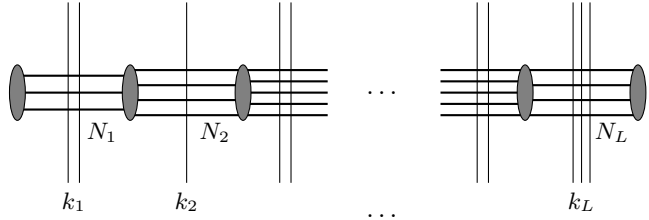

We now review the brane construction of the theories in (\ref{eq:quiver}), following \cite{Hanany:1996ie,Gaiotto:2008ak}, and discuss the limit in (\ref{eq:lim}) from that perspective.
Each $U(N_t)$ gauge node is represented by a stack of $N_t$ D3-branes suspended between NS5 branes as in fig.~\ref{fig:brane}, and the D5-branes represented by vertical lines in fig.~\ref{fig:brane} add fundamental matter.
The limit discussed in the previous section amounts to taking a large number of NS5-branes, $L+1$, with the numbers of D3-branes, $N_t$, of $\mathcal O(L^2)$ such that they fall off towards the boundary nodes.
Having a total of $\mathcal O(L)$ fundamental hypermultiplets distributed over the gauge nodes means that the total number of D5-branes is $\mathcal O(L)$.

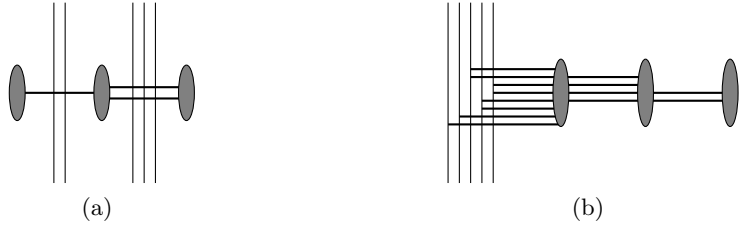
\begin{figure}
\subfigure[][]{
	\begin{tikzpicture}[scale=1.5]
		\draw[thick] (0,0) -- (0.75,0);
		\draw[thick] (0.75,0.05) -- (1.5,0.05);
		\draw[thick] (0.75,-0.05) -- (1.5,-0.05);		
		
		\foreach \i in {0,0.75,1.5}{ \draw[fill=gray] (\i,0) ellipse (2pt and 7pt);}

		\foreach \i in {-1/2,1/2} \draw ({0.375+\i/10},-0.8) -- +(0,1.6);
		\foreach \i in {-1,0,1} \draw ({1.125+\i/10},-0.8) -- +(0,1.6);		
	\end{tikzpicture}
}	\hskip 30mm
\subfigure[][]{\label{fig:branes-2a}
	\begin{tikzpicture}[scale=1.5]
		\draw[thick] (-0.6,0) -- (1.5,0);
		\draw[thick] (-0.6,0.07) -- (0.75,0.07);			
		
		\draw[thick] (-0.7,-0.07) -- (1.5,-0.07);						
		\draw[thick] (-0.7,-0.14) -- (0,-0.14);
		
		\draw[thick] (-0.8,0.14) -- (0.75,0.14);
		\draw[thick] (-0.8,0.21) -- (0,0.21);			
		
		\draw[thick] (-0.9,-0.21) -- (0,-0.21);
		\draw[thick] (-1.0,-0.28) -- (0,-0.28);			
		
		\foreach \i in {0,0.75,1.5}{ \draw[fill=gray] (\i,0) ellipse (2pt and 8.5pt);}
		
		\foreach \i in {0,...,4} \draw ({-1+\i/10},-0.8) -- +(0,1.6);
	\end{tikzpicture}
}
	\caption{Moving all D5-branes to one side, for the quiver $[2]-(1)-(2)-[3]$. Each D5-brane has as many D3-branes attached as it has crossed NS5-branes.
		The partitions are $\rho=[4,2,2]$, $\sigma=[2,2,2,1,1]$.
		The vertical positions of the D3-brane parametrize the Coulomb branch and do not affect the IR SCFT.
		\label{fig:branes-2}}
\end{figure}

One can bring all D5-branes to one side and all NS5-branes to the other side using Hanany-Witten transitions, as shown for an example in fig.~\ref{fig:branes-2}. The gauge theory data is now encoded in two Young tableaux, both encoding partitions of the total number of D3-branes stretched between the stack of D5-branes and the stack of NS5-branes.
One of them, $\rho$, encodes how the D3-branes end on the NS5-branes.
The other one, $\sigma$, encodes how the D3-branes end on the D5-branes. Explicit expressions for the quiver gauge theory data in terms of $(\rho,\sigma)$ can be found e.g.\ in \cite{Nishioka:2011dq,Cremonesi:2014uva}. 
The constraints for having a `good' theory amount to 
\begin{align}\label{eq:cond_good}
	\rho^T>\sigma~.
\end{align}
In words, the sum of the boxes in the first $i$ rows of $\rho^T$ is strictly larger than the same quantity for $\sigma$.
This has to hold for all $i$ up to the number of rows in $\rho^T$. It implies that $\sigma$ has more rows than $\rho^T$. The condition \eqref{eq:cond_good} is equivalent to $\sigma^T>\rho$ so that, when satisfied, both $T_\rho^\sigma [SU(N)]$ and $T_\sigma^\rho [SU(N)]$ are 'good' theories. They are related by mirror symmetry \cite{Intriligator:1996ex} and expected to flow to the same SCFT in the infrared. From the brane perspective, the exchange $\rho \leftrightarrow \sigma$ can be understood as S-duality.

To set the stage for the discussion of supergravity duals it will be useful to make the partitions more explicit: 
Suppose we have $p$ groups of D5-branes, labeled by $a=1,\ldots, p$, with $N_5^{(a)}$ D5-branes in the $a^{\rm th}$  group.
Let the total number of D3-branes ending on the $a^{\rm th}$ group be $N_{3}^{(a)}$, with $N_{3}^{(a)}/N_{5}^{(a)}$ D3-branes ending on each individual D5-brane in that group.
Then
\begin{align}\label{eq:rho}
	\sigma&=\left[\big(N_3^{(1)}/N_5^{(1)}\big)^{N_5^{(1)}},\ldots,\big(N_3^{(p)}/N_5^{(p)}\big)^{N_5^{(p)}}\right],
\end{align}
where the exponent denotes how often an entry is repeated.
One can similarly group the NS5-branes according to the net number of D3-branes ending on each (number of branes emerging to the left minus number of branes emerging to the right).
Let there be $\hat p$ groups of NS5-branes, labeled by $b=1,\ldots, \hat p$, with $\hat N_5^{(b)}$ NS5-branes in the $b^{\rm th}$ group. Let the total number of D3-branes ending on the $b^{\rm th}$ group be $\hat N_3^{(b)}$. Then
\begin{align}\label{eq:sigma}
	\rho&=\left[\big(\hat N_3^{(1)}/\hat N_5^{(1)}\big)^{\hat N_5^{(1)}},\ldots,\big(\hat N_3^{(\hat p)}/\hat N_5^{(\hat p)}\big)^{\hat N_5^{(\hat p)}}\right].
\end{align}
The total number of D3-branes suspended between D5 and NS5 branes, corresponding to $N$ in $T_\rho^\sigma[SU(N)]$,  is $N=\sum_{a=1}^p N_3^{(a)} = \sum_{b=1}^{\hat p}\hat N_3^{(b)}$.

The scaling discussed around (\ref{eq:lim}) can be characterized in terms of $\rho$ and $\sigma$ by having
\begin{align}\label{eq:lim-brane}
	N_5^{(a)}&=\mathcal O(L)~, & \hat N_5^{(b)}&=\mathcal O(L)~,
	\nonumber\\
	N_3^{(a)}&=\mathcal O(L^2)~, & \hat N_3^{(b)}&=\mathcal O(L^2)~.
\end{align}
That is, generically entries in $\rho$ and $\sigma$ are $\mathcal O(L)$ and appear $\mathcal O(L)$ times. We also take $p$ and $\hat p$ to be $\mathcal O(1)$.
The total number of gauge nodes is given by the number of NS5-branes minus one, which is $\mathcal O(L)$. The total number of flavors along the quiver is given by the total number of D5-branes, and also $\mathcal O(L)$. The D5-branes typically have a large number of D3-branes ending on them, so they realize flavors at nodes well in the interior of the quiver.

\subsection{Supergravity duals}\label{sec:sugra}

Holographic duals for the 3d $T_\rho^\sigma[SU(N)]$ theories were constructed in \cite{Assel:2011xz}, building on the general local Type IIB supergravity solutions of \cite{DHoker:2007zhm}.
The geometry is a warped product of $AdS_4$ and two 2-spheres, $S_1^2$ and $S_2^2$, over a Riemann surface $\Sigma$.
The Einstein frame metric and axio-dilaton are
\begin{align}
	ds^2&=f_4^2 ds^2_{AdS_4}+f_1^2 ds^2_{S_1^2}+f_2^2 ds^2_{S_2^2}+4\rho^2 |dz|^2~, 
	& \tau &=\sqrt{\frac{N_1}{N_2}}~.
\end{align}
The 3-form and 5-form field strengths are
\begin{align}
	H_{(3)}&=\vol_{S_1^2}\wedge db_1~, \qquad\qquad
	F_{(3)}=\vol_{S_2^2}\wedge db_2~,
	\nonumber\\
	F_{(5)}&=-4 \vol_{AdS_4}\wedge dj_1+4 f_1^2f_2^2f_4^{-4}\vol_{S_1^2}\wedge \vol_{S_2^2}\wedge (\star_2 dj_1)~,
\end{align}
where $\star_2$ denotes Poincar\'e duality on $\Sigma$.
The solutions are parametrized by a pair of harmonic functions $h_1$, $h_2$ on $\Sigma$.
The metric functions are
\begin{align}
	f_4^8&=16\frac{N_1N_2}{W^2}~, & f_1^8&=16h_1^8\frac{N_2 W^2}{N_1^3}~, & f_2^8&=16 h_2^8 \frac{N_1 W^2}{N_2^3}~,
	&
	\rho^8&=\frac{N_1N_2W^2}{h_1^4h_2^4}~,
\end{align}
where
\begin{align}
	W&=\partial\bar\partial (h_1 h_2)~, & N_i &=2h_1 h_2 |\partial h_i|^2 -h_i^2 W~.
\end{align}
The quantities $b_1$, $b_2$, $j_1$ appearing in the fluxes will not be needed here; they can be found in \cite{Assel:2011xz}.

For the holographic duals of the $T_\rho^\sigma[SU(N)]$ theories, $\Sigma$ is a strip, $\Sigma=\lbrace z| 0\leq \Im(z)\leq \frac{\pi}{2}\rbrace$,
and the harmonic functions are (sec.~4.1 of \cite{Assel:2012cp})
\begin{align}\label{eq:h1h2-gen}
	h_1&=-\sum_{a=1}^p \frac{\alpha^\prime}{4}N_5^{(a)}\ln\left[\tanh\left(\frac{i\pi}{4}-\frac{z-\delta_a}{2}\right)\right]+\rm{c.c.}
	\nonumber\\
	h_2&=-\sum_{b=1}^{\hat p} \frac{\alpha^\prime}{4}\hat N_5^{(b)}\ln\left[\tanh\left(\frac{z-\hat\delta_b}{2}\right)\right]+\rm{c.c.}
\end{align}
On each boundary component of $\Sigma$ one of the two spheres collapses, closing off the internal space smoothly:
$S_1^2$ shrinks at $\Im(z)=0$ and $S_2^2$ at $\Im(z)=\frac{\pi}{2}$.
The points $\Re(z)\rightarrow \pm\infty$ are regular.
The parameters encode the brane configurations as in (\ref{eq:rho}), (\ref{eq:sigma}).
The $p$ D5-brane stacks with $N_5^{(a)}$ D5-branes in the $a^{\rm th}$ stack are at $z=\delta_a+\frac{i\pi}{2}$;
the $\hat p$ NS5-brane stacks with $\hat N_5^{(b)}$ NS5-branes in the $b^{\rm th}$ stack are at $z=\hat \delta_b$.
The locations $\delta_a$ and $\hat\delta_b$ are determined from the conditions
\begin{align}\label{eq:sugra-reg}
	N_3^{(a)}&=N_5^{(a)}\sum_{b=1}^{\hat p}\hat N_5^{(b)}\frac{2}{\pi}\arctan e^{\hat \delta_b-\delta_a}~,
	&
	\hat N_3^{(b)}&=\hat N_5^{(b)}\sum_{a=1}^{p}N_5^{(a)}\frac{2}{\pi}\arctan e^{\hat \delta_b-\delta_a}~.
\end{align}
Summing the first set of conditions over $a$ is equivalent to summing the second set of conditions over $b$.
This ensures that the total number of D3-branes agrees.
With the scaling in (\ref{eq:lim-brane}), the left hand sides of the equations, $N_3^{(a)}$ and $\hat N_3^{(b)}$, are of $\mathcal O(L^2)$, and so are the coefficients $N_5^{(a)}\hat N_5^{(b)}$ on the right hand sides.
Thus, the $\arctan e^{\hat \delta_b-\delta_a}$ factors are generically $\mathcal O(1)$: the brane sources are at finite locations on the upper/lower boundary of the strip, and they are separated by finite amounts.
With $p$ and $\hat p$ of $\mathcal O(1)$ the solutions have a finite number of brane sources.

The free energy can be obtained holographically from the on-shell action. 
The general expression for the $AdS_4\times S_1^2\times S_2^2\times\Sigma$ solutions (see (4.39) and (4.40) of \cite{Assel:2012cp}) reads
\begin{align}\label{eq:SIIB}
	F_{\rm sugra}=	S_{\rm IIB}&=-\frac{32}{\pi^3{\alpha^\prime}^{4}}\int_\Sigma d^2\!z\, h_1h_2\partial\bar\partial(h_1h_2)~.
\end{align}
This will be used to discuss the free energies from the supergravity perspective in sec.~\ref{sec:sample}.
A general evaluation of the free energies obtained from this expression can also be found in  \cite{VanRaamsdonk:2020djx}.

\section{Free energies from localization}\label{sec:F-loc}

The free energy on $S^3$ for $T[SU(N)]$ for general $N$ (not necessarily large) was obtained in \cite{Benvenuti:2011ga,Nishioka:2011dq}. A formula for the more general $T_\rho^\sigma[SU(N)]$ theories was proposed in \cite{Nishioka:2011dq}, and passed several consistency checks. 
Here we work directly in the planar limit and derive explicit expressions for the free energies from the matrix models resulting from supersymmetric localization.

We start by spelling out the continuum formulation of the matrix models for generic long quiver theories of the form (\ref{eq:quiver}) on $S^3$, following \cite{Uhlemann:2019ypp,Coccia:2020cku}.
The matrix models will be formulated in $\cN=2$ language and we will allow for more general $R$-charge assignments than would be allowed by $\cN=4$ supersymmetry. 
We then derive the saddle point equations, with no assumption on the specific scaling of $N(z)$ other than that it is large,
and discuss the general solution for balanced quivers.
Differences to the 5d discussion in \cite{Uhlemann:2019ypp} arise for theories with unbalanced nodes, reflecting the differences in the flavor bounds.

The localized partition function for 3d $\mathcal N=2$ gauge theories was derived in \cite{Jafferis:2010un,Hama:2010av,Hama:2011ea} (for a review see \cite{Willett:2016adv}).
For a theory with $N_f$ chiral multiplets it is given by
\begin{align}\label{eqn:generalpartfunc}
	\cZ  &=
	\frac{1}{\left| W \right|} \int_{\rm Cartan} d\lambda\,
	\prod_{\alpha>0} \left(2 \sinh(\pi\alpha(\lambda))\right)^2\times \prod_{f=1}^{N_f} \prod_{\rho_{f}} e^{\ell\left(1-r_f+i\rho_{f}(\lambda)\right)} ~,
\end{align}
where $\alpha>0$ are the positive roots of the gauge group, $W$ is the Weyl group, $\rho_f$ are the weights of the representations of the chiral multiplets and $r_f$ is their R-charge.
The function $\ell$ is given by
\begin{align}
	\ell(z)&=- z\ln \left(1-e^{2\pi i z}\right)+\frac{i}{2}\left(\pi z^2+\frac{1}{\pi}\Li_2\left(e^{2\pi i z}\right)\right)-\frac{i\pi}{12}~.
\end{align}

In specializing to the theories in (\ref{eq:quiver}) we note that an $\mathcal N=4$ vector multiplet consists of an $\cN=2$ vector and an $\cN=2$ adjoint chiral multiplet, while the $\mathcal N=4$ hypermultiplets correspond to pairs of $\mathcal N=2$ chiral and anti-chiral multiplets.
Choosing a uniform $R$-charge $r$ for the bifundamental and fundamental fields fixes the $R$-charge of the adjoint chiral multiplets to $\tilde r=2(1-r)$.
Thus,
\begin{align}\label{eqn:generalpartfunc-2}
	\cZ &= \frac{1}{\left| W \right|} \int \left[ \prod_{t=1}^{L} \prod_{i=1}^{N_t} d \lambda_{i}^{(t)}  \right]
	e^{-\cF},
	\nonumber\\
	\cF &=
	\sum_{t=1}^L\sum_{\ell,m=1}^{N_t}F_V\big(\lambda_\ell^{(t)}-\lambda_m^{(t)}\big)
	+\sum_{t=1}^{L-1}\sum_{\ell=1}^{N_t}\sum_{m=1}^{N_{t+1}}F_H\big(\lambda_\ell^{(t)}-\lambda_m^{(t+1)}\big)
	+\sum_{t=1}^L\sum_{\ell=1}^{N_t} k_tF_H\big(\lambda_\ell^{(t)}\big)
	~,
\end{align}
where the $\cN=4$ vector and hypermultiplet contributions are collected in
\begin{align}
	F_V(\lambda)&=-\frac{1}{2}\ln\left(4\sinh^2(\pi \lambda)\right)-\frac{1}{2}\left[\ell(1-\tilde r+i\lambda)+\ell(1-\tilde r-i\lambda)\right]~,
	\nonumber\\
	F_H(\lambda)&=-\ell(1-r+i\lambda)-\ell(1-r-i\lambda)~,
\end{align}
with the first term in $F_V$ understood to vanish for argument zero to implement the product over positive roots in (\ref{eqn:generalpartfunc}).
For $R$-charge $r=\tfrac{1}{2}$, the functions simplify due to  $\ell(\tfrac{1}{2}+i\lambda)+\ell(\frac{1}{2}-i\lambda)=-\frac{1}{2}\ln(4\cosh^2(\pi\lambda))$.
To pass to the continuum description in (\ref{eq:cont-rep}), we introduce an eigenvalue density for each gauge node, $\rho_{t}(\lambda)$,
and a fuction of two effectively continuous variables $\rho(z,\lambda)\equiv \rho_{zL}(\lambda)$.
This allows to combine the contributions from the vector and bifundamental hypermultiplets to form derivatives along $z$.
In parallel to \cite{Uhlemann:2019ypp,Coccia:2020cku}, the integrand $\cF$ becomes
\begin{align}\label{eq:cF-gen-4}
	\cF =\, &
	L\int_0^1 dz \int d\lambda\,d\tilde\lambda\,\cL
	-\frac{1}{2}\!\sum_{z\in\lbrace 0,1\rbrace}\int d\lambda\,d\tilde\lambda\,N(z)^2\rho(z,\lambda)\rho(z,\tilde\lambda)F_H(\lambda-\tilde\lambda)
	\nonumber\\ & 
	+L\int_0^1dz\!\int\! d\lambda\, N(z)\rho(z,\lambda)k(z)F_H(\lambda)~,
\end{align}
where, with $F_0(x)\equiv (F_H(x)+F_V(x))/(2r)^2$,
\begin{align}\label{eq:cL-0}
	\cL&=N(z)^2\rho(z,\lambda)\rho(z,\tilde\lambda)(4r^2)F_0(\lambda-\tilde\lambda)
	-\frac{1}{2L^2}\partial_z\big(N(z)\rho(z,\lambda)\big)\partial_z\big(N(z)\rho(z,\tilde\lambda)\big)F_H(\lambda-\tilde\lambda) \ .
\end{align}
The free energy at large $N$ is given by $\cF$ evaluated on the dominant saddle point,
\begin{align}\label{eq:F-cF}
	F_{S^3}&=-\ln \cZ \approx \cF\big\vert_{\rho=\rho_s}~.
\end{align}
In the following only the behavior of $F_0$ and $F_H$ for large real argument will be needed.
It is given by (see \cite{Coccia:2020cku} for a detailed discussion)
\begin{align}\label{eq:F0FH-asympt}
	F_0(x)&=\frac{\pi}{2}(1-r)\delta(x)~,
	&
	F_H(x)&=2\pi(1-r)|x|~.
\end{align}

We can compare the expression for $\cF$ in (\ref{eq:cF-gen-4}) to the analogous expression for a 5d $SU(\cdot)$ quiver of the form (\ref{eq:quiver}), given in eq.~(2.20) of \cite{Uhlemann:2019ypp} (ignoring the Chern-Simons terms).
The form of the first line in (\ref{eq:cF-gen-4}) is identical to the one in 5d, up to an overall factor of $L^2$.
However, the functions $F_H$ and $F_0$ are different. 
In particular, compared to 5d the scaling of $F_H$ and $F_0$ in (\ref{eq:F0FH-asympt}) is reduced.
The contribution of flavors in the second line of (\ref{eq:cF-gen-4}) again takes the same general form as in 5d, up to the same factor $L^2$ that appeared in the first line.

Crucially, the relation between $F_H$ and $F_0$ is identical in 3d and 5d, and in both cases given by
\begin{align}\label{eq:F0-FH}
	8F_0(x)&=F_H^{\prime\prime}(x)~.
\end{align}
As a result, the scaling of the eigenvalues is identical: the scaling is determined by balancing the two terms in $\cL$, for which the relation between $F_H$ and $F_0$ is crucial. The scaling of $N(z)$ does not enter, since both terms in $\cL$ are quadratic in $N(z)$. The eigenvalues thus  scale linearly with $L$.
We introduce new variables $x$ of $\mathcal O(1)$, defined by
\begin{align}\label{eq:lambda-x-def}
	\lambda&=2r L x~,
\end{align}
where $\mathcal O(1)$ factors were included to simplify the dependence on $r$.
The properly normalized density for $x$ is defined by $\hat\rho(z,x)dx=\rho(z,\lambda)d\lambda$.
It is actually convenient to further introduce 
\begin{align}
	\varrho(z,x)\equiv N(z)\hat\rho(z,x)~,
\end{align}
which encodes the densities normalized to $N(z)$.
Then (\ref{eq:cF-gen-4}) with (\ref{eq:cL-0}) simplifies to
\begin{align}\label{eq:cF-gen-5}
	\frac{\cF}{2r} =\, &
	\int_0^1 dz \int dx \,dy\,\mathcal \cL 
	-\frac{1}{2}L\!\sum_{z\in\lbrace 0,1\rbrace}\int dx\,dy\,\varrho(z,x)\varrho(z,y)F_H(x-y)
	\nonumber\\ & 
	+L^2\int_0^1dz\,\left[\int\! dx\, \varrho(z,x)k(z)F_H(x)+\mu(z) \left(\int dx \varrho(z,x)-N(z)\right)\right],
\end{align}
where a Lagrange multiplier has been added to enforce the correct normalization of $\varrho$, and 
\begin{align}\label{eq:cL}
	\cL&=\varrho(z,x)\varrho(z,y)F_0(x-y)
	-\frac{1}{2}\partial_z \varrho(z,x)\partial_z\varrho(z,y)F_H(x-y)~.
\end{align}

The next step is to discuss the extremality conditions.
The identical relation between $F_H$ and $F_0$ in (\ref{eq:F0-FH}) implies that the local saddle point equation is identical to the one in 5d, and given by
\begin{align}\label{eq:varrho-saddle-eq}
	\frac{1}{4}\partial_x^2\varrho(z,x)+\partial_z^2\varrho(z,x)+L\delta(x)\sum_{t=1}^L k_t\delta(z-z_t)&=0~.
\end{align}
All three terms are of the same order for the scalings discussed around (\ref{eq:lim}), i.e.\ for $N(z)=\mathcal O(L^2)$ and $k_t$ of $\mathcal O(L)$.
The derivation of the boundary conditions at $z=0$ and $z=1$ also proceeds in parallel to the 5d case and we refer to \cite{Uhlemann:2019ypp} for details. Assuming that the number of flavors at the boundary nodes is of the same order as the rank of the boundary gauge group, they are given by
\begin{align}\label{eq:rho-bc-gen}
	\varrho(z_b,x)&=N(z_b)\delta(x)~, \qquad z_b\in\lbrace 0,1\rbrace~.
\end{align}
For the quivers described in sec.~\ref{sec:quivers}, $N(z)$ is subleading at the boundary nodes compared to generic $z$.
For the leading-order results it is thus sufficient to impose vanishing Dirichlet boundary conditions.

\subsection{Junction conditions}\label{sec:junction}

The remaining ingredient are the junction conditions at unbalanced nodes, which differ between 3d and 5d.
The expression for $\cF$ in terms of $F_H$ and $F_0$ is identical in 3d and 5d (up to an overall factor $L^2$).
The boundary term at an interior unbalanced node $z=z_t$ resulting from variation of $\cF$ is analogous to (2.46) of \cite{Uhlemann:2019ypp},
and given by
\begin{align}\label{eq:delta-cF-junction-2}
	\delta \cF&=2r\int dx\,\delta\varrho(z_t,x)\left[\int dy\,\left[\partial_z \varrho(z,y)\right]_{z=z_t-\epsilon}^{z=z_t+\epsilon}F_H(x-y)+  Lk_tF_H(x)+L\mu_t\right]\,.
\end{align}
From the requirement for this variation to vanish we find the necessary junction condition
\begin{align}\label{eq:junction-3}
	T(x)\equiv\int dy\,\left[\partial_z \varrho(z,y)\right]_{z=z_t-\epsilon}^{z=z_t+\epsilon}F_H(x-y)+  Lk_tF_H(x)+L\mu_t&\stackrel{!}{=}0~.
\end{align}
For large $|x|$ the condition simplifies due to  $F_H(x-y)\approx F_H(x)$, which grows linearly. 
The integral over $y$ can be performed in the leading term, and one finds that the condition $T(x)=0$ is consistent at large $|x|$ only if $[\partial_z N(z)]_{z_t-\epsilon}^{z_t+\epsilon}=-L k_t$. This is precisely the requirement for the node at $z_t$ to be balanced.
If the node is not balanced, the support of $\varrho$ needs to be constrained,
\begin{align}
	\varrho(z_t,x)=0 \qquad \text{for $x\notin (x_-,x_+)$}~.
\end{align}
This goes along with the variations being constrained to $(x_-,x_+)$ and only requires (\ref{eq:junction-3}) to be satisfied on that interval.

One way to evaluate the condition in (\ref{eq:junction-3}) is to note that upon taking three derivatives w.r.t.\ $x$ the left hand side vanishes, $T^{\prime\prime\prime}(x)=0$. This allows to express $T(x)$ as a polynomial of degree $2$ in $x$ and leads to $3$ conditions.
The condition (\ref{eq:junction-3}) takes the same general form as in 5d; however, since $F_H$ scales differently in 5d one obtains $5$ conditions.
This is to be contrasted with only one less parameter in 3d, due to a missing Lagrange multiplier (since the gauge nodes are $U(\cdot)$ in 3d as opposed to $SU(\cdot)$ in 5d).
We show in app.~\ref{app:junction} that the condition (\ref{eq:junction-3}) merely fixes $\mu_t$.
However, the condition (\ref{eq:junction-3}) in the interior of the interval $(x_-,x_+)$ is not in general sufficient for the variation (\ref{eq:delta-cF-junction-2}) to vanish.
Rather, (\ref{eq:junction-3}) has to be interpreted in a distributional sense -- what has to vanish is $T(x)$ is integrated against $\delta\varrho$ as in (\ref{eq:delta-cF-junction-2}). 
As also derived explicitly in app.~\ref{app:junction}, this leads to an additional requirement constraining the allowed singularities at the end points,
\begin{align}\label{eq:junction-sing}
	\lim_{x\rightarrow x_\pm}\varrho(z_t,x)\sqrt{x-x_\pm}&=0~.
\end{align}
This is a stronger requirement than in 5d, where the corresponding condition (see app.~\ref{app:junction}) is $\lim_{x\rightarrow x_\pm}(x-x_\pm)^{3/2}\varrho(z_t,x)=0$ and square root singularities are allowed. 
This adds a constraint in 3d compared to 5d and balances the counting.
The differences in the junction conditions reflect the different flavor bounds in 3d and 5d, as will be discussed in more detail below.

\subsection{Balanced quivers}\label{sec:balanced}

The saddle points and free energies for generic quivers with all nodes balanced can be obtained straightforwardly.
The derivation proceeds in parallel to the discussion for 5d theories in sec.~3 of \cite{Uhlemann:2019ypp}, leading up to the solution $\varrho_s$ encoding the saddle point eigenvalue densities given in (3.14) there. Assuming fundamental flavors at a finite number of nodes $z_t$,
\begin{align}\label{eq:varrho-s}
	\varrho_s(z,x) = \,&
	\frac{N(0)\sin(\pi z)}{\cosh(2\pi x)-\cos(\pi z)}+\frac{N(1)\sin(\pi z)}{\cosh(2\pi x)+\cos(\pi z)}
	\nonumber\\&
	-\frac{L}{2\pi}\sum_{t=2}^{L-1} k_t
	\ln \left(\frac{\cosh (2 \pi  x)-\cos\left(\pi(z-z_t)\right)}{\cosh (2 \pi  x)-\cos\left(\pi(z+z_t)\right)}\right)
	~, & z_t&=\frac{t}{L}~.
\end{align}
This encodes the normalized densities for $x$ defined in (\ref{eq:lambda-x-def}), $\hat\rho_s(z,x)$, via $\varrho_s(z,x)=N(z)\hat \rho_s(z,x)$.

Differences to the 5d discussion arise in the evaluation of the free energies, due to the different scalings of $F_H$ and $F_0$, which leads to subtleties at the boundary nodes.
Evaluating $\cF$ in (\ref{eq:cF-gen-4}) using integration by parts in $\cL$ and the saddle point equation (\ref{eq:varrho-saddle-eq}) leads to 
\begin{align}\label{eq:cF-balanced-0}
	\cF\big\vert_{\varrho=\varrho_s} &=
	-r\int dx\,\left[N(z)\partial_z\varrho_s(z,x)\right]_{z=0}^{z=1}F_H(x)
	+rL\sum_{t=2}^{L-1}k_t \int dx\, \varrho_s(z_t,x)F_H(x)~.
\end{align}
For the theories discussed in sec.~\ref{sec:quivers} with the scalings as in (\ref{eq:lim}), the first term in (\ref{eq:cF-balanced-0}) is $\mathcal O(L^3)$ and subleading with respect to the second term, which is $\mathcal O(L^4)$.
Among the contributions to $\varrho_s$ in (\ref{eq:varrho-s}), only the second line contributes to the leading-order result for $\cF$, which evaluates to
\begin{align}
	\label{eq:cF-balanced}
	\cF\big\vert_{\varrho=\varrho_s} &=
	-\frac{rL^2}{2\pi}\sum_{s,t=2}^{L-1}k_tk_s\int dx\,F_H(x) \ln \left(\frac{\cosh (2 \pi  x)-\cos\left(\pi(z_t-z_s)\right)}{\cosh (2 \pi  x)-\cos\left(\pi(z_t+z_s)\right)}\right)
	\nonumber\\
	&=
	-\frac{r(1-r)}{\pi^2}L^2\sum_{s,t=2}^{L-1}k_tk_s \Re\left[\Li_3\left(e^{i\pi (z_s+z_t)}\right) -\Li_3\left(e^{i\pi(z_s-z_t)}\right)\right]~.
\end{align}
Via (\ref{eq:F-cF}) this yields the general free energy for balanced quivers of the form (\ref{eq:quiver}) with the scaling (\ref{eq:lim}).
With the $k_t$ of $\mathcal O(L)$, $\cF$ scales like $L^4$, i.e.\ quartic in the length of the quiver and quadratic in the ranks of the largest gauge nodes.
The matrix models have $\mathcal O(L^3)$ integration variables, so the issue with the validity of the large-$N$ approximation discussed in \cite{Coccia:2020cku} does not arise.

The logarithmic scaling found for the theories of \cite{Assel:2012cp} can also be understood from the expressions above.
The theories of \cite{Assel:2012cp} have large numbers of flavors at the boundary nodes, and na\"ively the two terms in (\ref{eq:cF-balanced-0}) are of the same order.
However, as discussed in \cite{Coccia:2020cku} for the $T[SU(N)]$ theory, the integral in the first term of (\ref{eq:cF-balanced-0}) is actually divergent in that case.
The integral was regularized in \cite{Coccia:2020cku} by introducing a cut-off replacing $\left[\,\cdot\,\right]_{z=0}^{z=1}\rightarrow \left[\,\cdot\,\right]_{z=1/L}^{z=1-1/L}$, which results in a logarithmic enhancement of the naive scaling with the leading term independent of the precise choice of cut-off.
It makes the first term in (\ref{eq:cF-balanced-0}) the leading contribution and reproduces the result of \cite{Assel:2012cp}.
The results can also be recovered from the expression in (\ref{eq:cF-balanced}), as will be discussed in detail in sec.~\ref{sec:N2LogN-limit}.

\section{Topologically twisted indices}\label{sec:index}

Let us now consider the topologically twisted index, namely the partition function on  $\Sigma_{\mathfrak{g}} \times S^1$ with a topological twist on the Riemann surface $\Sigma_{\mathfrak{g}}$, where $\mathfrak{g}$ denotes the genus of the surface. 
The topologically twisted index is expressed in terms of complex fugacities $y$ for the global symmetries and a set of integer magnetic fluxes $\mathfrak{n}$ on  $\Sigma_{\mathfrak{g}}$, parametrizing inequivalent twists. 
Using the localization results of \cite{Benini:2015noa,Benini:2016hjo,Closset:2016arn}, the index for a $\mathcal{N}\ge 2$ theory with gauge group $G$ can be written as
\begin{equation}\label{eq:index_1}
	Z_{\Sigma_\mathfrak{g} \times S^1}=\frac{1}{\abs{W}}\sum_{\mathfrak{m}\in \Gamma}\oint_{\cC} \cZ(\lambda,y,\mathfrak{m},\mathfrak{n})\left(\det \frac{\partial^2 \log \cZ (\lambda,y,\mathfrak{m},\mathfrak{n})}{\partial i u \partial \mathfrak{m}} \right)^{\mathfrak{g}} \ ,
\end{equation}
where the sum is over magnetic fluxes in the co-root lattice $\Gamma$ of $G$. With $\lambda=e^{iu}$, we have
\begin{equation}
	\cZ = \prod_{\text{Cartan}} (i du) \prod_{\alpha \in G}(1-\lambda^{\alpha})^{1-\mathfrak{g}}\prod_I\prod_{\rho_I \in \mathfrak{R}_I} \left( \frac{\lambda^{\rho_I/2}y_I^{1/2}}{1-\lambda^{\rho_I}y_I}\right)^{\rho_I(\mathfrak{m})-\mathfrak{n}_I+1-\mathfrak{g}} \ .
\end{equation}
In this expression, $\alpha$ are the roots of $G$ and $I$ labels the chiral fields in the theory, transforming in the representation $\mathfrak{R}_I$ of $G$ which has weights $\rho_I$. The integration contour in \eqref{eq:index_1} can be formulated in terms of Jeffrey-Kirwan residues \cite{Benini:2015noa,Benini:2016hjo,Closset:2016arn}. We also choose the parametrization for which, to each chiral field, we associate a fugacity $y_I$ and a flux $\mathfrak{n}_I$. More precisely, in terms of an assignment $\mathfrak{m}^f_a$ for background global symmetries, we have the relation
\begin{equation} \label{eq:cond_flux_gen}
	\mathfrak{n}_I=\mathfrak{m}^f_I+(1-\mathfrak{g})r_I
\end{equation}
with $r_I$ the R-charge of the chiral field. Therefore, the requirement for the superpotential to be invariant under the global symmetries and to have charge $2$ under R-symmetry results in
\begin{equation}
	\sum_I \mathfrak{n}_I=2(1-\mathfrak{g})~,
\end{equation}
where the sum runs over each monomial term in the superpotential. Following \cite{Benini:2016hjo,Closset:2016arn}, one can rewrite the index as
\begin{equation}\label{eq:Z_general}
	Z_{\Sigma_\mathfrak{g} \times S^1}=\frac{(-1)^{\text{rank} \ G}}{\abs{W}}\sum_{\text{saddle}} \cZ \lvert _{\mathfrak{m}=0} \left(\det \frac{\partial^2 \log \cZ}{ \partial \mathfrak{m}\partial iu}\right)^{\mathfrak{g}-1} ,
\end{equation}
where the sum is over the saddle points of the two-dimensional twisted superpotential $\mathcal{W}$, obtained by compactifying the theory on a finite-size circle. We will shortly review the expression of $\mathcal{W}$ for the theories in (\ref{eq:quiver}). The formulation of the twisted index as a sum over critical points of $\mathcal{W}$ has been first derived in the context of the  Gauge/Bethe correspondence \cite{Nekrasov:2009uh,Okuda:2012nx,Okuda:2013fea,Nekrasov:2014xaa,Okuda:2015yea}. See also \cite{Gukov:2015sna, Closset:2017zgf,Closset:2017bse,Closset:2018ghr}.

The $\cN=4$ theories in (\ref{eq:quiver}) have R-symmetry $SU(2)_H \times SU(2)_C$, so that different topological twists can be realized, with different choices of the $\mathcal{N}=2$  R-symmetry  $U(1)$. They leave an additional global $U(1)$ symmetry with an associated fugacity and a flux (see \cite{Closset:2016arn} for a discussion). 
We will keep $\mathfrak{n}_I$ generic, without fixing the R-charge in \eqref{eq:cond_flux_gen}, so that we do not restrict to a particular twist. More precisely our choice is to assign a uniform flux $\mathfrak{n}$ to each $\cN=2$ chiral coming from the $\mathcal{N}=4$ hypermultiplets and a flux $\tilde{\mathfrak{n}}$ to each adjoint chiral, constrained by
\begin{equation} \label{eq:cond_flux}
	2\mathfrak{n}+\tilde{\mathfrak{n}}=2(1-\mathfrak{g}) \ .
\end{equation}
Similarly, we associate a uniform fugacity $y=e^{i\Delta}$ to each chiral coming from the $\mathcal{N}=4$ hypermultiplets. 
Due to the $\mathcal{N}=4$ superpotential this fixes the fugacity $\tilde{y}$ for the adjoint chirals:
\begin{align}
	y^2 \tilde{y}&=1 \ ,& \tilde{\Delta}+2 \Delta&=2 \pi \ ,
\end{align}
where we choose the phases such that $0 \le \Delta, \tilde{\Delta} < 2 \pi$.

In principle, more general assignments of fugacities and fluxes are possible. For example, we could associate fugacities $y_f$ and $y_f'$ to the fundamentals and antifundamentals appearing at nodes with flavors, together with fluxes $\mathfrak{n}_f$ and $\mathfrak{n}_f'$. However, those contributions would be subleading in the large $N$ limit, due to the conditions $y_f y_f'=1$ and $\mathfrak{n}_f=-\mathfrak{n}_f'$ imposed by the superpotential.\footnote{Indeed, writing $y_f=e^{i\Delta_f}$, $y_f'=e^{i\Delta_f'}$ at each node, in the large $N$ limit and for each pair of fundamental and anti-fundamental (see for example Eq. (A.34) in \cite{Hosseini:2016tor}), the leading contribution to the twisted superpotential is
	\begin{equation}
		\mathcal{W}^{\text{(anti)-fund}}\sim \int dx \rho_t(x)\abs{x}\left[\Delta_f+\Delta_f' \right]=\int dx \rho_t(x)\abs{x}\left[\Delta_f-\Delta_f \right]=0 \ .
	\end{equation}
	A similar argument holds for the contribution to the topologically twisted index, using the relation $\mathfrak{n}_f=-\mathfrak{n}_f'$.}

With these conventions, the twisted superpotential for the long quivers in (\ref{eq:quiver}) can be written in a form analogous to the form found for the free energy on $S^3$ in \eqref{eq:cF-gen-5}. Indeed, retracing the procedure discussed in Section \ref{sec:F-loc} and, performing the scaling
\begin{equation}
	u=2 i L \Delta x~,
\end{equation}
leads to \cite{Coccia:2020cku} 
\begin{equation}\label{eq:W}
	\begin{split}
		\frac{\mathcal{W}}{i\Delta}=& \int_0^1 dz\int dx dy\left[\varrho(z,x)\varrho(z,y) V_0(x-y)-\frac{1}{2} V_H(x-y) \partial_z\varrho(z,x) \partial_z \varrho(z,y)  \right] \\
		&-\frac{L}{2}\sum_{z \in \{ 0, 1 \} } \int dx dy \varrho(z,x) \varrho(z,y)V_H(x-y) +L^2\!\int dz\! \int dx  \varrho(z,x) k(z) V_H(x)
	\end{split}
\end{equation}
with
\begin{equation}
	V_0(x) =\frac{\pi}{2}\left(1-\frac{\Delta}{\pi}\right)\delta(x) \ , \qquad  \qquad V_H(x) =2\pi\left(1-\frac{\Delta}{\pi}\right)\abs{x} \ .
\end{equation}
The expression for the twisted superpotential can be obtained from the expression \eqref{eq:cF-gen-5} for the free energy  by the substitution $r \to \Delta/\pi$ as follows
\begin{align}\label{eq:cF-W}
	\mathcal{W}&=\frac{i\pi}{2}\cF\Big\vert_{r\rightarrow \Delta/\pi}~.
\end{align}
Analogous relations between free energy and twisted superpotential have been found in \cite{Hosseini:2016tor,Jain:2019lqb,Jain:2019euv}. 
The expression for $\cW$ in (\ref{eq:W}) depends on $\Delta$ only through an overall factor, since $V_0$ and $V_H$ have the same dependence on $\Delta$.
Similarly, $\cF$ in \eqref{eq:cF-gen-5} depends on $r$ only through an overall factor.
Therefore, the saddle points of $\cW$ and $\cF$ are independent of $r$ and $\Delta$, and identical
\begin{align}\label{eq:rho-cF-W}
	\varrho_{s,\cW}(z,x)&=\varrho_{s,\cF}(z,x)=\varrho_s(z,x)~.
\end{align}

The topologically twisted index can now be obtained from \eqref{eq:Z_general}. The meaningful quantity in the large $N$ limit is the logarithm of the absolute value of the index. Moreover, the determinant in \eqref{eq:Z_general}, for the theories we are considering, turns out to be subleading, using the same argument as in \cite{Coccia:2020cku}. So the expression for the index in the continuum limit can be written, in a straightforward generalization of the results in \cite{Coccia:2020cku}, as
\begin{equation}\label{eq:logZ}
	\begin{split}
		&\ln \abs{Z}_{\Sigma_\mathfrak{g} \times S^1}= \int_0^1 dz \int dx dy \left[ \varrho_s(z,x)\varrho_s(z,y) Z_0(x-y) -\frac{1}{2}\partial_z \varrho_s(z,x) \partial_z \varrho_s(z,y)Z_H (x-y)\right]  \\
		&\hskip 16mm -\frac{L}{2} \sum_{z \in \{ 0,1 \} } \int dx dy \ \varrho_s(z,x) \varrho_s(z,y)Z_H(x-y)+L^2 \int dz \ k(z) \int dx \ \varrho_s (z,x) Z_H(x)
	\end{split}
\end{equation}
with 
\begin{align} 
	Z_0(x) = \frac{1}{2}\left[\mathfrak{n}(-2 \pi+3 \Delta)-\Delta(1-\mathfrak{g})\right]\delta(x) \ , \qquad \qquad 
	Z_H(x) =2\Delta\left[\mathfrak{n}-(1-\mathfrak{g})\right]\abs{x} \ .
\end{align}

\subsection{Index theorem}

Evaluating the twisted index starting from \eqref{eq:logZ} is, in general, not trivial. However, in the spirit of \cite{Hosseini:2016tor}, we can derive an ``index theorem", extending results obtained for other 3d $\mathcal{N} \ge 2$ theories \cite{Hosseini:2016tor, Hosseini:2016ume, Azzurli:2017kxo, Jain:2019lqb,Jain:2019euv} to the theories considered here. This theorem relates $\ln \abs{Z}_{\Sigma_{\mathfrak{g}}\times S^1}$ to the twisted superpotential evaluated on the saddle point configuration (denoted by $\overline{\mathcal{W}}$) by
\begin{equation}\label{eq:index_th}
	\ln \abs{Z}_{\Sigma_{\mathfrak{g}}\times S^1}=(1-\mathfrak{g})\left(\frac{2 i}{\pi}\overline{\mathcal{W}}(\Delta)+i\left(\frac{\mathfrak{n}}{1-\mathfrak{g}}-\frac{\Delta}{\pi} \right)\frac{\partial \overline{\mathcal{W}}(\Delta)}{\partial \Delta} \right) \ .
\end{equation}
The argument of  \cite{Hosseini:2016tor} is based on promoting, in $\overline{\mathcal{W}}$, the explicit factors of $\pi$ to a formal variable. 
The important observation for the theories in \cite{Hosseini:2016tor} is that, as a function of $\pi$ and the chemical potentials, $\overline{\cW}$ is homogenous of degree 2. This is in general not true for the theories considered here. But we can follow a similar approach to establish the relation (\ref{eq:index_th}) for the theories discussed in sec.~\ref{sec:quivers}. 

We start from the expressions \eqref{eq:W} and \eqref{eq:logZ}, and define a ``deformed" version of the twisted superpotential \eqref{eq:W}, with a parameter $a$ such that $a=1$ corresponds to the expression in  \eqref{eq:W},
\begin{equation}\label{eq:W_deg2} 
	\begin{split}
		\mathcal{W}_a= i\Delta(\pi a-&\Delta)\Bigg[ \int_0^1 dz\int dx dy\left(\frac{1}{2} \varrho(z,x)\varrho(z,y)\delta(x-y)- \abs{x-y} \partial_z \varrho(z,x) \partial_z \varrho(z,y)  \right) \\
		& -L\sum_{z \in \{ 0, 1 \} } \int dx dy \ \varrho(z,x)\varrho(z,y)\abs{x-y} +2L^2\int dz \ k(z) \int dx \ \varrho(z,x)\abs{x}  \Bigg] \ . 
	\end{split}
\end{equation}
The dependence on $\Delta$ and $a$ has been isolated as an overall factor. The saddle point configuration $\varrho_{s,\cW_a}$ is therefore independent of $\Delta$ and $a$ (analogously to (\ref{eq:rho-cF-W})). The twisted superpotential $\cW_a$ evaluated on $\varrho_{s,\cW_a}$ thus is a homogeneous function of degree $2$ in $a$ and $\Delta$, and satisfies
\begin{equation}\label{eq:hom_deg_2}
	\Delta\frac{\partial \overline{\mathcal{W}}_a}{\partial \Delta}+a\frac{\partial \overline{\mathcal{W}}_a}{\partial a}=2 \overline{\mathcal{W}}_a \ .
\end{equation}

With this relation in hand we move on to rescale $x \to x/2 \Delta$, resulting in
\begin{align} \label{eq:W_deformed}
	\frac{\mathcal{W}_a}{i}=&\, \int_0^1 dz\int dx dy\left[\hat{\varrho}(z,x)\hat{\varrho}(z,y) V_0^{(a)}(x-y)-\frac{1}{2} V_H^{(a)}(x-y) \partial_z\hat{\varrho}(z,x) \partial_z \hat{\varrho}(z,y)  \right] 
	\nonumber\\
	&-\frac{L}{2}\!\sum_{z \in \{ 0, 1 \} } \int\! dx dy \, \hat{\varrho}(z,x) \hat{\varrho}(z,y)V_H^{(a)}(x-y) +L^2\!\int \!dz\!  \int \! dx \, \hat{\varrho}(z,x) k(z) V_H^{(a)}(x)  
\end{align}
where $\hat{\varrho}(z,x)$ is the eigenvalue density in the new variables and 
\begin{align}
	V_0^{(a)}(x) &= \pi \Delta^2 \left(a-\frac{\Delta}{\pi}\right)\delta(x) \ , 
	&
	V_H^{(a)}(x) &= \pi \left(a-\frac{\Delta}{\pi}\right)\abs{x} \ .
\end{align}
With the rescaling $x \to x/2 \Delta$ the expression for $\ln\abs{Z}_{\Sigma_{\mathfrak{g}}\times S^1}$ in \eqref{eq:logZ} similarly becomes
\begin{align}\label{eq:logZ_reparametr}
	\ln\abs{Z}_{\Sigma_{\mathfrak{g}}\times S^1}=\,& \int_0^1 dz\int dx dy\Bigg[\hat{\varrho}_s(z,x)\hat{\varrho}_s(z,y) \hat{Z}_0(x-y)-\frac{1}{2} \hat{Z}_H(x-y) \partial_z\hat{\varrho}_s (z,x) \partial_z \hat{\varrho}_s(z,y)  \Bigg] 
	\nonumber\\
	&-\frac{L}{2}\!\sum_{z \in \{ 0, 1 \} } \int\! dx dy\, \hat{\varrho}_s(z,x) \hat{\varrho}_s(z,y)\hat{Z}_H(x-y) +L^2\!\int \! dz \, k(z)\! \int\! dx \, \hat{Z}_H(x) \hat{\varrho}_s(z,x) 
\end{align}
with
\begin{align} 
	\hat{Z}_0(x) &= \left[\Delta(-2\pi+3\Delta)\mathfrak{n} -\Delta^2(1-\mathfrak{g})\right] \delta(x) \ , 
&
	\hat{Z}_H(x) &=(\mathfrak{n}-1+\mathfrak{g})\abs{x} \ .
\end{align}
The functions $V_0^{(a)}$, $V_H^{(a)}$ entering the twisted superpotential $\cW_a$ are related to the functions $\hat Z_0$, $\hat Z_H$ entering the expression for $\ln\abs{Z}_{\Sigma_{\mathfrak{g}}\times S^1}$ by
\begin{align}
	(1-\mathfrak{g})\left[-\frac{1}{\pi}\frac{\partial V_{0/H}^{(a)} }{\partial a}-\left(\frac{\mathfrak{n}}{1-\mathfrak{g}}\right)\frac{\partial V_{0/H}^{(a)}}{\partial \Delta}\right]_{a=1}= \hat{Z}_{0/H}(x) \ ,
\end{align}
so that on shell (namely, on the saddle point configuration)
\begin{equation}
	(1-\mathfrak{g})\left(\frac{i}{\pi}\frac{\partial \mathcal{W}_a }{\partial a}+i\left(\frac{\mathfrak{n}}{1-\mathfrak{g}}\right)\frac{\partial \mathcal{W}_a}{\partial \Delta}\right)_{a=1}= \log \abs{Z}_{\Sigma_{\mathfrak{g}}\times S^1}+\underbrace{\frac{i}{\pi}\frac{\partial \mathcal{W}}{\partial \hat{\varrho}}\frac{\partial \hat{\varrho}}{\partial a}}_{=0 \ \text{on shell}} +i\left(\frac{\mathfrak{n}}{1-\mathfrak{g}}\right)\underbrace{\frac{\partial \mathcal{W}}{\partial \hat{\varrho}}\frac{\partial \hat{\varrho}}{\partial \Delta}}_{=0 \ \text{on shell}} \ .
\end{equation}
Using, in the left hand side, eq. \eqref{eq:hom_deg_2}, we can finally write
\begin{equation}
	\ln \abs{Z}_{\Sigma_{\mathfrak{g}}\times S^1}=(1-\mathfrak{g})\left(\frac{2 i}{\pi}\overline{\mathcal{W}}(\Delta)+i\left(\frac{\mathfrak{n}}{1-\mathfrak{g}}-\frac{\Delta}{\pi} \right)\frac{\partial \overline{\mathcal{W}}}{\partial \Delta} \right) \ .
\end{equation}
For $T[SU(N)]$ this relation has been verified in \cite{Coccia:2020cku}. 
Note that we did not use the explicit form of the saddle point eigenvalue density. Finally, using (\ref{eq:cF-W}) and (\ref{eq:rho-cF-W}), the index theorem can be expressed in terms of the free energy.
With the dependence of the free energy on $r$ explicitly indicated as $F_{S^3}=F_{S^3}(r)$, the relation takes the form
\begin{equation}\label{eq:index-thm}
	\ln \abs{Z}_{\Sigma_{\mathfrak{g}}\times S^1}=(1-\mathfrak{g})\left(-F_{S^3}\left(\frac{\Delta}{\pi}\right)-\frac{\pi}{2}\left(\frac{\mathfrak{n}}{1-\mathfrak{g}}-\frac{\Delta}{\pi} \right)\frac{\partial F_{S^3}(\Delta/\pi)}{\partial \Delta} \right) \ .
\end{equation}
Moreover, as clear from the expression \eqref{eq:cF-gen-5}, the free energy only depends on $r$ through an overall factor $r(1-r)$. 
Hence, we can rewrite the relation \eqref{eq:index-thm} as
\begin{equation}
	\ln \abs{Z}_{\Sigma_{\mathfrak{g}}\times S^1}=\frac{\pi\left(\mathfrak{n}(\pi-2\Delta)+\Delta(1-\mathfrak{g}) \right)}{2\Delta(\Delta-\pi)}F_{S^3}\left(\frac{\Delta}{\pi} \right) \ .
\end{equation}
This general expression will be applied in further concrete theories in sec.~\ref{sec:sample}. Note that for a particular choice of $\Delta, \mathfrak{n}$, the so-called universal twist
\begin{equation}
	\mathfrak{n}=\frac{\bar{\Delta}}{\pi}(1-\mathfrak{g}) \ , \qquad \bar{\Delta}=\frac{\pi}{2}~,
\end{equation}
the relation between twisted index and the free energy simplifies to 
\begin{equation}\label{eq:Z-F-universal}
	\ln \abs{Z}_{\Sigma_{\mathfrak{g}}\times S^1}=(\mathfrak{g}-1) F_{S^3}\left(\frac{\bar{\Delta}}{\pi}\right)~.
\end{equation}

This relation can be interpreted from the holographic perspective. 
Following the insight of \cite{Benini:2015eyy}, the index is expected to account for the entropy of magnetically charged  $AdS_4$ black holes (see \cite{ Zaffaroni:2019dhb} for a review), while the $S^3$ free energy corresponds to the action of a vanilla $AdS_4$ solution. The two are related as in (\ref{eq:Z-F-universal}) for 3d SCFTs whose holographic duals admit a consistent trunction to 4d gauged supergravity \cite{Azzurli:2017kxo}. For the holographic duals of the $T_\rho^\sigma[SU(N)]$ theories, a consistent truncation has not been constructed to our knowledge. But we certainly expect it to exist, in line with the general conjecture of \cite{Gauntlett:2007sm}. This would allow to uplift the solutions of \cite{Romans:1991nq,Caldarelli:1998hg} to asymptotically-$AdS_4$ black hole solutions of Type IIB supergravity and 
the relation (\ref{eq:Z-F-universal}) would explain their entropy.

\section{\texorpdfstring{$T_\rho^\sigma[SU(N)]$}{T[rho,sigma,SU(N)]} case studies}\label{sec:sample}

In this section we study a sample of concrete theories, including theories which have at least one quiver gauge theory description with all nodes balanced, and theories in which neither of the two mirror-dual gauge theory descriptions has all nodes balanced.
When there is a gauge theory description with all nodes balanced, the field theory results can be taken from sec.~\ref{sec:case-balanced}.
For theories with unbalanced nodes, where the difference in the flavor bounds in 3d compared to 5d is crucial, we will illustrate how the differences are reflected in the localization computations.

\subsection{General balanced quivers}\label{sec:case-balanced}

Balanced quivers can be obtained by taking $T_\rho^\sigma[SU(N)]$ with
$N=R_1R_2$, and with $\rho$ and $\sigma$ partitions of $N$ that are given  by
\begin{align}\label{eq:rho-sigma-balanced}
	\rho&=[R_1^{R_2}]~, & \sigma&=[\mathsf{t}_1^{\mathsf{k}_1},\ldots ,\mathsf{t}_\ell^{\mathsf{k}_\ell}]~,
\end{align}
with $\mathsf{t}_a< R_2$ for all $a=1,\ldots \ell$.
That is, there are $R_2$ NS5-branes and each of them has a net number of $R_1$ D3-branes ending on it.
There are $\ell$ groups of D5-branes, with $\mathsf{k}_a$ D5-branes in the $a^{th}$ group on each of which a net number of $\mathsf{t}_a$ D3-branes end.
These brane configurations generally realize balanced quivers.\footnote{%
	In the configuration as in fig.~\ref{fig:branes-2a}, there are $(R_2-t)R_1$ D3-branes between the $t^{\rm th}$ and $(t+1)^{\rm th}$ NS5-branes, and each node considered in isolation would be balanced.
	Moving the D5-branes to the node where they realize flavors does not change the flavor balance:
	If D5-branes pass a pair of NS5-branes, the ranks of the node and its neighbors are reduced equally. 
	If D5-branes end up between a pair of NS5-branes, the rank of the neighboring node to one side is reduced, and a corresponding number of bifundamentals is converted to fundamental flavors.}
If $\ell=1$ the mirror dual is balanced as well; in general the mirror duals are not balanced. 
The limit (\ref{eq:lim}) is realized by taking  $R_1,R_2$ and $\mathsf{t}_a,\mathsf{k}_a$ all of the same order.

To avoid confusion with the notation for similar but not identical quantities in (\ref{eq:quiver}) we have set $(\mathsf{k}_a,\mathsf{t}_a)$ in a different font.
The gauge theory for (\ref{eq:rho-sigma-balanced}) can be characterized as follows:
The total number of nodes is $R_2-1$.
The nodes which have fundamental flavors, in the notation as in (\ref{eq:quiver}), are at $t=\mathsf{t}_a$.
There are $\mathsf{k}_a$ flavors at the node with $t=\mathsf{t}_a$.
The quiver takes the form
\begin{align}\label{eq:balanced-quiver}
	U(\sum_{a=1}^\ell \mathsf{k}_a-R_1)-\ldots-U(&N_{\mathsf{t}_1})-\ldots - U(N_{\mathsf{t}_2})- \ldots \hphantom{-} \ldots  - U(N_{\mathsf{t}_\ell})-\ldots -U(R_1)&&
	\nonumber\\[-3mm]
	&\ \,|\hskip 24mm | \hskip 13.5mm \cdots \hskip 14mm |
	\\
	&[\mathsf{k}_1]\hskip 19mm [\mathsf{k}_2] \hskip 11mm \cdots \hskip 12mm [\mathsf{k}_\ell]
	\nonumber
\end{align}
Between the node with rank $N_{\mathsf{t}_a}$ at $t=\mathsf{t}_a$ and the node with rank $N_{\mathsf{t}_{a+1}}$ at $t=\mathsf{t}_{a+1}$, the rank changes in increments of 
$\Delta N_a$ with
\begin{align}
	\Delta N_a&=-R_1+\sum_{i=a+1}^{\ell}\mathsf{k}_i~.
\end{align}
$\Delta N_a$ can be positive or negative.
Between the first node at $t=1$ and the node at $t=\mathsf{t}_1$ the rank changes by $\Delta N_0$ with the definition above;
between the node at $t=\mathsf{t}_\ell$ and the last node it changes in increments of $-R_1$. With the given assumptions the $\mathsf{t}_a$ are generically well in the interior of the quiver. However, one can allow for $\mathsf{t}_1=1$ and  $\mathsf{t}_\ell=R_2-1$, i.e.\ flavors at boundary nodes.
The tails on the left and right of (\ref{eq:balanced-quiver}) are then absent.

The leading-order saddle point eigenvalue distribution, with the scalings assumed in (\ref{eq:lim}), i.e.\ no $\mathcal O(N^2)$ flavors at the boundary nodes, is given by the second line of (\ref{eq:varrho-s}), 
\begin{align}\label{eq:rho-balanced-rho-sigma}
	\varrho_s(z,x) &=
	-\frac{R_2}{2\pi}\sum_{a=1}^{\ell} \mathsf{k}_a
	\ln \left(\frac{\cosh (2 \pi  x)-\cos\left(\pi(z-\mathsf{z}_a)\right)}{\cosh (2 \pi  x)-\cos\left(\pi(z+\mathsf{z}_a)\right)}\right)
	~, & \mathsf{z}_a&=\frac{\mathsf{t}_a}{R_2}~.
\end{align}
The free energy is given by (\ref{eq:cF-balanced})
\begin{align}\label{eq:cF-balanced-rho-sigma}
	F_{S^3} &=-\frac{r R_2^2}{\pi^2}(1-r)\sum_{a,b=1}^{\ell}\mathsf{k}_a\mathsf{k}_b \Re \left[\Li_3\left(e^{i\pi (\mathsf{z}_a+\mathsf{z}_b)}\right) -\Li_3\left(e^{i\pi(\mathsf{z}_a-\mathsf{z}_b)}\right)\right]~.
\end{align}
With all $\mathsf{k}_a$  of $\mathcal O(R_2)$, the free energy is $\mathcal O(R_2^4)$, which is quartic in the length of the quiver.
In terms of the rank of the gauge group in the 4d $\cN=4$ SYM theory on an interval, $N$, the scaling of the free energy is quadratic.
The topologically twisted index is obtained from (\ref{eq:index-thm}) as
\begin{align}
	\ln \abs{Z}_{\Sigma_{\mathfrak{g}}\times S^1}&=\frac{R_2^2}{2 \pi^3}\left(\Delta(1-\mathfrak{g})+\mathfrak{n}(\pi-2\Delta)\right)\sum_{a,b=1}^{\ell}\mathsf{k}_a\mathsf{k}_b \Re \left[\Li_3\left(e^{i\pi (\mathsf{z}_a+\mathsf{z}_b)}\right) -\Li_3\left(e^{i\pi(\mathsf{z}_a-\mathsf{z}_b)}\right)\right]  .
\end{align}
It exhibits the same scaling and the same dependence on the flavor locations as the free energy, and differs only in an overall factor.

\subsubsection{Supergravity duals}\label{sec:balanced-sugra}

The general supergravity duals for the $T_\rho^\sigma[SU(N)]$ theories with at least one balanced quiver description, as in (\ref{eq:rho-sigma-balanced}), can be spelled out explicitly starting from (\ref{eq:h1h2-gen}). 
The solutions involve one NS5-brane source on the real line, $\hat p=1$, and $\ell$ D5-brane sources, $p=\ell$, on the second boundary component with $\Im(z)=\pi/2$.
The appropriate brane charges are realized for
\begin{align}
	\hat N_5^{(1)}&=R_2~, & \hat N_3^{(1)}&=R_1R_2~,
	\nonumber\\
	N_5^{(a)}&=\mathsf{k}_a~, & N_3^{(a)}&=\mathsf{k}_a \mathsf{t}_a~, & a&=1,\ldots \ell~.
\end{align}
The regularity conditions in (\ref{eq:sugra-reg}) are solved by
\begin{align}
	\hat\delta_1&=0~, &
	\delta_a&=\ln\tan\frac{\pi \mathsf{t}_a}{2R_2}~.
\end{align}
For $\mathsf{t}_a$ and $R_2$ of the same order, the $\delta_a$ are at finite locations.
The functions $h_1$, $h_2$ are given by
\begin{align}\label{eq:hi-balanced}
	h_1&=-\sum_{a=1}^\ell \frac{\alpha^\prime}{4}\mathsf{k}_a\ln\left[\tanh\left(\frac{i\pi}{4}-\frac{z-\delta_a}{2}\right)\right]+\rm{c.c.},
	&
	h_2&=-\frac{\alpha^\prime}{4}R_2\ln\tanh\left(\frac{z}{2}\right)+\rm{c.c.}
\end{align}
The free energy obtained from (\ref{eq:SIIB}) can be expressed as
\begin{align}\label{eq:sugra-F-balanced}
	F_{\rm sugra}&=-\frac{R_2^2}{8\pi^3}\sum_{a,b=1}^\ell \mathsf{k}_a \mathsf{k}_b \int d^2\!z \, X_a \partial\bar\partial X_b~, & X_a&=\ln\left|\tanh\left(\frac{i\pi}{4}-\frac{z-\delta_a}{2}\right)\right|^2 \ln\left|\tanh\frac{z}{2}\right|^2.
\end{align}
With all brane sources at finite locations, i.e.\ $\delta_a$ finite, and the $k_a$ of $\mathcal O(R_2)$, the free energy is $\mathcal O(R_2^4)$.
The matching of this supergravity free energy to (\ref{eq:cF-balanced-rho-sigma}) for $r=\tfrac{1}{2}$
amounts to
\begin{align}\label{eq:sugra-F-match}
	\int d^2\!z \, X_a \partial\bar\partial X_b =2\pi \Re \left[\Li_3\left(e^{i\pi (\mathsf{z}_a+\mathsf{z}_b)}\right) -\Li_3\left(e^{i\pi(\mathsf{z}_a-\mathsf{z}_b)}\right)\right]~,
\end{align}
with $X_a$ defined in (\ref{eq:sugra-F-balanced}) and $\mathsf{z}_a$ in (\ref{eq:rho-balanced-rho-sigma}).
The left hand side is a function of two variables, $\mathsf{t}_a/R_2, \mathsf{t}_b/R_2\in(0,1)$, which one can evaluate numerically to compare to the right hand side. This shows that the relation (\ref{eq:sugra-F-match}) is indeed satisfied.

We thus find that the supergravity free energy computed from (\ref{eq:SIIB}) with $h_1$ and $h_2$ in (\ref{eq:hi-balanced}) precisely matches the field theory free energy in (\ref{eq:cF-balanced-rho-sigma}), including the $L^4$ scaling and the coefficient functions involving trilogarithms.
More generally, 
with the scalings assumed in (\ref{eq:lim-brane}), the functions $h_1$ and $h_2$ are $\mathcal O(L)$, so the free energy in (\ref{eq:SIIB}) is $\mathcal O(L^4)$.
Regarding the validity of supergravity approximation, we note that the residues of $\partial h_i$ at the brane singularities are $\mathcal O(L)$. 
So following the comments in sec.~4.5 of \cite{Assel:2012cp} we expect corrections to the supergravity approximation of $\mathcal O(L^2)$. 
These corrections are strongly subleading now.

\subsection{\texorpdfstring{$T_{R,M,k}[SU(N)]$}{T[R1,R2,M,k]} theories}\label{sec:pretty}

The theories discussed in the previous section have at least one balanced quiver description.
In this section we take a special case of the theories in (\ref{eq:rho-sigma-balanced}) and discuss the localization computations for the unbalanced mirror dual.
This will illustrate how the difference in the flavor bounds in 3d compared to 5d is implemented in the localization computation through the junction conditions.

A minimally non-trivial example, in the sense that the mirror dual has one unbalanced node, is when $\sigma$ involves two groups of D5-branes with different numbers of D3-branes ending on them.
Take $T_\rho^\sigma[SU(N)]$ with $N=RkM$ ($R_1=R$ and $R_2=kM$ in the notation of (\ref{eq:rho-sigma-balanced})) and
\begin{align}
	\rho&=[R^{k M}]~, & \sigma&=[((k-1)M)^{R},M^{R}]~.
\end{align}
We consider the limit where $R$ and $M$ are homogeneously large while $k$ is of order one.
For $k=2$ the partitions become equal and the mirror is balanced as well.
The constraint $\rho^T=[(kM)^{R}]>\sigma$ is satisfied if $k\geq 2$, with $\rho^T-\sigma$ large.

We will discuss the balanced quiver first, and the unbalanced one afterwards.
The balanced UV quiver is
\begin{align}\label{eq:quiver-1}
	U(R)-U(2R)-\ldots -U(&MR)-\ldots -U(MR)-U((M-1)R)-\ldots - U(R)
	\nonumber\\
	&\ \vert\hskip 28mm |
	\nonumber\\
	&\![R\,] \hskip 23mm [R\,]
\end{align}
Along the first ellipsis the rank increases in increments of $R$. Along the second ellipsis the rank is constant, with a total of $(k-2)M+1$ gauge nodes of rank $MR$. Along the third ellipsis the rank decreases in steps of $R$.

For the theory in (\ref{eq:quiver-1}), the number of nodes is $L=kM-1$. It has two flavor contributions from $\mathsf{z}_1=1/k$ and $\mathsf{z}_2=1-1/k$, with $R$ flavors each.
From (\ref{eq:cF-balanced-rho-sigma}), the free energy is
\begin{align}\label{eq:F_quiver-1}
	F_{S^3}&=\frac{ (1-r) r}{2 \pi ^2} (k M R)^2\Re\left(4\Li_3\left(-e^{\frac{2 i \pi }{k}}\right)-4\Li_3\left(e^{\frac{2 i \pi}{k}}\right)+7 \zeta (3)\right) \ .
\end{align}
This shows the quartic scaling with the length of the quiver and the quadratic scaling with $N$.
For $k\in\lbrace 2,3,4,6\rbrace$ the result can be expressed in terms of $\zeta(3)$, but in general this is not possible.
Applying the index theorem (\ref{eq:index-thm}) for the theory in \eqref{eq:quiver-1} yields the topologically twisted index
\begin{equation} \label{eq:Z_quiver-bal}
	\ln \abs{Z}_{\Sigma_{\mathfrak{g}}\times S^1}=\frac{(k M R)^2}{4 \pi^3}\left(\mathfrak{n}(2\Delta-\pi)-(1-\mathfrak{g})\Delta \right)\Re\left(4\Li_3\left(-e^{\frac{2 i \pi }{k}}\right)-4\Li_3\left(e^{\frac{2 i \pi}{k}}\right)+7 \zeta (3)\right) .
\end{equation}
We verified that this expression agrees with the result obtained by directly evaluating \eqref{eq:logZ}.

Before moving on to the mirror dual, we note that a quiver very similar to (\ref{eq:quiver-1}) has been discussed in 5d in \cite{Chaney:2018gjc}:
Aside from the gauge nodes being $U(\cdot)$ instead of $SU(\cdot)$, the quivers for the 5d $+_{N,M,j,k}$ theories take the same form.
Concretely, (\ref{eq:quiver-1}) matches to the 5d $+_{N,M,j,k}$ quiver with $N^{(5d)}=R M$, $M^{(5d)}=kM$, $j^{(5d)}=k^{(5d)}=M$.
Compared to 5d, where the natural planar limit amounts to $N^{(5d)}$ and $M^{(5d)}$ large and of the same order while $j^{(5d)}=k^{(5d)}$ are $\mathcal O(1)$,
the scaling we considered in 3d is different. 
This will be discussed further in sec.~\ref{sec:5d}.

\subsubsection{Localization for mirror dual}\label{sec:bal_unbal}

We now turn to the mirror-dual theory, which is described by the UV quiver
\begin{align}\label{eq:quiver-2}
	U(M)-U(2M)-\ldots - U((R-1)M) - U&(RM) - U((R-1)M) -\ldots U(2M)-U(M)
	\nonumber\\
	&\ \ \vert\nonumber\\
	&\![kM]
\end{align}
The central node is not balanced for $k>2$ and has a flavor excess, as appropriate for `good' theories in the sense of \cite{Gaiotto:2008ak}.
The number of excess flavors is small compared to the rank of the gauge group at the central node and the rank function is concave.

At the central node of the theory in (\ref{eq:quiver-2}) the junction conditions for the saddle point eigenvalue densities corresponding to $z>1/2$ and $z<1/2$ are dictated by (\ref{eq:delta-cF-junction-2}).
The large-$|x|$ behavior of the terms in the square brackets of (\ref{eq:delta-cF-junction-2}) for $k>2$ requires the support of the eigenvalue density at $z=1/2$ to be constrained. The problem is symmetric in $x\rightarrow -x$.
For a saddle point we therefore seek a function $\varrho_s(z,x)$ satisfying the bulk saddle point equation (\ref{eq:varrho-saddle-eq}) with the appropriate sources and the condition $\varrho_s(1/2,x)=0$ for $|x|>x_1$, for some $x_1$ to be determined, along with (\ref{eq:junction-sing}).
It is convenient to map the strip $\lbrace (z,x) \in [0,1]\times\RR\rbrace$ to the complex plane with coordinate $v$ as follows,
\begin{align}\label{eq:v-def}
	v&=\frac{u e^{4\pi x_1}+1}{u+e^{4\pi x_1}}~, & u&=e^{4\pi x+2\pi i z}~.
\end{align}
In the $v$ coordinate $\varrho_s$ has to vanish on the positive real axis, and there is a source at $v=-1$.
It is convenient to unfold this, by considering the upper half plane with coordinate $i\sqrt{-v}$, and vanishing Dirichlet boundary condition on the entire real axis.
This leads to the general expression (cf.\ (3.13) in \cite{Uhlemann:2019ypp})
\begin{align}\label{eq:rho-unbal-gen}
	\varrho_s&=-\frac{1}{2}L \sum_{t=2}^{L-1}k_tG(i\sqrt{-v},i\sqrt{-v_t})~, & G(u,v)&=\frac{1}{\pi}\ln\left|\frac{u-v}{u-\bar v}\right|^2~.
\end{align}
For the theory in (\ref{eq:quiver-2}), with $v$ as defined in (\ref{eq:v-def}),
\begin{align}\label{eq:rho-T-mirror}
	\varrho_s&=-RMk\, G(i\sqrt{-v},i)=-\frac{RMk}{\pi}\ln\left|\frac{1-\sqrt{-v}}{1+\sqrt{-v}}\right|~.
\end{align}
This $\varrho_s$ satisfies the saddle point equation and the constraint in (\ref{eq:junction-sing}).
The value of $x_1$ is fixed by normalization.
One can only find acceptable solutions for $k\geq 2$, which is the localization manifestation of the 3d flavor bound.
For $k\geq 2$,
\begin{align}\label{eq:x1}
	2\pi x_1&=\ln\tan\left(\frac{\pi}{4}+\frac{\pi}{2k}\right)~.
\end{align}
For $k=2$ the quiver is balanced and $x_1=+\infty$.
With this choice $\varrho_s$ satisfies the junction condition (\ref{eq:junction-3}) with
\begin{align}\label{eq:mu-mirror}
	\mu_R&=-\frac{2(1-r)kM}{\pi} D_2\left(i\cot\left(\frac{\pi}{k}\right)\right)~,
\end{align}
where $D_2(z)=\Im(\Li_2(z)+\ln(1-z)\ln|z|)$ is the Bloch-Wigner function.

The free energy is obtained by evaluating $\cF$ in (\ref{eq:cF-gen-5}) on the saddle point $\varrho_s$.
Using integration by parts in $\cL$, one obtains 
\begin{align}
	\cF&=2r\int dx \,\varrho_s\big(\tfrac{1}{2},x\big)\left[
	\int dy\,\frac{1}{2}\left[\partial_z \varrho_s(z,y)\right]_{z=\tfrac{1}{2}-\epsilon}^{z=\tfrac{1}{2}+\epsilon}F_H(x-y)
	+Lk_R F_H(x)\right]~.
\end{align}
Using further the junction condition, this can be simplified to
\begin{align}
	\cF&=r\int dx \varrho_s\big(\tfrac{1}{2},x\big)\left[Lk_R F_H(x)-L\mu_R\right]
	=
	rLk_R\int dx \varrho_s\big(\tfrac{1}{2},x\big) F_H(x)-rL\,N\big(\tfrac{1}{2}\big)\,\mu_R
	~.
\end{align}
With $\mu_R$ in (\ref{eq:mu-mirror}) one finally obtains
\begin{align}
	F_{S^3}&=\frac{k^2 M^2 R^2 (1-r) r}{2 \pi ^2} \left(4\Re\Li_3\left(-e^{\frac{2 i \pi }{k}}\right)-4\Re\Li_3\left(e^{\frac{2 i \pi}{k}}\right)+7 \zeta (3)\right)\,.
\end{align}
This is identical to the result for the mirror-dual in (\ref{eq:F_quiver-1}), as it should be,
thus validating the discussion of the saddle point conditions in sec.~\ref{sec:F-loc} also for unbalanced theories.
The supergravity dual for the mirror theory is given by an S-duality transformation of the solution for the theory in (\ref{eq:quiver-1}), which is a special case of the solutions discussed in sec.~\ref{sec:balanced-sugra}.

The computation shown above highlights the differences to the 5d case, which we discuss briefly.
The theory in (\ref{eq:quiver-2}) has a flavor excess which is small compared to the rank of the gauge group at the central node.
This may be compared to the 5d $X_N$ theory, which has a small flavor deficit at the central node and otherwise a rank function of similar shape.
The free energy for the 5d $X_N$ theory was derived in app.~A of \cite{Uhlemann:2020bek}.
The eigenvalue density ansatz for the 5d $X_N$ theory is
\begin{align}
	\varrho_{X_N}&=\frac{a(1-v)}{\sqrt{-v}}+\mathrm{c.c.}~,
\end{align}
with real $a$. This $\varrho_{X_N}$ satisfies the saddle point equation (\ref{eq:varrho-saddle-eq}) with no sources.
It would thus seem that $\varrho_{X_N}$ could be added to $\varrho_s$ in (\ref{eq:rho-T-mirror}), giving a flat direction: the normalization constraint would now fix $x_1$ in terms of $a$, and allow for $x_1$ to move further inwards at the expense of increasing the admixture of $\varrho_{X_N}$.
However, this is not correct; the free energy would depend on $a$ and there is no flat direction.
The key is the constraint in (\ref{eq:junction-sing}), which, as discussed below (\ref{eq:junction-sing}), is more restrictive than in 5d.
It is not satisfied by $\varrho_{X_N}$, forcing its coefficient to zero for a saddle point in 3d.
In turn, in 5d the junction condition (\ref{eq:junction-3}) is stronger than in 3d and prevents a flavor excess.

\subsection{Unbalanced quiver pairs}\label{sec:unbalanced_couple}
As a further application we consider theories in which both of the mirror-dual gauge theories include unbalanced nodes. In particular, we consider the subclass of $T_\rho^\sigma[SU(N)]$ defined by 
\begin{equation}
	\label{eq:part_unbal-unbal}
	\rho=[R_1^{M_1},R_2^{M_2}] \ , \qquad \sigma=[M_3^{R_3},M_4^{R_4}] \ ,
\end{equation}
with $R_1,\ldots, R_4$ and $M_1,\ldots, M_4$ homogeneously large and $R_1 M_1+R_2M_2=R_3 M_3+R_4M_4 = N$.
As explained in section \ref{sec:3d-branes}, in order to have a good theory we need $\sigma^T > \rho$, with
\begin{equation}
	\sigma^T=\left[(R_3+R_4)^{M_4},R_3^{M_3-M_4} \right] \ .
\end{equation}

The partitions \eqref{eq:part_unbal-unbal} correspond to a quiver of length  $L=M_1+M_2-1$, with an unbalanced node at $t=M_1$ and fundamental flavors which in the notation of (\ref{eq:quiver}) correspond to $k_{M_3}=R_3$, $k_{M_4}=R_4$. 
In the notation of sec.~\ref{sec:case-balanced}, the flavors correspond to
\begin{equation} \label{eq:zandk}
	\begin{split}
		&\mathsf{k}_1=R_4 \qquad \text{at} \quad \mathsf{z}_1=M_4/L \ , \\
		&\mathsf{k}_2=R_3 \qquad \text{at} \quad \mathsf{z}_2=M_3/L \ .
	\end{split}
\end{equation}  
The relative positions of the unbalanced node and the flavored nodes depend on the choice of parameters.
The rank of the gauge group for $t < M_{1,3,4}$, is $N_t=(R_3+R_4-R_1)t$. 
Along the other tail of the quiver, identified by $t > M_{1,3,4}$, the rank decreases with rate $\Delta N= -R_2$ until reaching the boundary rank $N_L=R_2$.  
Since general interior nodes have rank of $\mathcal O(N)$, boundary terms are subleading in the free energy. The general structure of the mirror dual, obtained by swapping $\rho$ and $\sigma$, is completely analogous. In particular the mirror dual has an unbalanced node at $t=R_3$.

In order to find a solution of the saddle point equation which, at the unbalanced node, vanishes outside of an interval $(-x_1,x_1)$, we follow the procedure of the previous section.  Denoting by $\tilde{z}=M_1/L$ the position of the unbalanced node, we perform the change of variables
\begin{equation}
	v=\frac{ue^{4 \pi x_1}+1}{u+e^{4 \pi x_1}} \ , \qquad \qquad u=e^{4 \pi x+2\pi i w} \ ,  \qquad \qquad w(z)=\frac{z(1-\tilde{z})}{z+\tilde{z}-2z \tilde{z}}
\end{equation}
where we chose $w$ in such a way that $w(0)=0, w(1)=1 $ and $ w(\tilde{z})=1/2$. Now the problem is analogous to the one solved in sec.~\ref{sec:bal_unbal} but with sources at
\begin{equation} \label{eq:vt}
	v_a=\frac{u_a e^{4 \pi x_1}+1}{u_a+e^{4 \pi x_1}} \ , \qquad \qquad \qquad  u_a=e^{2\pi i w (\mathsf{z}_a)} 
\end{equation}
and $\mathsf{z}_a$ given in equation \eqref{eq:zandk}.
As before, it is then convenient to consider the half upper plane, using the coordinate $i\sqrt{-v}$, and to read off the density from (\ref{eq:rho-unbal-gen})
\begin{equation} \label{eq:varrho_unb_unb}
	\varrho_s=-\frac{(M_1+M_2)}{2 \pi} \sum_{ a\in \{1, 2 \}} \mathsf{k}_a \ln \abs{\frac{i \sqrt{-v}-i \sqrt{-v_a}}{i \sqrt{-v}+i\sqrt{-\bar v_a}}}^2 \ .
\end{equation} 

The parameter $x_1$ can be fixed by considering the derivative of the rank function on the left end of the quiver, which is
\begin{equation}
	\partial_z N(z) \lvert_{z=0}=\int dx \ \partial_z \varrho_s(z,x) \bigg \lvert _{z=0} =L(R_3+R_4-R_1) \ .
\end{equation}
Solving this integral, one obtains a relation fixing the value of $x_1$
\begin{equation}\label{eq:fix_x1}
	i\frac{(1-\tilde{z})}{\tilde{z}}\sum_{a\in\{ 1,2 \} }\mathsf{k}_a \ln\left(\frac{i\sqrt{-v_a}+e^{2 \pi  x_1}}{1+i e^{2 \pi  x_1}\sqrt{-v_a} } \right)=\pi (R_3+R_4-R_1) \ .
\end{equation}
Since $v_a$ is a function of $x_1$ (see eq \eqref{eq:vt}) this equation can be complicated, and we will provide an explicit solution in particular cases.\footnote{We note that upon substituting the data of the theory in sec.~\ref{sec:bal_unbal}, eq.~\eqref{eq:fix_x1} is correctly solved by \eqref{eq:x1}.}

Once $x_1$ is fixed, the Lagrange multiplier $\mu$ can by determined from the junction condition, as explained in the appendix, and we can compute the free energy via the expression
\begin{equation}\label{eq:F-unb}
	\mathcal{F}=2r \int dx \,dy\, \varrho_s(\tilde{z},x) \frac{1}{2}\left[\partial_z \varrho_s(z,y) \right]^{z=\tilde{z}+\varepsilon}_{z=\tilde{z}-\varepsilon}F_H(x-y)+r L^2 \int dz \, k(z) \int dx \, \varrho_s(z,x) F_H(x)\,.
\end{equation}

\subsubsection{A mirror pair of unbalanced theories} 
To apply the results just obtained in an example, we consider the theory defined by the partitions
\begin{align}
	\rho&=\left[\left(\frac{4}{3}R \right)^M,\left(\frac{2}{3}R \right)^M\right] \ , 
	&
	\sigma&=\left[\left(\frac{3}{2}M \right)^R,\left(\frac{1}{2}M \right)^R \right] \ .
\end{align}
In this case the theory is symmetric with respect to the central node $\tilde{z}=\frac{1}{2}$, which is unbalanced. There are $R $ flavors at $\mathsf{z}_1=1/4$ and $\mathsf{z}_2=3/4$ so that the quiver is
\begin{align}\label{eq:quiver-unb_1}
	U\left(\frac{2 R}{3}\right)-\ldots -U&\left(\frac{MR}{3}\right)-\ldots -U\left(\frac{MR}{6}\right)-\ldots - U\left(\frac{MR}{3}\right)-\ldots - U\left(\frac{2R}{3}\right)
	\nonumber\\
	&\quad \ \  \vert\hskip 62mm |
	\nonumber\\
	&\quad  [R\,] \hskip 58mm [R\,]
\end{align}
Along the first ellipsis the rank of the group increases in steps of $2R/3$, along the second ellipsis it decreases in steps of $R/3$. The other ellipses follow by symmetry.  
Equation \eqref{eq:fix_x1} is solved by
\begin{equation}
	x_1=\frac{1}{2\pi}\ln \left(1+\sqrt{2} \right).
\end{equation}
The free energy obtained from \eqref{eq:F-unb} is given by
\begin{align} \label{eq:F_unb_ex}
	F_{S^3}= \frac{r(1-r)}{2\pi ^2}R^2M^2 \Big[&
16 \cL_3\left(1+\sqrt[6]{-1}\right)-8 \cL_3\left(\sqrt[6]{-1}\right)+16 \cL_3\big(\sqrt[6]{-1} \sqrt{3}\big)+8 \cL_3\big((2+\sqrt{3})i\big)
\nonumber\\&
+8 \cL_3\big(\sqrt[3]{-1} (2+\sqrt{3})\big)-2\cL_3\big(7+4 \sqrt{3}\big)-5 \zeta (3)\Big]\,,
\end{align}
with the single-valued trilogarithm $\cL_3(z)=\Re\left[\Li_3(z)-\ln|z|\,\Li_2(z)-\frac{1}{3}\ln^2\!|z|\,\ln(1-z)\right]$ \cite{Zagier2007}.

Similarly, the mirror dual theory, obtained by exchanging $\rho$ and $\sigma$, has an unbalanced node at $\tilde{z}=\frac{1}{2}$.
The quiver is again symmetric with respect to the unbalanced central node. There are $M$ flavors at $z=1/3$ and  at $z=2/3$, so that we have
\begin{align}\label{eq:quiver-unb_2}
	U\left(\frac{M}{2}\right)-\ldots -U&\left(\frac{MR}{3}\right)-\ldots -U\left(\frac{MR}{6}\right)-\ldots - U\left(\frac{MR}{3}\right)-\ldots - U\left(\frac{M}{2}\right)
	\nonumber\\
	&\quad \ \  \vert\hskip 62mm |
	\nonumber\\
	&\quad  [M\,] \hskip 56mm [M\,]
\end{align}
Along the first ellipses the rank increases in steps of $M/2$ until reaching $z=1/3$; then it decreases in steps of $-M/2$. Equation \eqref{eq:fix_x1} is now solved choosing
\begin{equation}
	x_1=\frac{1}{4 \pi}\ln \left( 2+\sqrt{3}\right),
\end{equation}
and computing the free energy via \eqref{eq:F-unb} reproduces \eqref{eq:F_unb_ex}.

The supergravity duals for these theories can be obtained from the formulas in sec.~\ref{sec:sugra}. The brane charges are
\begin{align}
	N_{3}^{(1)}&=\frac{3}{2}MR~, & 	N_{3}^{(2)}&=\frac{1}{2}MR~, & N_5^{(1)}=N_5^{(2)}&=R~,\nonumber\\
	\hat N_{3}^{(1)}&=\frac{4}{3}MR~, & 	\hat N_{3}^{(2)}&=\frac{1}{3}MR~, & \hat N_5^{(1)}=\hat N_5^{(2)}&=M~.
\end{align}	
The regularity conditions in (\ref{eq:sugra-reg}) are solved by
\begin{align}
	\delta_1&=-\delta_2=\ln\big(\sqrt{3}-\sqrt{2}\big)~, & \hat\delta_1&=-\hat\delta_2=\ln\big(1+\sqrt{2}\big)~.
\end{align}
With these parameters and the functions $h_1$, $h_2$ in (\ref{eq:h1h2-gen}), the expression for the supergravity free energy in (\ref{eq:SIIB}) matches the field theory free energy in (\ref{eq:F_unb_ex}) for $r=1/2$.

\subsection{Theories with \texorpdfstring{$N^2\ln N$}{N**2 ln(N)} scaling}\label{sec:N2LogN-limit}

We close this part with a more detailed discussion of how the general results of \cite{Assel:2012cp} for the theories with $N^2 \ln N$ scaling can be obtained as a limiting case of the free energy for balanced quivers in \eqref{eq:cF-balanced-rho-sigma}.
We start with the $T[SU(N)]$ theory, corresponding to 
\begin{align}
	\rho&=\sigma=[1^N]~.
\end{align}
The total number of D3-branes suspended between the D5 and NS5 branes is $N$,
and the gauge theory reads
\begin{align}
	[N]-(N-1)-(N-2)-\ldots -(1)~.
\end{align}
This theory can be seen as special case of (\ref{eq:rho-sigma-balanced}) with $R_1=1$, $R_2=N$ and $\ell=1$, $\mathsf{t}_1=1$ and $\mathsf{k}_1=N$.
The scaling of the rank function is not of the form (\ref{eq:lim}), but the free energy can nevertheless be recovered from the general expression in (\ref{eq:cF-balanced-rho-sigma}) as a limiting case.
Namely, by setting $\mathsf{z}_1=1/N$ and expanding $\Li_3$ for small argument.
For small real $x$,
\begin{align}\label{eq:Li3-exp}
	\Re\left[\Li_3(e^{ix})\right]&=\zeta(3)+\frac{1}{4}x^2\ln x^2+\mathcal O(x^2)~.
\end{align}
With this expansion the expression in (\ref{eq:cF-balanced-rho-sigma}) leads to
\begin{align}
	F_{S^3}^{T[SU(N)]}&=\frac{1}{2}N^2\ln N~,
\end{align}
in agreement with the result of \cite{Assel:2012cp}. This result was recovered in \cite{Coccia:2020cku} by introducing a cut-off on the quiver coordinate (see the comments at the end of sec.~\ref{sec:balanced}).
The way the result is recovered here may be seen as an alternative regularization procedure, in which the logarithmic enhancement arises as the flavors approach the boundary node.

In the supergravity duals spelled out in sec.~\ref{sec:balanced-sugra}, there is one D5-brane source at $z=\delta_1$, whose location is fixed by the regularity conditions to $\delta_1=\ln \tan(\pi/(2N))$. As $N$ becomes large, $\delta_1\rightarrow -\infty$. In the large-$N$ limit the brane source approaches the point where the two boundary components, on which different $S^2$ collapse,  connect. The volume of the internal space contributes a factor $\ln N$ in that limit, leading to the enhanced scaling also in the holographic result.

The more general theories considered in \cite{Assel:2012cp} can be discussed accordingly. They are $T_\rho^\sigma[SU(N)]$ theories with
\begin{align}\label{eq:rho-sigma-log}
	\rho&=[\hat l^{N\hat\gamma}]~, &
	\sigma&=[(N^{\kappa_1}\lambda^{(1)})^{N^{1-\kappa_1}\gamma_1},\ldots,(N^{\kappa_\ell}\lambda^{(\ell)})^{N^{1-\kappa_\ell}\gamma_\ell}]~,
\end{align}
where $0 \le \kappa_a < 1$ for all $a=1, \dots , \ell$ and  $\hat l\hat\gamma=1=\sum_a \gamma_a\lambda^{(a)}$. In the limit considered in \citep{Assel:2012cp}, $N$ is taken large and the other quantities are finite. 
These theories can be obtained by the following replacements and scalings in the partitions \eqref{eq:rho-sigma-balanced}
\begin{align} \label{eq:subst_assel}
	R_1 &\to \hat{l} \ , & R_2 &\to N \hat{\gamma} \ , & \mathsf{t}_a &\to N^{\kappa_a}\lambda^{(a)} \ , & \mathsf{k}_a &\to N^{1-\kappa_a}\gamma_a \ .
\end{align}
The free energies for the theories in (\ref{eq:rho-sigma-log}) can again be recovered from the general result for balanced quivers in \eqref{eq:cF-balanced-rho-sigma}.
The crucial point for the scaling is that the quiver description of \eqref{eq:rho-sigma-log}, in the notation of (\ref{eq:quiver}), has $\mathsf{k}_a=N^{(1-\kappa_a)}\gamma_a$ flavors at nodes $\mathsf{z}_a$ with
\begin{equation}\label{eq:za-log}
	\mathsf{z}_a=N^{\kappa_a-1}\lambda^{(a)}\hat{l}~.
\end{equation}
Since $\kappa_a<1$, all $\mathsf{z}_a$ approach zero in the large-$N$ limit, although at different rates dictated by $\kappa_a$. That is, all flavors accumulate at one end of the quiver. The free energies can be recovered from \eqref{eq:cF-balanced-rho-sigma} by expanding the trilogarithm functions using (\ref{eq:Li3-exp}).
This leads to
\begin{equation}
	F_{S^3} =-\frac{r(1-r)(N\hat{\gamma})^2}{2} \sum_{a,b=1}^{\ell}\mathsf{k}_a \mathsf{k}_b\left((\mathsf{z}_a+\mathsf{z}_b)^2\ln|\mathsf{z}_a+\mathsf{z}_b|-(\mathsf{z}_a-\mathsf{z}_b)^2 \ln|\mathsf{z}_a-\mathsf{z}_b|\right)\,,
\end{equation}
where $(\mathsf{z}_a-\mathsf{z}_b)^2 \ln|\mathsf{z}_a-\mathsf{z}_b|$ is understood to be zero if $a=b$.
With the substitutions \eqref{eq:subst_assel} and \eqref{eq:za-log}, and using that $\sum_a \gamma_a\lambda^{(a)}=1$, one arrives at
\begin{equation}
	F_{S^3}=2r(1-r)N^2\ln N \left(1-\sum_{a=1}^\ell (\gamma_a \lambda^{(a)})^2 \kappa_a-2\sum_{a < b} (\gamma_a \lambda^{(a)}\gamma_b \lambda^{(b)})\kappa_a\right).
\end{equation}
For $r=1/2$ this is exactly the result given in (3.30) of \cite{Assel:2012cp}.
If all $\kappa_\alpha$ are zero one recovers the $T[SU(N)]$ theory.

In the supergravity solutions spelled out in sec.~\ref{sec:balanced-sugra}, the replacement (\ref{eq:subst_assel}) leads to the locations of the D5-brane sources at $z=\delta_a$ with
\begin{align}
	\delta_a&=\ln\tan\left(N^{\kappa_a-1}\frac{\pi \lambda^{(a)}}{2\hat\gamma}\right)~.
\end{align}
The position of the sources thus depends on $N$. 
For $\kappa_a<1$ all sources accumulate at $z=-\infty$, which again is the cause of the logarithmic scaling from the supergravity perspective.

\section{Comparison to 5d}\label{sec:5d}

In this section we summarize and discuss relations between long quiver theories with all nodes balanced in 3d and 5d, first at the level of the matrix models and then of their supergravity duals.

The 5d theories discussed in \cite{Uhlemann:2019ypp} are the strong-coupling limits of linear quiver gauge theories with $SU(\cdot)$ nodes and possibly Chern-Simons terms, whose levels are denoted $c_t$,
of the form
\begin{align}
	5d:&&SU&(N_1)_{c_1}-SU(N_2)_{c_2}-\ldots -SU(N_{L-1})_{c_{L-1}}-SU(N_L)_{c_L} &&
	\nonumber\\
	&&&\hskip 2mm |\hskip 20mm |\hskip 32mm |\hskip 26mm |
	\label{eq:5d-balanced-quiver}	\\
	&&&[k_1] \hskip 15mm [k_2] \hskip 25mm [k_{L-1}] \hskip 19mm [k_L]
	\nonumber
\end{align}
The flavor excess compared to a balanced node with $2N_t$ flavors is captured by $N_{t+1}-N_{t-1}+k_t-2N_t$.
While non-negative for `good' theories in 3d, this quantity is bounded from above in 5d, with the bound depending on the Chern-Simons level.\footnote{The bounds of \cite{Intriligator:1997pq} make the flavor excess non-positive. However, these bounds can be relaxed (see e.g.\ \cite{Bergman:2015dpa}).}
Theories with no Chern-Simons terms and all nodes balanced are admissible in 3d and in 5d, and we focus on these in the following.

For the planar limit in 5d the rank function $N^{(5d)}(z)$ was $\mathcal O(L)$ in \cite{Uhlemann:2019ypp}, with order-one flavor numbers at interior nodes and up to $\mathcal O(L)$ flavors at boundary nodes.
For the 3d theories, on the other hand, we took $N^{(3d)}(z)$ of $\mathcal O(L^2)$ with $\mathcal O(L)$ flavors at interior nodes (other scalings will be discussed below).
For each balanced 3d theory a 5d partner can thus be identified by
\begin{align}\label{eq:Nk-3d-5d}
	N^{(3d)}(z)&=L N^{(5d)}(z)~, & k^{(3d)}(z)&=L k^{(5d)}(z)~.
\end{align}
The matrix models for these theories take an identical general form (compare (\ref{eq:cF-gen-5}) in 3d to (2.27) of \cite{Uhlemann:2019ypp} for 5d theories), with different $F_0$ and $F_H$ but with the same relation (\ref{eq:F0-FH}).
The saddle point eigenvalue densities are obtained from the same electrostatics problem, and are both encoded in (\ref{eq:varrho-s}).
Many observables are thus related between the 3d and 5d theories, in particular quantities which depend on the eigenvalues only, like expectation values of Wilson loops.\footnote{Wilson loops in 5d were discussed in \cite{Uhlemann:2020bek}. The example theories considered there have no flavors at interior nodes; they include the 5d $T_N$ theories and are related to 3d theories with $N^2\ln N$ scaling like the $T[SU(N)]$ theory.}

The free energies for 3d theories with no large flavor numbers at boundary nodes are given by (\ref{eq:cF-balanced}); for the corresponding 5d theories they can be obtained from (3.17) of \cite{Uhlemann:2019ypp}:
\begin{align}\label{eq:F-3d-5d}
	F_{S^3}^{(3d)}&=
	-\frac{L^2}{4\pi^2}\sum_{s,t=1}^{L}k_tk_s \Re\left[\Li_3\left(e^{i\pi (z_s+z_t)}\right) -\Li_3\left(e^{i\pi(z_s-z_t)}\right)\right]~,
	\nonumber\\
	F_{S^5}^{(5d)}&=
	\frac{27L^4}{16\pi^4}\sum_{s,t=1}^{L}k_tk_s \Re\left[\Li_5\left(e^{i\pi (z_s+z_t)}\right) -\Li_5\left(e^{i\pi(z_s-z_t)}\right)\right]~.
\end{align}
The flavors at interior nodes are $\mathcal O(L)$ in 3d and $\mathcal O(1)$ in 5d, so the scaling is $\mathcal O(L^4)$ for both.
The free energies are thus related by a rescaling and an adjustment of the weights of the polylogarithms, resulting from the different scalings of the functions $F_0$ and $F_H$ in the matrix models.
This relation extends to the topologically twisted indices, since these are related to the free energy in the planar limit in both cases, as shown for 3d in sec.~\ref{sec:index} and for a sample of 5d theories in \cite{Fluder:2019szh}.

3d theories with large flavor numbers at the boundary nodes, like $T[SU(N)]$, are also related to 5d theories.
The discussion of the saddle points proceeds as before, and the free energies can be understood from the expressions in (\ref{eq:F-3d-5d}):
For flavors at boundary nodes, e.g.\ non-zero $k_t$ such that $z_t$ is $\mathcal O(1/L)$, one can expand the corresponding terms in (\ref{eq:F-3d-5d}) accordingly. 
For a 5d theory with $\mathcal O(L)$ boundary flavors this leads to a combination of $\Li_3$, $\Li_4$ and $\Li_5$ terms contributing at the same order (cf.\ (3.17) in \cite{Uhlemann:2019ypp}).
In 3d, the expansion of $\Li_3$ produces a logarithmically enhanced contribution, as discussed in sec.~\ref{sec:N2LogN-limit} (the subleading term in (\ref{eq:Li3-exp}) is enhanced compared to the analogous term in $\Li_5$).
Thus, terms corresponding to large flavor numbers at boundary nodes dominate the free energy in 3d, leading to expressions in terms of logarithms only. 

The relation between SCFT and gauge theory description is different in 3d and 5d: In 3d the SCFT arises as IR fixed point and in 5d as UV fixed point.
In 3d the Yang-Mills terms are exact and $F_{S^3}$ is constant along the flow, while in 5d $F_{S^5}$ grows towards the UV and takes the form given in (\ref{eq:F-3d-5d}) at the fixed points.
The relations between the  gauge theory matrix models imply relations between the SCFTs for quantities that can be computed from the zero-instanton matrix models, such as the planar free energies and topologically twisted indices.
For such relations it is sufficient for one of the perhaps multiple (S-dual) gauge theory descriptions in 5d and mirror-dual descriptions in 3d to be balanced. 
From other perspectives, however, different pairings between 3d and 5d theories are more natural.
For example, the relations discussed above connect the gauge theory description of the 3d $T[SU(N)]$ theory with $SU(N)^2$ global symmetry to the gauge theory description of the 5d $T_N$ theory with $SU(N)^3$ global symmetry, while we have not discussed the star-shaped quiver for the 3d $T_N$ theory with $SU(N)^3$ symmetry \cite{Benini:2010uu}.

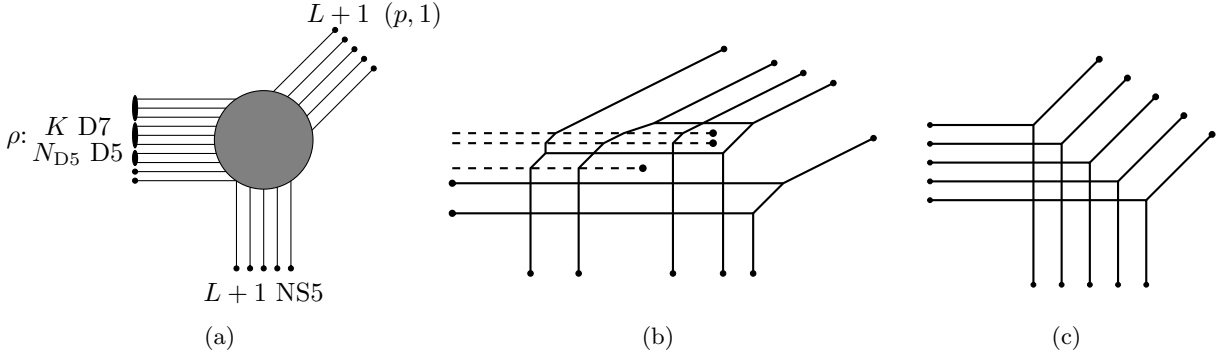
\begin{figure}
	\subfigure[][]{\label{fig:5d-balanced-1}
		\begin{tikzpicture}[scale=0.95]
			\foreach \i in {-3,-3/2,0,3/2,3}{
				\draw (0.127*\i,-0.4) -- (0.127*\i,-1.8) [fill=black] circle (1pt);
				\draw (0.28+0.09*\i,0.28-0.09*\i) -- (1.27+0.09*\i,1.27-0.09*\i) [fill=black] circle (1pt) ;
			}
			\foreach \i in {-9/2,-7/2,-5/2,-3/2,-1/2,1/2,3/2,5/2,7/2,9/2}{	
				\draw (-0,0.127*\i) -- (-1.8,0.127*\i);
			}
			\draw[fill=black] (-1.8,0.127*7/2) ellipse (1pt and 5pt);
			\draw[fill=black] (-1.8,0.127*1/2) ellipse (1pt and 5pt);
			\draw[fill=black] (-1.8,-0.127*2) ellipse (1pt and 3pt);
			\draw[fill=black] (-1.8,-0.127*7/2) circle (1pt);
			\draw[fill=black] (-1.8,-0.127*9/2) circle (1pt);
			\draw[fill=gray] (0,0) circle (0.69);    
			\node at (0,-2.15) {\small $L+1$ NS5};
			
			\node at (-2.6,0.2) {\small $K$ D7};
			\node at (-2.6,-0.2) {\small $N_{\rm D5}$ D5};
			\node at (-3.45,0){\small $\rho$:};
			
			\node at (1.55,1.75) {\small $L+1$\, $(p,1)$};
		\end{tikzpicture}
	}
	\hspace*{-5mm}
	\subfigure[][]{\label{fig:5d-balanced-2}
		\begin{tikzpicture}[scale=0.8]
			\draw[thick] (0,-1) -- (0,0) -- ++(0.5,0.5) -- ++(1.5,0.75);
			\draw[thick] (0,0) -- (-5,0) [fill=black] circle (1.2pt);
			\draw[thick] (0.5,0.5) -- (-5,0.5) [fill=black] circle (1.2pt);
			
			\draw[thick] (-0.5,-1) -- (-0.5,1) -- ++(0.5,0.5) -- ++(1.333,0.666);
			\draw[thick] (-0.5,1) -- (-3.7+0.25,1);
			\draw[thick] (0,1.5) -- (-2.9+0.25+0.166+0.333+0.5,1.5);
			
			\draw[thick] (-1.333,-1) -- (-1.333,1.1666) -- ++(0.1666,0.1666) -- ++(2,1);
			\draw[dashed,thick] (-5,1.1666) -- (-0.666,1.1666); 		\draw[fill=black,thick] (-0.666,1.1666) circle (1.3pt);
			\draw[dashed,thick] (-5,1.333) -- (-0.666,1.333); 		\draw[fill=black,thick] (-0.666,1.333) circle (1.3pt);
			
			\draw[thick] (-2.9,-1) -- (-2.9,0.75) -- ++(0.25,0.25) -- ++(0.166,0.166) -- ++(0.333,0.166) -- ++(0.5,0.166) -- ++(2,1);
			\draw[dashed,thick] (-5,0.75) -- (-1.833,0.75); 		\draw[fill=black,thick] (-1.833,0.75) circle (1.3pt);
			
			\draw[thick] (-3.7,-1) -- (-3.7,0.75) -- ++(0.25,0.25) -- ++ (0,0.166) -- ++ (0.166,0.166) -- ++ (2.8,1.4);
			
			\foreach \i in {0,-0.5,-1.333,-2.9,-3.7} { \draw (\i,-1) [fill=black] circle (1.3pt);}
			\foreach \i in {(2.0,1.25),(-2.9+0.25+0.5+0.5+2,2.5),(-1.333+0.1666+2,1.3333+1),(-3.7+0.25+0.166+2.8,1.333+1.4),(1.333,1.5+0.666)} { \draw \i [fill=black] circle (1.3pt);}
			
			\node at (0,-1.5) {};
		\end{tikzpicture}
	}
	\hspace*{2mm}
	\subfigure[][]{\label{fig:5d-TN}
		\begin{tikzpicture}[xscale=0.5,yscale=0.5]
			\draw[thick] (-4,0.75) -- (-0.5,0.75) -- (-0.5,-3);
			\draw[thick] (-0.5,0.75) -- +(1.75,1.75) [fill=black] circle (1.8pt);
			
			\draw[thick] (-4,0.25) -- (0.25,0.25) -- (0.25,-3);
			\draw[thick] (0.25,0.25) -- +(1.75,1.75) [fill=black] circle (1.8pt);
			
			\draw[thick] (-4,-0.25) -- (1.0,-0.25) -- (1.0,-3);
			\draw[thick] (1.0,-0.25) -- +(1.75,1.75) [fill=black] circle (1.8pt);
			
			\draw[thick] (-4,-0.75) -- (1.75,-0.75) -- (1.75,-3);
			\draw[thick] (1.75,-0.75) -- +(1.75,1.75) [fill=black] circle (1.8pt);
			
			\draw[thick] (-4,1.25) -- (-1.25,1.25) -- (-1.25,-3);
			\draw[thick] (-1.25,1.25) -- +(1.75,1.75) [fill=black] circle (1.8pt);
			\node at (0,-3.5) {};
			
			\foreach \i in {0.75,0.25,-0.25,-0.75,1.25}{\draw (-4,\i) [fill=black] circle (1.8pt);}
			\foreach \i in {-0.5,0.25,1.0,1.75,-1.25}{\draw (\i,-3) [fill=black] circle (1.8pt);}		
		\end{tikzpicture}
	}
	
	\caption{Left: Junction of $L+1$ NS5-branes, $L+1$ $(p,1)$ 5-branes, and $N_{\rm D5}$ D5-branes ending on $K$ D7-branes as specified by $\rho$ in (\ref{eq:rho-5-brane}).
		Center: Gauge theory for $\rho=[3^2,2,1^2]$ and $p=2$, after moving D7-branes into the web (the dashed lines show the branch cuts).
		Right: 5d $T_N$ theory, corresponding to $\rho=[1^N]$.
		\label{fig:5d-balanced}}
\end{figure}

\subsection{Supergravity duals}

We now spell out a brane construction for balanced $SU(\cdot)$ quiver gauge theories in 5d and briefly discuss the relation to 3d from the holographic perspective.
5d quiver gauge theories can be engineered by $(p,q)$ 5-brane webs \cite{Aharony:1997ju,Aharony:1997bh}.
We take the $(p,q)$ 5-branes to span the $(01234)$ directions and a line in the $(56)$ plane determined by $\Delta x_5+i\Delta x_6 = p+i q$.
$(p,q)$ 5-branes may end on $[p,q]$ 7-branes spanning the $(01234789)$ directions,
and if multiple 5-branes end on the same 7-brane their junctions with other 5-branes are constrained by the s-rule \cite{DeWolfe:1999hj,Benini:2009gi}.

Balanced 5d quiver gauge theories with no Chern-Simons terms can be engineered by 5-brane junctions of the form shown in fig.~\ref{fig:5d-balanced-1}, where the gray disc schematically represents the internal structure of the web.
The junction involves $L+1$ NS5-branes and $L+1$ $(p,1)$ 5-branes which are unconstrained by the $s$-rule, and a number $N_{\rm D5}$ of D5-branes whose distribution over $K$ D7-branes is specified by a Young tableau $\rho$.
For a balanced quiver of the form (\ref{eq:5d-balanced-quiver}), let
\begin{align}\label{eq:rho-5-brane}
	\rho&=\Big[L^{k_L}, \ldots, 2^{k_2}, 1^{k_1}\Big]~, & 	N_{\rm D5}&=\sum_{t=1}^L  t k_t~.
\end{align}
That is, there are $L$ groups of D7-branes (some of which may be empty), with $k_t$ D7-branes in the $t^{\rm th}$ group and with $t$ D5-branes ending on each D7-brane.
The charge $p$ of the $(p,1)$ 5-branes is
\begin{align}
	p&=\frac{N_{\rm D5}}{L+1}=N_t+\sum_{s=t+1}^L k_s-N_{t+1}~.
\end{align}
The first expression follows from overall charge conservation in fig.~\ref{fig:5d-balanced-1}, the second from considering sub-webs for individual gauge nodes. 
Due to the balancing condition $N_{t+1}+N_{t-1}-2N_t-k_t=0$ the second expression is independent of $t$.

Similarly to the discussion for 3d theories in sec.~\ref{sec:3d-branes}, one can move the D7-branes into the brane web,
to faces where they have no D5-branes attached and describe flavors at the corresponding gauge node.
An example is shown in fig.~\ref{fig:5d-balanced-2}, for a gauge theory with $L=4$ and $(N_1,k_1)=(3,2)$, $(N_2,k_2)=(4,1)$, $(N_3,k_3)=(4,2)$, $(N_4,k_4)=(2,0)$.
Fig.~\ref{fig:5d-TN} shows the gauge theory deformation of the 5d $T_N$ theory, $[N]-SU(N-1)-\ldots -SU(2)-[2]$ (see \cite{Bergman:2014kza}).

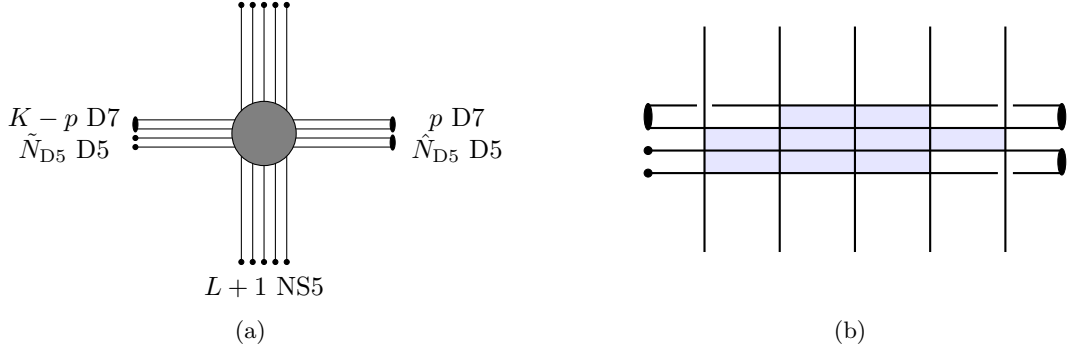
\begin{figure}
	\subfigure[][]{\label{fig:balanced-plus}
		\begin{tikzpicture}[scale=0.95]
			\foreach \i in {-5/2,-5/4,0,5/4,5/2}{
				\draw (0.127*\i,0) -- (0.127*\i,-1.8) [fill=black] circle (1pt);
				\draw (0.127*\i,0) -- (0.127*\i,1.8) [fill=black] circle (1pt);
			}
			\foreach \i in {-3/2,-1/2,1/2,3/2}{
				\draw (0,0.127*\i) -- (-1.8,0.127*\i) [fill=black];
				\draw (0,0.127*\i) -- (1.8,0.127*\i) [fill=black];			
			}
			\draw[fill=black] (-1.8,0.127*1) ellipse (1pt and 3pt);
			\draw[fill=black] (-1.8,-0.127*1/2) circle (1pt);
			\draw[fill=black] (-1.8,-0.127*3/2) circle (1pt);	
			
			\draw[fill=black] (1.8,0.127*1) ellipse (1pt and 3pt);
			\draw[fill=black] (1.8,-0.127*1) ellipse (1pt and 3pt);				
			
			\draw[fill=gray] (0,0) circle (0.45);    
			\node at (0,-2.15) {\small $L+1$ NS5};
			
			\node at (-2.8,0.2) {\small $K-p$ D7};
			\node at (-2.8,-0.2) {$\tilde N_{\rm D5}$ D5};
			
			\node at (2.7,0.2) {\small $p$ D7};
			\node at (2.7,-0.2) {$\hat N_{\rm D5}$ D5};
		\end{tikzpicture}
	}
	\qquad\qquad
	\subfigure[][]{
		\begin{tikzpicture}
			
			\draw[fill=blue!10] (-1,-0.45) rectangle (1,0.45);
			\draw[fill=blue!10] (-2,-0.45) rectangle (-1,0.15);		
			\draw[fill=blue!10] (1,-0.15) rectangle (2,0.15);				
			
			\foreach \i in {-2,-1,0,1,2} {\draw[thick] (\i,-1.5) -- (\i,1.5);}
			\draw[thick] (-2.75, -0.45) -- (1.9, -0.45); \draw[thick] (2.1,-0.45) -- (2.75,-0.45);
			\draw[thick] (-2.75,-0.15) -- (2.75,-0.15);
			\draw[thick] (-2.75, 0.15) -- (2.75, 0.15);
			\draw[thick] (-2.75, 0.45) -- (-2.1, 0.45); \draw[thick] (-1.9, 0.45) -- (1.9, 0.45); \draw[thick] (2.1,0.45) -- (2.75,0.45);
			
			\draw[fill=black] (-2.75,-0.45) circle (1.5pt);
			\draw[fill=black] (-2.75,-0.15) circle (1.5pt);		
			\draw[fill=black] (-2.75,0.3) ellipse (1.5pt and 5pt);		
			
			\draw[fill=black] (2.75,0.3) ellipse (1.5pt and 5pt);		
			\draw[fill=black] (2.75,-0.3) ellipse (1.5pt and 5pt);		
			
			\node at (0,-2.1) {};
		\end{tikzpicture}
	}
	\caption{Left: The junction shown in fig.~\ref{fig:5d-balanced-1} after moving $p$ D7-branes to the right and rotating their branch cuts accordingly. 
		Right: The gauge theory of fig.~\ref{fig:5d-balanced-2}. The broken lines are avoided intersections due to the $s$-rule. The closed (gauge theory) faces are shaded.\label{fig:5d-balanced-web-2}}
\end{figure}

The brane web in fig.~\ref{fig:5d-balanced-1} can be brought into the form shown in fig.~\ref{fig:balanced-plus} by moving $p$ of the D7-branes from the left to the right, while rotating their branch cuts clockwise from pointing West to pointing East.
A D7-brane which initially has $t$ D5-branes attached from the right ends up with $L+1-t$ D5-branes attached from the left,
while rotating the branch cuts turns the $(p,1)$ 5-branes into NS5-branes.

In the form in fig.~\ref{fig:5d-balanced-web-2} the relation between 5d balanced gauge theories and 3d balanced $T_\rho^\sigma[SU(N)]$ gauge theories amounts to replacing (D7,D5,NS5) branes by (D5,D3,NS5) branes, which can be achieved by T-dualizing two of the (01234) directions.
Upon taking appropriate scaling limits this relates the brane constructions for 3d and 5d theories with related matrix models as discussed above.
Relating fixed-point theories with identical global symmetries is more involved. For the 5d $T_N$ theory with $SU(N)^3$ global symmetry, for example, the 3d version was identified in \cite{Benini:2010uu}, by separately treating the three `arms' of the $T_N$ 5-brane junction with $SL(2,\ZZ)$ and T-duality, leading to a $T[SU(N)]$ theory for each arm, joined to form a star-shaped quiver. 
This picture extends to the supergravity duals.
Supergravity solutions for 5-brane webs that are unconstrained by the $s$-rule were constructed in \cite{DHoker:2016ujz,DHoker:2016ysh,DHoker:2017mds}, and solutions for 5-brane webs with mutually local 7-branes in \cite{DHoker:2017zwj}. The latter include general junctions of the form in fig.~\ref{fig:5d-balanced-1} and \ref{fig:balanced-plus} (some examples were discussed in \cite{Gutperle:2018vd,Chaney:2018gjc}).
The supergravity duals represent the features of the SCFTs e.g.\ regarding the global symmetries, and have qualitative differences in 3d and 5d.
But certain quantities, like on-shell actions and black hole entropies, can be related between the duals for 3d SCFTs and 5d SCFTs arising as fixed points of gauge theories that have related (planar) matrix models.

\section{Discussion}\label{sec:disc}

We discussed a planar limit of the 3d $T_\rho^\sigma[SU(N)]$ theories in which the free energy shows polynomial scaling.
It is the standard $N^2$ scaling from the perspective of the $\cN=4$ SYM theory on an interval from which these theories can be derived, 
and a quartic scaling in terms of the length of their 3d quiver gauge theory description.
This scaling arises naturally in the supergravity duals, in which the brane sources remain fixed and well separated in the planar limit, and we have shown for a sample of (classes of) theories that the free energies match perfectly.
For the topologically twisted index we have shown that the leading-order expression is related in a universal way to the sphere free energy, in line with the `index theorem' of \cite{Hosseini:2016tor} and the recent discussion in \cite{Hosseini:2020mut}. 
For theories with all nodes balanced we discussed relations to 5d gauge theories, which relate certain quantities between SCFTs obtained as IR fixed points in 3d and as UV fixed points in 5d, and connect their supergravity duals.
We have not considered squashed spheres, but certainly expect a universal dependence of the free energy on squashing parameters, much like for the 5d theories in \cite{Uhlemann:2019ypp}. 
This would imply that the central charge $C_T$, which can be obtained from the squashed-sphere free energy \cite{Closset:2012ru}, is also related to $F_{S^3}$ in a universal way.

We close with a discussion of future directions. It would be interesting to extend the localization computations to circular quivers, for which holographic duals were constructed in \cite{Assel:2012cj}. Similarly, it would be interesting to discuss theories with Chern-Simons terms, for which the discussion in \cite{Assel:2013lpa} may be a good starting point,
or generalizations of the S-fold theories of \cite{Assel:2018vtq}.
More general supergravity solutions may be constructed by incorporating orientifold planes, e.g.\ to realize $T_\rho^\sigma[SO(N)]$ and $T_\rho^\sigma[Sp(N)]$ theories. More generally, long quivers are studied in other dimensions, e.g.\ \cite{Lozano:2020txg,Lozano:2019emq,Lozano:2019jza,Lozano:2019zvg,Aharony:2012tz,Nunez:2019gbg,Gaiotto:2014lca,Apruzzi:2015wna,Bergman:2020bvi,Heckman:2020otd,Baume:2020ure}, and it would be interesting to apply similar localization methods to gain further insights.

A different class of 3d theories with quartic scaling of the free energies are the theories of class~$\cF$ obtained by compactifying 5d SCFTs engineered by $(p,q)$ 5-brane webs on Riemann surfaces. Their holographic duals can be obtained by uplifting the 6d $AdS_4\times\Sigma$ solution of \cite{Naka:2002jz} to Type IIB solutions based on \cite{DHoker:2016ujz,DHoker:2016ysh,DHoker:2017mds,DHoker:2017zwj}, using the uplifts of \cite{Hong:2018amk,Malek:2018zcz}. 
It would be interesting to develop a detailed field theory understanding of these theories (and of the black holes constructed recently in \cite{Hosseini:2020wag}).

\let\oldaddcontentsline\addcontentsline
\renewcommand{\addcontentsline}[3]{}

\begin{acknowledgments}
LC thanks Gabriele Lo Monaco, Matteo Sacchi and Alberto Zaffaroni for useful discussions and explanations.
LC is supported by the INFN and by the MIUR-PRIN contract 2017CC72MK003.
CFU is supported, in part, by the US Department of Energy under Grant No.~DE-SC0007859 	and by the Leinweber Center for Theoretical Physics.
\end{acknowledgments}

\let\addcontentsline\oldaddcontentsline

\appendix

\section{Junction condition}\label{app:junction}

In this Appendix we discuss the junction condition (\ref{eq:delta-cF-junction-2}) for unbalanced nodes.
The matrix models are invariant under $\lambda\rightarrow - \lambda$ (individually for each node), so we can assume the same for the saddle point.

Consider an unbalanced node $z_t$ with $k_t$ flavors. We first show that the condition (\ref{eq:junction-3}) is solved by fixing $\mu_t$.
With $x^\epsilon_\pm = x_\pm\mp \epsilon$ and $z_\pm=z_t\pm \epsilon$,
\begin{align}
 \int_{x^\epsilon_-}^{x^\epsilon_+} dy\,\left[\partial_z \varrho(z,y)\right]_{z=z_-}^{z=z_+}F_H(x-y)
 &=
 \int_{x^\epsilon_-}^{x^\epsilon_+} dy\int_{z_-}^{z_+} dz\,F_H(x-y)\partial_z^2\varrho(z,y)
 \nonumber\\
 &=
 -Lk_t F_H(x)
 -\frac{1}{4} \int_{x^\epsilon_-}^{x^\epsilon_+} dy\int_{z_-}^{z_+} dz\,F_H(x-y)\partial_y^2\varrho(z,y)
\end{align}
where the bulk saddle point equation has been used for the second line.
Successive integrations by parts in the last term lead to
\begin{align}
 &\int_{x^\epsilon_-}^{x^\epsilon_+} dy\int_{z_-}^{z_+} dz\,F_H(x-y)\partial_y^2\varrho(z,y)
 =
 \nonumber\\
&\int_{z_-}^{z_+} dz\left(\left[F_H(x-y)\partial_y\varrho(z,y)-F_H^\prime(x-y)\varrho(z,y)\right]_{y=x^\epsilon_-}^{y=x^\epsilon_+}+\int_{x^\epsilon_-}^{x^\epsilon_+}dy\, F_H^{\prime\prime}(x-y)\varrho(z,y)\right).
\end{align}
Using that in the definition of $T(x)$ gives
\begin{align}
 T(x)&=L\mu_t+\int_{y\in\RR\setminus (x^\epsilon_-,x^\epsilon_+)} dy\,\left[F_H(x-y)\partial_z \varrho(z,y)\right]_{z=z_-}^{z=z_+}
 -\frac{1}{4}\int_{z_-}^{z_+} dz \,\left[F_H(x-y)\partial_y\varrho(z,y)\right]_{y=x^\epsilon_-}^{y=x^\epsilon_+}
 \nonumber\\
 &\hphantom{=}+\frac{1}{4}\int_{z_-}^{z_+} dz\left(\left[F_H^\prime(x-y)\varrho(z,y)\right]_{y=x^\epsilon_-}^{y=x^\epsilon_+}-\int_{x^\epsilon_-}^{x^\epsilon_+}dy\, F_H^{\prime\prime}(x-y)\varrho(z,y)\right).
\end{align}
Now assume that $x^\epsilon_-<x<x^\epsilon_+$. Using  $\varrho(z,y)=\varrho(z,-y)$,
\begin{align}
 T(x)&=L\mu_t+4\pi(1-r)\int_{x^\epsilon_+}^\infty dy\,y\left[\partial_z \varrho(z,y)\right]_{z=z_-}^{z=z_+}
 -\pi(1-r)x^\epsilon_+\int_{z_-}^{z_+} dz \,\partial_y\varrho(z,y)\vert_{y=x^\epsilon_+}
 \nonumber\\
 &\hphantom{=}+\frac{\pi}{2}(1-r)\int_{z_-}^{z_+} dz\left(2\varrho(z,x^\epsilon_+)-\varrho(z,x)\right).
\end{align}
Only the last term in the second line depends on $x$, and it is $\mathcal O(\epsilon)$.
Thus, $T(x)$ is a constant, and the junction condition $T(x)=0$ is solved by fixing $\mu_t$.

We now discuss the allowed behavior of $\varrho(z_t,x)$ at the end points of the interval on which $\varrho$ has support, $x_\pm$.
Assume that $\varrho$ satisfies the bulk saddle point equation, so $T(x)$ vanishes for $x\in(x_-,x_+)$.
Assume that, near the end points,
\begin{align}
 \varrho(z_t,x)\sim (x-x_\pm)^a~.
\end{align}
To allow variations of the end points, $\delta\varrho \sim \partial_{x_\pm}\varrho$ should be allowed as variation.
Thus, we have to allow for $\delta\varrho \sim (x-x_\pm)^{a-1}$ in (\ref{eq:delta-cF-junction-2}).
To avoid a $\delta$-function contribution from $x=x_\pm$ we need
\begin{align}\label{eq:app-delta-cF}
  \delta \cF&=\int dx\,\delta\varrho(z_t,x) T(x)
  \sim \int dx\,(x-x_\pm)^{a-1} T(x)
\end{align}
to be continuous across the end points. 
With the assumed behavior of $\varrho$ we find, for $x$ approaching the end points from outside of $(x_-,x_+)$,
\begin{align}
 T''(x)&=2\pi(1-r) \left[\partial_z \varrho(z,x)\right]_{z=z_t-\epsilon}^{z=z_t+\epsilon}\sim (x-x_\pm)^{a-1}~.
\end{align}
This leads to $T(x)\sim (x-x_\pm)^{a+1}$, and to avoid a $\delta$-function contribution in (\ref{eq:app-delta-cF}) we need $a>-1/2$.

The constraint on the behavior near $x_\pm$ is different in 5d: Due to the steeper scaling of $F_H$ in 5d, we have $T^{(4)}(x)\sim \left[\partial_z \varrho(z,x)\right]_{z=z_t-\epsilon}^{z=z_t+\epsilon}$, leading to  $T(x)\sim (x-x_\pm)^{a+3}$ and $a>-3/2$.

\bibliography{3d}

\begin{thebibliography}{83}%
\makeatletter
\providecommand \@ifxundefined [1]{%
 \@ifx{#1\undefined}
}%
\providecommand \@ifnum [1]{%
 \ifnum #1\expandafter \@firstoftwo
 \else \expandafter \@secondoftwo
 \fi
}%
\providecommand \@ifx [1]{%
 \ifx #1\expandafter \@firstoftwo
 \else \expandafter \@secondoftwo
 \fi
}%
\providecommand \natexlab [1]{#1}%
\providecommand \enquote  [1]{``#1''}%
\providecommand \bibnamefont  [1]{#1}%
\providecommand \bibfnamefont [1]{#1}%
\providecommand \citenamefont [1]{#1}%
\providecommand \href@noop [0]{\@secondoftwo}%
\providecommand \href [0]{\begingroup \@sanitize@url \@href}%
\providecommand \@href[1]{\@@startlink{#1}\@@href}%
\providecommand \@@href[1]{\endgroup#1\@@endlink}%
\providecommand \@sanitize@url [0]{\catcode `\\12\catcode `\$12\catcode
  `\&12\catcode `\#12\catcode `\^12\catcode `\_12\catcode `\%12\relax}%
\providecommand \@@startlink[1]{}%
\providecommand \@@endlink[0]{}%
\providecommand \url  [0]{\begingroup\@sanitize@url \@url }%
\providecommand \@url [1]{\endgroup\@href {#1}{\urlprefix }}%
\providecommand \urlprefix  [0]{URL }%
\providecommand \Eprint [0]{\href }%
\providecommand \doibase [0]{https://doi.org/}%
\providecommand \selectlanguage [0]{\@gobble}%
\providecommand \bibinfo  [0]{\@secondoftwo}%
\providecommand \bibfield  [0]{\@secondoftwo}%
\providecommand \translation [1]{[#1]}%
\providecommand \BibitemOpen [0]{}%
\providecommand \bibitemStop [0]{}%
\providecommand \bibitemNoStop [0]{.\EOS\space}%
\providecommand \EOS [0]{\spacefactor3000\relax}%
\providecommand \BibitemShut  [1]{\csname bibitem#1\endcsname}%
\let\auto@bib@innerbib\@empty
%</preamble>
\bibitem [{\citenamefont {Aharony}\ \emph
  {et~al.}(2008{\natexlab{a}})\citenamefont {Aharony}, \citenamefont {Bergman},
  \citenamefont {Jafferis},\ and\ \citenamefont {Maldacena}}]{Aharony:2008ug}%
  \BibitemOpen
  \bibfield  {author} {\bibinfo {author} {\bibfnamefont {O.}~\bibnamefont
  {Aharony}}, \bibinfo {author} {\bibfnamefont {O.}~\bibnamefont {Bergman}},
  \bibinfo {author} {\bibfnamefont {D.~L.}\ \bibnamefont {Jafferis}},\ and\
  \bibinfo {author} {\bibfnamefont {J.}~\bibnamefont {Maldacena}},\ }\bibfield
  {title} {\bibinfo {title} {{N=6 superconformal Chern-Simons-matter theories,
  M2-branes and their gravity duals}},\ }\href
  {https://doi.org/10.1088/1126-6708/2008/10/091} {\bibfield  {journal}
  {\bibinfo  {journal} {JHEP}\ }\textbf {\bibinfo {volume} {10}},\ \bibinfo
  {pages} {091}},\ \Eprint {https://arxiv.org/abs/0806.1218} {arXiv:0806.1218
  [hep-th]} \BibitemShut {NoStop}%
\bibitem [{\citenamefont {Aharony}\ \emph
  {et~al.}(2008{\natexlab{b}})\citenamefont {Aharony}, \citenamefont
  {Bergman},\ and\ \citenamefont {Jafferis}}]{Aharony:2008gk}%
  \BibitemOpen
  \bibfield  {author} {\bibinfo {author} {\bibfnamefont {O.}~\bibnamefont
  {Aharony}}, \bibinfo {author} {\bibfnamefont {O.}~\bibnamefont {Bergman}},\
  and\ \bibinfo {author} {\bibfnamefont {D.~L.}\ \bibnamefont {Jafferis}},\
  }\bibfield  {title} {\bibinfo {title} {{Fractional M2-branes}},\ }\href
  {https://doi.org/10.1088/1126-6708/2008/11/043} {\bibfield  {journal}
  {\bibinfo  {journal} {JHEP}\ }\textbf {\bibinfo {volume} {11}},\ \bibinfo
  {pages} {043}},\ \Eprint {https://arxiv.org/abs/0807.4924} {arXiv:0807.4924
  [hep-th]} \BibitemShut {NoStop}%
\bibitem [{\citenamefont {Drukker}\ \emph {et~al.}(2011)\citenamefont
  {Drukker}, \citenamefont {Marino},\ and\ \citenamefont
  {Putrov}}]{Drukker:2010nc}%
  \BibitemOpen
  \bibfield  {author} {\bibinfo {author} {\bibfnamefont {N.}~\bibnamefont
  {Drukker}}, \bibinfo {author} {\bibfnamefont {M.}~\bibnamefont {Marino}},\
  and\ \bibinfo {author} {\bibfnamefont {P.}~\bibnamefont {Putrov}},\
  }\bibfield  {title} {\bibinfo {title} {{From weak to strong coupling in ABJM
  theory}},\ }\href {https://doi.org/10.1007/s00220-011-1253-6} {\bibfield
  {journal} {\bibinfo  {journal} {Commun. Math. Phys.}\ }\textbf {\bibinfo
  {volume} {306}},\ \bibinfo {pages} {511} (\bibinfo {year} {2011})},\ \Eprint
  {https://arxiv.org/abs/1007.3837} {arXiv:1007.3837 [hep-th]} \BibitemShut
  {NoStop}%
\bibitem [{\citenamefont {Schwarz}(2004)}]{Schwarz:2004yj}%
  \BibitemOpen
  \bibfield  {author} {\bibinfo {author} {\bibfnamefont {J.~H.}\ \bibnamefont
  {Schwarz}},\ }\bibfield  {title} {\bibinfo {title} {{Superconformal
  Chern-Simons theories}},\ }\href
  {https://doi.org/10.1088/1126-6708/2004/11/078} {\bibfield  {journal}
  {\bibinfo  {journal} {JHEP}\ }\textbf {\bibinfo {volume} {11}},\ \bibinfo
  {pages} {078}},\ \Eprint {https://arxiv.org/abs/hep-th/0411077}
  {arXiv:hep-th/0411077} \BibitemShut {NoStop}%
\bibitem [{\citenamefont {Gaiotto}\ and\ \citenamefont
  {Tomasiello}(2010)}]{Gaiotto:2009mv}%
  \BibitemOpen
  \bibfield  {author} {\bibinfo {author} {\bibfnamefont {D.}~\bibnamefont
  {Gaiotto}}\ and\ \bibinfo {author} {\bibfnamefont {A.}~\bibnamefont
  {Tomasiello}},\ }\bibfield  {title} {\bibinfo {title} {{The gauge dual of
  Romans mass}},\ }\href {https://doi.org/10.1007/JHEP01(2010)015} {\bibfield
  {journal} {\bibinfo  {journal} {JHEP}\ }\textbf {\bibinfo {volume} {01}},\
  \bibinfo {pages} {015}},\ \Eprint {https://arxiv.org/abs/0901.0969}
  {arXiv:0901.0969 [hep-th]} \BibitemShut {NoStop}%
\bibitem [{\citenamefont {Guarino}\ \emph {et~al.}(2015)\citenamefont
  {Guarino}, \citenamefont {Jafferis},\ and\ \citenamefont
  {Varela}}]{Guarino:2015jca}%
  \BibitemOpen
  \bibfield  {author} {\bibinfo {author} {\bibfnamefont {A.}~\bibnamefont
  {Guarino}}, \bibinfo {author} {\bibfnamefont {D.~L.}\ \bibnamefont
  {Jafferis}},\ and\ \bibinfo {author} {\bibfnamefont {O.}~\bibnamefont
  {Varela}},\ }\bibfield  {title} {\bibinfo {title} {{String Theory Origin of
  Dyonic N=8 Supergravity and Its Chern-Simons Duals}},\ }\href
  {https://doi.org/10.1103/PhysRevLett.115.091601} {\bibfield  {journal}
  {\bibinfo  {journal} {Phys. Rev. Lett.}\ }\textbf {\bibinfo {volume} {115}},\
  \bibinfo {pages} {091601} (\bibinfo {year} {2015})},\ \Eprint
  {https://arxiv.org/abs/1504.08009} {arXiv:1504.08009 [hep-th]} \BibitemShut
  {NoStop}%
\bibitem [{\citenamefont {Gaiotto}\ and\ \citenamefont
  {Witten}(2009)}]{Gaiotto:2008ak}%
  \BibitemOpen
  \bibfield  {author} {\bibinfo {author} {\bibfnamefont {D.}~\bibnamefont
  {Gaiotto}}\ and\ \bibinfo {author} {\bibfnamefont {E.}~\bibnamefont
  {Witten}},\ }\bibfield  {title} {\bibinfo {title} {{S-Duality of Boundary
  Conditions In N=4 Super Yang-Mills Theory}},\ }\href
  {https://doi.org/10.4310/ATMP.2009.v13.n3.a5} {\bibfield  {journal} {\bibinfo
   {journal} {Adv. Theor. Math. Phys.}\ }\textbf {\bibinfo {volume} {13}},\
  \bibinfo {pages} {721} (\bibinfo {year} {2009})},\ \Eprint
  {https://arxiv.org/abs/0807.3720} {arXiv:0807.3720 [hep-th]} \BibitemShut
  {NoStop}%
\bibitem [{\citenamefont {Assel}\ \emph {et~al.}(2011)\citenamefont {Assel},
  \citenamefont {Bachas}, \citenamefont {Estes},\ and\ \citenamefont
  {Gomis}}]{Assel:2011xz}%
  \BibitemOpen
  \bibfield  {author} {\bibinfo {author} {\bibfnamefont {B.}~\bibnamefont
  {Assel}}, \bibinfo {author} {\bibfnamefont {C.}~\bibnamefont {Bachas}},
  \bibinfo {author} {\bibfnamefont {J.}~\bibnamefont {Estes}},\ and\ \bibinfo
  {author} {\bibfnamefont {J.}~\bibnamefont {Gomis}},\ }\bibfield  {title}
  {\bibinfo {title} {{Holographic Duals of D=3 N=4 Superconformal Field
  Theories}},\ }\href {https://doi.org/10.1007/JHEP08(2011)087} {\bibfield
  {journal} {\bibinfo  {journal} {JHEP}\ }\textbf {\bibinfo {volume} {08}},\
  \bibinfo {pages} {087}},\ \Eprint {https://arxiv.org/abs/1106.4253}
  {arXiv:1106.4253 [hep-th]} \BibitemShut {NoStop}%
\bibitem [{\citenamefont {D'Hoker}\ \emph
  {et~al.}(2007{\natexlab{a}})\citenamefont {D'Hoker}, \citenamefont {Estes},\
  and\ \citenamefont {Gutperle}}]{DHoker:2007zhm}%
  \BibitemOpen
  \bibfield  {author} {\bibinfo {author} {\bibfnamefont {E.}~\bibnamefont
  {D'Hoker}}, \bibinfo {author} {\bibfnamefont {J.}~\bibnamefont {Estes}},\
  and\ \bibinfo {author} {\bibfnamefont {M.}~\bibnamefont {Gutperle}},\
  }\bibfield  {title} {\bibinfo {title} {{Exact half-BPS Type IIB interface
  solutions. I. Local solution and supersymmetric Janus}},\ }\href
  {https://doi.org/10.1088/1126-6708/2007/06/021} {\bibfield  {journal}
  {\bibinfo  {journal} {JHEP}\ }\textbf {\bibinfo {volume} {06}},\ \bibinfo
  {pages} {021}},\ \Eprint {https://arxiv.org/abs/0705.0022} {arXiv:0705.0022
  [hep-th]} \BibitemShut {NoStop}%
\bibitem [{\citenamefont {D'Hoker}\ \emph
  {et~al.}(2007{\natexlab{b}})\citenamefont {D'Hoker}, \citenamefont {Estes},\
  and\ \citenamefont {Gutperle}}]{DHoker:2007hhe}%
  \BibitemOpen
  \bibfield  {author} {\bibinfo {author} {\bibfnamefont {E.}~\bibnamefont
  {D'Hoker}}, \bibinfo {author} {\bibfnamefont {J.}~\bibnamefont {Estes}},\
  and\ \bibinfo {author} {\bibfnamefont {M.}~\bibnamefont {Gutperle}},\
  }\bibfield  {title} {\bibinfo {title} {{Exact half-BPS Type IIB interface
  solutions. II. Flux solutions and multi-Janus}},\ }\href
  {https://doi.org/10.1088/1126-6708/2007/06/022} {\bibfield  {journal}
  {\bibinfo  {journal} {JHEP}\ }\textbf {\bibinfo {volume} {06}},\ \bibinfo
  {pages} {022}},\ \Eprint {https://arxiv.org/abs/0705.0024} {arXiv:0705.0024
  [hep-th]} \BibitemShut {NoStop}%
\bibitem [{\citenamefont {Assel}\ \emph
  {et~al.}(2012{\natexlab{a}})\citenamefont {Assel}, \citenamefont {Estes},\
  and\ \citenamefont {Yamazaki}}]{Assel:2012cp}%
  \BibitemOpen
  \bibfield  {author} {\bibinfo {author} {\bibfnamefont {B.}~\bibnamefont
  {Assel}}, \bibinfo {author} {\bibfnamefont {J.}~\bibnamefont {Estes}},\ and\
  \bibinfo {author} {\bibfnamefont {M.}~\bibnamefont {Yamazaki}},\ }\bibfield
  {title} {\bibinfo {title} {{Large N Free Energy of 3d N=4 SCFTs and
  $AdS_4/CFT_3$}},\ }\href {https://doi.org/10.1007/JHEP09(2012)074} {\bibfield
   {journal} {\bibinfo  {journal} {JHEP}\ }\textbf {\bibinfo {volume} {09}},\
  \bibinfo {pages} {074}},\ \Eprint {https://arxiv.org/abs/1206.2920}
  {arXiv:1206.2920 [hep-th]} \BibitemShut {NoStop}%
\bibitem [{\citenamefont {Uhlemann}(2019)}]{Uhlemann:2019ypp}%
  \BibitemOpen
  \bibfield  {author} {\bibinfo {author} {\bibfnamefont {C.~F.}\ \bibnamefont
  {Uhlemann}},\ }\bibfield  {title} {\bibinfo {title} {{Exact results for 5d
  SCFTs of long quiver type}},\ }\href
  {https://doi.org/10.1007/JHEP11(2019)072} {\bibfield  {journal} {\bibinfo
  {journal} {JHEP}\ }\textbf {\bibinfo {volume} {11}},\ \bibinfo {pages}
  {072}},\ \Eprint {https://arxiv.org/abs/1909.01369} {arXiv:1909.01369
  [hep-th]} \BibitemShut {NoStop}%
\bibitem [{\citenamefont {Coccia}(2021)}]{Coccia:2020cku}%
  \BibitemOpen
  \bibfield  {author} {\bibinfo {author} {\bibfnamefont {L.}~\bibnamefont
  {Coccia}},\ }\bibfield  {title} {\bibinfo {title} {{Topologically twisted
  index of $T[SU(N)]$ at large $N$}},\ }\href
  {https://doi.org/10.1007/JHEP05(2021)264} {\bibfield  {journal} {\bibinfo
  {journal} {JHEP}\ }\textbf {\bibinfo {volume} {05}},\ \bibinfo {pages}
  {264}},\ \Eprint {https://arxiv.org/abs/2006.06578} {arXiv:2006.06578
  [hep-th]} \BibitemShut {NoStop}%
\bibitem [{\citenamefont {Hosseini}\ and\ \citenamefont
  {Zaffaroni}(2016)}]{Hosseini:2016tor}%
  \BibitemOpen
  \bibfield  {author} {\bibinfo {author} {\bibfnamefont {S.~M.}\ \bibnamefont
  {Hosseini}}\ and\ \bibinfo {author} {\bibfnamefont {A.}~\bibnamefont
  {Zaffaroni}},\ }\bibfield  {title} {\bibinfo {title} {{Large $N$ matrix
  models for 3d ${\cal N}=2$ theories: twisted index, free energy and black
  holes}},\ }\href {https://doi.org/10.1007/JHEP08(2016)064} {\bibfield
  {journal} {\bibinfo  {journal} {JHEP}\ }\textbf {\bibinfo {volume} {08}},\
  \bibinfo {pages} {064}},\ \Eprint {https://arxiv.org/abs/1604.03122}
  {arXiv:1604.03122 [hep-th]} \BibitemShut {NoStop}%
\bibitem [{\citenamefont {Gutperle}\ \emph {et~al.}(2017)\citenamefont
  {Gutperle}, \citenamefont {Marasinou}, \citenamefont {Trivella},\ and\
  \citenamefont {Uhlemann}}]{Gutperle:2017tjo}%
  \BibitemOpen
  \bibfield  {author} {\bibinfo {author} {\bibfnamefont {M.}~\bibnamefont
  {Gutperle}}, \bibinfo {author} {\bibfnamefont {C.}~\bibnamefont {Marasinou}},
  \bibinfo {author} {\bibfnamefont {A.}~\bibnamefont {Trivella}},\ and\
  \bibinfo {author} {\bibfnamefont {C.~F.}\ \bibnamefont {Uhlemann}},\
  }\bibfield  {title} {\bibinfo {title} {{Entanglement entropy vs. free energy
  in IIB supergravity duals for 5d SCFTs}},\ }\href
  {https://doi.org/10.1007/JHEP09(2017)125} {\bibfield  {journal} {\bibinfo
  {journal} {JHEP}\ }\textbf {\bibinfo {volume} {09}},\ \bibinfo {pages}
  {125}},\ \Eprint {https://arxiv.org/abs/1705.01561} {arXiv:1705.01561
  [hep-th]} \BibitemShut {NoStop}%
\bibitem [{\citenamefont {Fluder}\ and\ \citenamefont
  {Uhlemann}(2018)}]{Fluder:2018chf}%
  \BibitemOpen
  \bibfield  {author} {\bibinfo {author} {\bibfnamefont {M.}~\bibnamefont
  {Fluder}}\ and\ \bibinfo {author} {\bibfnamefont {C.~F.}\ \bibnamefont
  {Uhlemann}},\ }\bibfield  {title} {\bibinfo {title} {{Precision Test of
  AdS$_6$/CFT$_5$ in Type IIB String Theory}},\ }\href
  {https://doi.org/10.1103/PhysRevLett.121.171603} {\bibfield  {journal}
  {\bibinfo  {journal} {Phys. Rev. Lett.}\ }\textbf {\bibinfo {volume} {121}},\
  \bibinfo {pages} {171603} (\bibinfo {year} {2018})},\ \Eprint
  {https://arxiv.org/abs/1806.08374} {arXiv:1806.08374 [hep-th]} \BibitemShut
  {NoStop}%
\bibitem [{\citenamefont {Cremonesi}\ \emph {et~al.}(2015)\citenamefont
  {Cremonesi}, \citenamefont {Hanany}, \citenamefont {Mekareeya},\ and\
  \citenamefont {Zaffaroni}}]{Cremonesi:2014uva}%
  \BibitemOpen
  \bibfield  {author} {\bibinfo {author} {\bibfnamefont {S.}~\bibnamefont
  {Cremonesi}}, \bibinfo {author} {\bibfnamefont {A.}~\bibnamefont {Hanany}},
  \bibinfo {author} {\bibfnamefont {N.}~\bibnamefont {Mekareeya}},\ and\
  \bibinfo {author} {\bibfnamefont {A.}~\bibnamefont {Zaffaroni}},\ }\bibfield
  {title} {\bibinfo {title} {{T$_{\rho}^{\sigma}$ (G) theories and their
  Hilbert series}},\ }\href {https://doi.org/10.1007/JHEP01(2015)150}
  {\bibfield  {journal} {\bibinfo  {journal} {JHEP}\ }\textbf {\bibinfo
  {volume} {01}},\ \bibinfo {pages} {150}},\ \Eprint
  {https://arxiv.org/abs/1410.1548} {arXiv:1410.1548 [hep-th]} \BibitemShut
  {NoStop}%
\bibitem [{\citenamefont {Hanany}\ and\ \citenamefont
  {Witten}(1997)}]{Hanany:1996ie}%
  \BibitemOpen
  \bibfield  {author} {\bibinfo {author} {\bibfnamefont {A.}~\bibnamefont
  {Hanany}}\ and\ \bibinfo {author} {\bibfnamefont {E.}~\bibnamefont
  {Witten}},\ }\bibfield  {title} {\bibinfo {title} {{Type IIB superstrings,
  BPS monopoles, and three-dimensional gauge dynamics}},\ }\href
  {https://doi.org/10.1016/S0550-3213(97)00157-0} {\bibfield  {journal}
  {\bibinfo  {journal} {Nucl. Phys. B}\ }\textbf {\bibinfo {volume} {492}},\
  \bibinfo {pages} {152} (\bibinfo {year} {1997})},\ \Eprint
  {https://arxiv.org/abs/hep-th/9611230} {arXiv:hep-th/9611230} \BibitemShut
  {NoStop}%
\bibitem [{\citenamefont {Nishioka}\ \emph {et~al.}(2011)\citenamefont
  {Nishioka}, \citenamefont {Tachikawa},\ and\ \citenamefont
  {Yamazaki}}]{Nishioka:2011dq}%
  \BibitemOpen
  \bibfield  {author} {\bibinfo {author} {\bibfnamefont {T.}~\bibnamefont
  {Nishioka}}, \bibinfo {author} {\bibfnamefont {Y.}~\bibnamefont
  {Tachikawa}},\ and\ \bibinfo {author} {\bibfnamefont {M.}~\bibnamefont
  {Yamazaki}},\ }\bibfield  {title} {\bibinfo {title} {{3d Partition Function
  as Overlap of Wavefunctions}},\ }\href
  {https://doi.org/10.1007/JHEP08(2011)003} {\bibfield  {journal} {\bibinfo
  {journal} {JHEP}\ }\textbf {\bibinfo {volume} {08}},\ \bibinfo {pages}
  {003}},\ \Eprint {https://arxiv.org/abs/1105.4390} {arXiv:1105.4390 [hep-th]}
  \BibitemShut {NoStop}%
\bibitem [{\citenamefont {Intriligator}\ and\ \citenamefont
  {Seiberg}(1996)}]{Intriligator:1996ex}%
  \BibitemOpen
  \bibfield  {author} {\bibinfo {author} {\bibfnamefont {K.~A.}\ \bibnamefont
  {Intriligator}}\ and\ \bibinfo {author} {\bibfnamefont {N.}~\bibnamefont
  {Seiberg}},\ }\bibfield  {title} {\bibinfo {title} {{Mirror symmetry in
  three-dimensional gauge theories}},\ }\href
  {https://doi.org/10.1016/0370-2693(96)01088-X} {\bibfield  {journal}
  {\bibinfo  {journal} {Phys. Lett. B}\ }\textbf {\bibinfo {volume} {387}},\
  \bibinfo {pages} {513} (\bibinfo {year} {1996})},\ \Eprint
  {https://arxiv.org/abs/hep-th/9607207} {arXiv:hep-th/9607207} \BibitemShut
  {NoStop}%
\bibitem [{\citenamefont {Van~Raamsdonk}\ and\ \citenamefont
  {Waddell}(2021)}]{VanRaamsdonk:2020djx}%
  \BibitemOpen
  \bibfield  {author} {\bibinfo {author} {\bibfnamefont {M.}~\bibnamefont
  {Van~Raamsdonk}}\ and\ \bibinfo {author} {\bibfnamefont {C.}~\bibnamefont
  {Waddell}},\ }\bibfield  {title} {\bibinfo {title} {{Holographic and
  localization calculations of boundary F for $ \mathcal{N} $ = 4 SUSY
  Yang-Mills theory}},\ }\href {https://doi.org/10.1007/JHEP02(2021)222}
  {\bibfield  {journal} {\bibinfo  {journal} {JHEP}\ }\textbf {\bibinfo
  {volume} {02}},\ \bibinfo {pages} {222}},\ \Eprint
  {https://arxiv.org/abs/2010.14520} {arXiv:2010.14520 [hep-th]} \BibitemShut
  {NoStop}%
\bibitem [{\citenamefont {Benvenuti}\ and\ \citenamefont
  {Pasquetti}(2012)}]{Benvenuti:2011ga}%
  \BibitemOpen
  \bibfield  {author} {\bibinfo {author} {\bibfnamefont {S.}~\bibnamefont
  {Benvenuti}}\ and\ \bibinfo {author} {\bibfnamefont {S.}~\bibnamefont
  {Pasquetti}},\ }\bibfield  {title} {\bibinfo {title} {{3D-partition functions
  on the sphere: exact evaluation and mirror symmetry}},\ }\href
  {https://doi.org/10.1007/JHEP05(2012)099} {\bibfield  {journal} {\bibinfo
  {journal} {JHEP}\ }\textbf {\bibinfo {volume} {05}},\ \bibinfo {pages}
  {099}},\ \Eprint {https://arxiv.org/abs/1105.2551} {arXiv:1105.2551 [hep-th]}
  \BibitemShut {NoStop}%
\bibitem [{\citenamefont {Jafferis}(2012)}]{Jafferis:2010un}%
  \BibitemOpen
  \bibfield  {author} {\bibinfo {author} {\bibfnamefont {D.~L.}\ \bibnamefont
  {Jafferis}},\ }\bibfield  {title} {\bibinfo {title} {{The Exact
  Superconformal R-Symmetry Extremizes Z}},\ }\href
  {https://doi.org/10.1007/JHEP05(2012)159} {\bibfield  {journal} {\bibinfo
  {journal} {JHEP}\ }\textbf {\bibinfo {volume} {05}},\ \bibinfo {pages}
  {159}},\ \Eprint {https://arxiv.org/abs/1012.3210} {arXiv:1012.3210 [hep-th]}
  \BibitemShut {NoStop}%
\bibitem [{\citenamefont {Hama}\ \emph
  {et~al.}(2011{\natexlab{a}})\citenamefont {Hama}, \citenamefont {Hosomichi},\
  and\ \citenamefont {Lee}}]{Hama:2010av}%
  \BibitemOpen
  \bibfield  {author} {\bibinfo {author} {\bibfnamefont {N.}~\bibnamefont
  {Hama}}, \bibinfo {author} {\bibfnamefont {K.}~\bibnamefont {Hosomichi}},\
  and\ \bibinfo {author} {\bibfnamefont {S.}~\bibnamefont {Lee}},\ }\bibfield
  {title} {\bibinfo {title} {{Notes on SUSY Gauge Theories on Three-Sphere}},\
  }\href {https://doi.org/10.1007/JHEP03(2011)127} {\bibfield  {journal}
  {\bibinfo  {journal} {JHEP}\ }\textbf {\bibinfo {volume} {03}},\ \bibinfo
  {pages} {127}},\ \Eprint {https://arxiv.org/abs/1012.3512} {arXiv:1012.3512
  [hep-th]} \BibitemShut {NoStop}%
\bibitem [{\citenamefont {Hama}\ \emph
  {et~al.}(2011{\natexlab{b}})\citenamefont {Hama}, \citenamefont {Hosomichi},\
  and\ \citenamefont {Lee}}]{Hama:2011ea}%
  \BibitemOpen
  \bibfield  {author} {\bibinfo {author} {\bibfnamefont {N.}~\bibnamefont
  {Hama}}, \bibinfo {author} {\bibfnamefont {K.}~\bibnamefont {Hosomichi}},\
  and\ \bibinfo {author} {\bibfnamefont {S.}~\bibnamefont {Lee}},\ }\bibfield
  {title} {\bibinfo {title} {{SUSY Gauge Theories on Squashed Three-Spheres}},\
  }\href {https://doi.org/10.1007/JHEP05(2011)014} {\bibfield  {journal}
  {\bibinfo  {journal} {JHEP}\ }\textbf {\bibinfo {volume} {05}},\ \bibinfo
  {pages} {014}},\ \Eprint {https://arxiv.org/abs/1102.4716} {arXiv:1102.4716
  [hep-th]} \BibitemShut {NoStop}%
\bibitem [{\citenamefont {Willett}(2017)}]{Willett:2016adv}%
  \BibitemOpen
  \bibfield  {author} {\bibinfo {author} {\bibfnamefont {B.}~\bibnamefont
  {Willett}},\ }\bibfield  {title} {\bibinfo {title} {{Localization on
  three-dimensional manifolds}},\ }\href
  {https://doi.org/10.1088/1751-8121/aa612f} {\bibfield  {journal} {\bibinfo
  {journal} {J. Phys. A}\ }\textbf {\bibinfo {volume} {50}},\ \bibinfo {pages}
  {443006} (\bibinfo {year} {2017})},\ \Eprint
  {https://arxiv.org/abs/1608.02958} {arXiv:1608.02958 [hep-th]} \BibitemShut
  {NoStop}%
\bibitem [{\citenamefont {Benini}\ and\ \citenamefont
  {Zaffaroni}(2015)}]{Benini:2015noa}%
  \BibitemOpen
  \bibfield  {author} {\bibinfo {author} {\bibfnamefont {F.}~\bibnamefont
  {Benini}}\ and\ \bibinfo {author} {\bibfnamefont {A.}~\bibnamefont
  {Zaffaroni}},\ }\bibfield  {title} {\bibinfo {title} {{A topologically
  twisted index for three-dimensional supersymmetric theories}},\ }\href
  {https://doi.org/10.1007/JHEP07(2015)127} {\bibfield  {journal} {\bibinfo
  {journal} {JHEP}\ }\textbf {\bibinfo {volume} {07}},\ \bibinfo {pages}
  {127}},\ \Eprint {https://arxiv.org/abs/1504.03698} {arXiv:1504.03698
  [hep-th]} \BibitemShut {NoStop}%
\bibitem [{\citenamefont {Benini}\ and\ \citenamefont
  {Zaffaroni}(2017)}]{Benini:2016hjo}%
  \BibitemOpen
  \bibfield  {author} {\bibinfo {author} {\bibfnamefont {F.}~\bibnamefont
  {Benini}}\ and\ \bibinfo {author} {\bibfnamefont {A.}~\bibnamefont
  {Zaffaroni}},\ }\bibfield  {title} {\bibinfo {title} {{Supersymmetric
  partition functions on Riemann surfaces}},\ }\href@noop {} {\bibfield
  {journal} {\bibinfo  {journal} {Proc. Symp. Pure Math.}\ }\textbf {\bibinfo
  {volume} {96}},\ \bibinfo {pages} {13} (\bibinfo {year} {2017})},\ \Eprint
  {https://arxiv.org/abs/1605.06120} {arXiv:1605.06120 [hep-th]} \BibitemShut
  {NoStop}%
\bibitem [{\citenamefont {Closset}\ and\ \citenamefont
  {Kim}(2016)}]{Closset:2016arn}%
  \BibitemOpen
  \bibfield  {author} {\bibinfo {author} {\bibfnamefont {C.}~\bibnamefont
  {Closset}}\ and\ \bibinfo {author} {\bibfnamefont {H.}~\bibnamefont {Kim}},\
  }\bibfield  {title} {\bibinfo {title} {{Comments on twisted indices in 3d
  supersymmetric gauge theories}},\ }\href
  {https://doi.org/10.1007/JHEP08(2016)059} {\bibfield  {journal} {\bibinfo
  {journal} {JHEP}\ }\textbf {\bibinfo {volume} {08}},\ \bibinfo {pages}
  {059}},\ \Eprint {https://arxiv.org/abs/1605.06531} {arXiv:1605.06531
  [hep-th]} \BibitemShut {NoStop}%
\bibitem [{\citenamefont {Nekrasov}\ and\ \citenamefont
  {Shatashvili}(2009)}]{Nekrasov:2009uh}%
  \BibitemOpen
  \bibfield  {author} {\bibinfo {author} {\bibfnamefont {N.~A.}\ \bibnamefont
  {Nekrasov}}\ and\ \bibinfo {author} {\bibfnamefont {S.~L.}\ \bibnamefont
  {Shatashvili}},\ }\bibfield  {title} {\bibinfo {title} {{Supersymmetric vacua
  and Bethe ansatz}},\ }\href
  {https://doi.org/10.1016/j.nuclphysbps.2009.07.047} {\bibfield  {journal}
  {\bibinfo  {journal} {Nucl. Phys. B Proc. Suppl.}\ }\textbf {\bibinfo
  {volume} {192-193}},\ \bibinfo {pages} {91} (\bibinfo {year} {2009})},\
  \Eprint {https://arxiv.org/abs/0901.4744} {arXiv:0901.4744 [hep-th]}
  \BibitemShut {NoStop}%
\bibitem [{\citenamefont {Okuda}\ and\ \citenamefont
  {Yoshida}(2012)}]{Okuda:2012nx}%
  \BibitemOpen
  \bibfield  {author} {\bibinfo {author} {\bibfnamefont {S.}~\bibnamefont
  {Okuda}}\ and\ \bibinfo {author} {\bibfnamefont {Y.}~\bibnamefont
  {Yoshida}},\ }\bibfield  {title} {\bibinfo {title} {{G/G gauged WZW model and
  Bethe Ansatz for the phase model}},\ }\href
  {https://doi.org/10.1007/JHEP11(2012)146} {\bibfield  {journal} {\bibinfo
  {journal} {JHEP}\ }\textbf {\bibinfo {volume} {11}},\ \bibinfo {pages}
  {146}},\ \Eprint {https://arxiv.org/abs/1209.3800} {arXiv:1209.3800 [hep-th]}
  \BibitemShut {NoStop}%
\bibitem [{\citenamefont {Okuda}\ and\ \citenamefont
  {Yoshida}(2014)}]{Okuda:2013fea}%
  \BibitemOpen
  \bibfield  {author} {\bibinfo {author} {\bibfnamefont {S.}~\bibnamefont
  {Okuda}}\ and\ \bibinfo {author} {\bibfnamefont {Y.}~\bibnamefont
  {Yoshida}},\ }\bibfield  {title} {\bibinfo {title} {{G/G gauged WZW-matter
  model, Bethe Ansatz for q-boson model and Commutative Frobenius algebra}},\
  }\href {https://doi.org/10.1007/JHEP03(2014)003} {\bibfield  {journal}
  {\bibinfo  {journal} {JHEP}\ }\textbf {\bibinfo {volume} {03}},\ \bibinfo
  {pages} {003}},\ \Eprint {https://arxiv.org/abs/1308.4608} {arXiv:1308.4608
  [hep-th]} \BibitemShut {NoStop}%
\bibitem [{\citenamefont {Nekrasov}\ and\ \citenamefont
  {Shatashvili}(2015)}]{Nekrasov:2014xaa}%
  \BibitemOpen
  \bibfield  {author} {\bibinfo {author} {\bibfnamefont {N.~A.}\ \bibnamefont
  {Nekrasov}}\ and\ \bibinfo {author} {\bibfnamefont {S.~L.}\ \bibnamefont
  {Shatashvili}},\ }\bibfield  {title} {\bibinfo {title} {{Bethe/Gauge
  correspondence on curved spaces}},\ }\href
  {https://doi.org/10.1007/JHEP01(2015)100} {\bibfield  {journal} {\bibinfo
  {journal} {JHEP}\ }\textbf {\bibinfo {volume} {01}},\ \bibinfo {pages}
  {100}},\ \Eprint {https://arxiv.org/abs/1405.6046} {arXiv:1405.6046 [hep-th]}
  \BibitemShut {NoStop}%
\bibitem [{\citenamefont {Okuda}\ and\ \citenamefont
  {Yoshida}(2015)}]{Okuda:2015yea}%
  \BibitemOpen
  \bibfield  {author} {\bibinfo {author} {\bibfnamefont {S.}~\bibnamefont
  {Okuda}}\ and\ \bibinfo {author} {\bibfnamefont {Y.}~\bibnamefont
  {Yoshida}},\ }\bibfield  {title} {\bibinfo {title} {{Gauge/Bethe
  correspondence on $S^1 \times \Sigma_h$ and index over moduli space}},\
  }\href@noop {} {\  (\bibinfo {year} {2015})},\ \Eprint
  {https://arxiv.org/abs/1501.03469} {arXiv:1501.03469 [hep-th]} \BibitemShut
  {NoStop}%
\bibitem [{\citenamefont {Gukov}\ and\ \citenamefont
  {Pei}(2017)}]{Gukov:2015sna}%
  \BibitemOpen
  \bibfield  {author} {\bibinfo {author} {\bibfnamefont {S.}~\bibnamefont
  {Gukov}}\ and\ \bibinfo {author} {\bibfnamefont {D.}~\bibnamefont {Pei}},\
  }\bibfield  {title} {\bibinfo {title} {{Equivariant Verlinde formula from
  fivebranes and vortices}},\ }\href
  {https://doi.org/10.1007/s00220-017-2931-9} {\bibfield  {journal} {\bibinfo
  {journal} {Commun. Math. Phys.}\ }\textbf {\bibinfo {volume} {355}},\
  \bibinfo {pages} {1} (\bibinfo {year} {2017})},\ \Eprint
  {https://arxiv.org/abs/1501.01310} {arXiv:1501.01310 [hep-th]} \BibitemShut
  {NoStop}%
\bibitem [{\citenamefont {Closset}\ \emph
  {et~al.}(2017{\natexlab{a}})\citenamefont {Closset}, \citenamefont {Kim},\
  and\ \citenamefont {Willett}}]{Closset:2017zgf}%
  \BibitemOpen
  \bibfield  {author} {\bibinfo {author} {\bibfnamefont {C.}~\bibnamefont
  {Closset}}, \bibinfo {author} {\bibfnamefont {H.}~\bibnamefont {Kim}},\ and\
  \bibinfo {author} {\bibfnamefont {B.}~\bibnamefont {Willett}},\ }\bibfield
  {title} {\bibinfo {title} {{Supersymmetric partition functions and the
  three-dimensional A-twist}},\ }\href
  {https://doi.org/10.1007/JHEP03(2017)074} {\bibfield  {journal} {\bibinfo
  {journal} {JHEP}\ }\textbf {\bibinfo {volume} {03}},\ \bibinfo {pages}
  {074}},\ \Eprint {https://arxiv.org/abs/1701.03171} {arXiv:1701.03171
  [hep-th]} \BibitemShut {NoStop}%
\bibitem [{\citenamefont {Closset}\ \emph
  {et~al.}(2017{\natexlab{b}})\citenamefont {Closset}, \citenamefont {Kim},\
  and\ \citenamefont {Willett}}]{Closset:2017bse}%
  \BibitemOpen
  \bibfield  {author} {\bibinfo {author} {\bibfnamefont {C.}~\bibnamefont
  {Closset}}, \bibinfo {author} {\bibfnamefont {H.}~\bibnamefont {Kim}},\ and\
  \bibinfo {author} {\bibfnamefont {B.}~\bibnamefont {Willett}},\ }\bibfield
  {title} {\bibinfo {title} {{$ \mathcal{N} $ = 1 supersymmetric indices and
  the four-dimensional A-model}},\ }\href
  {https://doi.org/10.1007/JHEP08(2017)090} {\bibfield  {journal} {\bibinfo
  {journal} {JHEP}\ }\textbf {\bibinfo {volume} {08}},\ \bibinfo {pages}
  {090}},\ \Eprint {https://arxiv.org/abs/1707.05774} {arXiv:1707.05774
  [hep-th]} \BibitemShut {NoStop}%
\bibitem [{\citenamefont {Closset}\ \emph {et~al.}(2018)\citenamefont
  {Closset}, \citenamefont {Kim},\ and\ \citenamefont
  {Willett}}]{Closset:2018ghr}%
  \BibitemOpen
  \bibfield  {author} {\bibinfo {author} {\bibfnamefont {C.}~\bibnamefont
  {Closset}}, \bibinfo {author} {\bibfnamefont {H.}~\bibnamefont {Kim}},\ and\
  \bibinfo {author} {\bibfnamefont {B.}~\bibnamefont {Willett}},\ }\bibfield
  {title} {\bibinfo {title} {{Seifert fibering operators in 3d $\mathcal{N}=2$
  theories}},\ }\href {https://doi.org/10.1007/JHEP11(2018)004} {\bibfield
  {journal} {\bibinfo  {journal} {JHEP}\ }\textbf {\bibinfo {volume} {11}},\
  \bibinfo {pages} {004}},\ \Eprint {https://arxiv.org/abs/1807.02328}
  {arXiv:1807.02328 [hep-th]} \BibitemShut {NoStop}%
\bibitem [{\citenamefont {Jain}\ and\ \citenamefont
  {Ray}(2019)}]{Jain:2019lqb}%
  \BibitemOpen
  \bibfield  {author} {\bibinfo {author} {\bibfnamefont {D.}~\bibnamefont
  {Jain}}\ and\ \bibinfo {author} {\bibfnamefont {A.}~\bibnamefont {Ray}},\
  }\bibfield  {title} {\bibinfo {title} {{3d $\mathcal{N}=2$ $\widehat{ADE}$
  Chern-Simons quivers}},\ }\href {https://doi.org/10.1103/PhysRevD.100.046007}
  {\bibfield  {journal} {\bibinfo  {journal} {Phys. Rev. D}\ }\textbf {\bibinfo
  {volume} {100}},\ \bibinfo {pages} {046007} (\bibinfo {year} {2019})},\
  \Eprint {https://arxiv.org/abs/1902.10498} {arXiv:1902.10498 [hep-th]}
  \BibitemShut {NoStop}%
\bibitem [{\citenamefont {Jain}(2019)}]{Jain:2019euv}%
  \BibitemOpen
  \bibfield  {author} {\bibinfo {author} {\bibfnamefont {D.}~\bibnamefont
  {Jain}},\ }\bibfield  {title} {\bibinfo {title} {{Twisted Indices of more 3d
  Quivers}},\ }\href@noop {} {\  (\bibinfo {year} {2019})},\ \Eprint
  {https://arxiv.org/abs/1908.03035} {arXiv:1908.03035 [hep-th]} \BibitemShut
  {NoStop}%
\bibitem [{\citenamefont {Hosseini}\ and\ \citenamefont
  {Mekareeya}(2016)}]{Hosseini:2016ume}%
  \BibitemOpen
  \bibfield  {author} {\bibinfo {author} {\bibfnamefont {S.~M.}\ \bibnamefont
  {Hosseini}}\ and\ \bibinfo {author} {\bibfnamefont {N.}~\bibnamefont
  {Mekareeya}},\ }\bibfield  {title} {\bibinfo {title} {{Large $N$
  topologically twisted index: necklace quivers, dualities, and Sasaki-Einstein
  spaces}},\ }\href {https://doi.org/10.1007/JHEP08(2016)089} {\bibfield
  {journal} {\bibinfo  {journal} {JHEP}\ }\textbf {\bibinfo {volume} {08}},\
  \bibinfo {pages} {089}},\ \Eprint {https://arxiv.org/abs/1604.03397}
  {arXiv:1604.03397 [hep-th]} \BibitemShut {NoStop}%
\bibitem [{\citenamefont {Azzurli}\ \emph {et~al.}(2018)\citenamefont
  {Azzurli}, \citenamefont {Bobev}, \citenamefont {Crichigno}, \citenamefont
  {Min},\ and\ \citenamefont {Zaffaroni}}]{Azzurli:2017kxo}%
  \BibitemOpen
  \bibfield  {author} {\bibinfo {author} {\bibfnamefont {F.}~\bibnamefont
  {Azzurli}}, \bibinfo {author} {\bibfnamefont {N.}~\bibnamefont {Bobev}},
  \bibinfo {author} {\bibfnamefont {P.~M.}\ \bibnamefont {Crichigno}}, \bibinfo
  {author} {\bibfnamefont {V.~S.}\ \bibnamefont {Min}},\ and\ \bibinfo {author}
  {\bibfnamefont {A.}~\bibnamefont {Zaffaroni}},\ }\bibfield  {title} {\bibinfo
  {title} {{A universal counting of black hole microstates in AdS$_{4}$}},\
  }\href {https://doi.org/10.1007/JHEP02(2018)054} {\bibfield  {journal}
  {\bibinfo  {journal} {JHEP}\ }\textbf {\bibinfo {volume} {02}},\ \bibinfo
  {pages} {054}},\ \Eprint {https://arxiv.org/abs/1707.04257} {arXiv:1707.04257
  [hep-th]} \BibitemShut {NoStop}%
\bibitem [{\citenamefont {Benini}\ \emph {et~al.}(2016)\citenamefont {Benini},
  \citenamefont {Hristov},\ and\ \citenamefont {Zaffaroni}}]{Benini:2015eyy}%
  \BibitemOpen
  \bibfield  {author} {\bibinfo {author} {\bibfnamefont {F.}~\bibnamefont
  {Benini}}, \bibinfo {author} {\bibfnamefont {K.}~\bibnamefont {Hristov}},\
  and\ \bibinfo {author} {\bibfnamefont {A.}~\bibnamefont {Zaffaroni}},\
  }\bibfield  {title} {\bibinfo {title} {{Black hole microstates in AdS$_{4}$
  from supersymmetric localization}},\ }\href
  {https://doi.org/10.1007/JHEP05(2016)054} {\bibfield  {journal} {\bibinfo
  {journal} {JHEP}\ }\textbf {\bibinfo {volume} {05}},\ \bibinfo {pages}
  {054}},\ \Eprint {https://arxiv.org/abs/1511.04085} {arXiv:1511.04085
  [hep-th]} \BibitemShut {NoStop}%
\bibitem [{\citenamefont {Zaffaroni}(2020)}]{Zaffaroni:2019dhb}%
  \BibitemOpen
  \bibfield  {author} {\bibinfo {author} {\bibfnamefont {A.}~\bibnamefont
  {Zaffaroni}},\ }\bibfield  {title} {\bibinfo {title} {{AdS black holes,
  holography and localization}},\ }\href
  {https://doi.org/10.1007/s41114-020-00027-8} {\bibfield  {journal} {\bibinfo
  {journal} {Living Rev. Rel.}\ }\textbf {\bibinfo {volume} {23}},\ \bibinfo
  {pages} {2} (\bibinfo {year} {2020})},\ \Eprint
  {https://arxiv.org/abs/1902.07176} {arXiv:1902.07176 [hep-th]} \BibitemShut
  {NoStop}%
\bibitem [{\citenamefont {Gauntlett}\ and\ \citenamefont
  {Varela}(2008)}]{Gauntlett:2007sm}%
  \BibitemOpen
  \bibfield  {author} {\bibinfo {author} {\bibfnamefont {J.~P.}\ \bibnamefont
  {Gauntlett}}\ and\ \bibinfo {author} {\bibfnamefont {O.}~\bibnamefont
  {Varela}},\ }\bibfield  {title} {\bibinfo {title} {{D=5 SU(2) x U(1) Gauged
  Supergravity from D=11 Supergravity}},\ }\href
  {https://doi.org/10.1088/1126-6708/2008/02/083} {\bibfield  {journal}
  {\bibinfo  {journal} {JHEP}\ }\textbf {\bibinfo {volume} {02}},\ \bibinfo
  {pages} {083}},\ \Eprint {https://arxiv.org/abs/0712.3560} {arXiv:0712.3560
  [hep-th]} \BibitemShut {NoStop}%
\bibitem [{\citenamefont {Romans}(1992)}]{Romans:1991nq}%
  \BibitemOpen
  \bibfield  {author} {\bibinfo {author} {\bibfnamefont {L.}~\bibnamefont
  {Romans}},\ }\bibfield  {title} {\bibinfo {title} {{Supersymmetric, cold and
  lukewarm black holes in cosmological Einstein-Maxwell theory}},\ }\href
  {https://doi.org/10.1016/0550-3213(92)90684-4} {\bibfield  {journal}
  {\bibinfo  {journal} {Nucl. Phys. B}\ }\textbf {\bibinfo {volume} {383}},\
  \bibinfo {pages} {395} (\bibinfo {year} {1992})},\ \Eprint
  {https://arxiv.org/abs/hep-th/9203018} {arXiv:hep-th/9203018} \BibitemShut
  {NoStop}%
\bibitem [{\citenamefont {Caldarelli}\ and\ \citenamefont
  {Klemm}(1999)}]{Caldarelli:1998hg}%
  \BibitemOpen
  \bibfield  {author} {\bibinfo {author} {\bibfnamefont {M.~M.}\ \bibnamefont
  {Caldarelli}}\ and\ \bibinfo {author} {\bibfnamefont {D.}~\bibnamefont
  {Klemm}},\ }\bibfield  {title} {\bibinfo {title} {{Supersymmetry of Anti-de
  Sitter black holes}},\ }\href {https://doi.org/10.1016/S0550-3213(98)00846-3}
  {\bibfield  {journal} {\bibinfo  {journal} {Nucl. Phys. B}\ }\textbf
  {\bibinfo {volume} {545}},\ \bibinfo {pages} {434} (\bibinfo {year}
  {1999})},\ \Eprint {https://arxiv.org/abs/hep-th/9808097}
  {arXiv:hep-th/9808097} \BibitemShut {NoStop}%
\bibitem [{\citenamefont {Chaney}\ and\ \citenamefont
  {Uhlemann}(2018)}]{Chaney:2018gjc}%
  \BibitemOpen
  \bibfield  {author} {\bibinfo {author} {\bibfnamefont {A.}~\bibnamefont
  {Chaney}}\ and\ \bibinfo {author} {\bibfnamefont {C.~F.}\ \bibnamefont
  {Uhlemann}},\ }\bibfield  {title} {\bibinfo {title} {{On minimal Type IIB
  AdS$_{6}$ solutions with commuting 7-branes}},\ }\href
  {https://doi.org/10.1007/JHEP12(2018)110} {\bibfield  {journal} {\bibinfo
  {journal} {JHEP}\ }\textbf {\bibinfo {volume} {12}},\ \bibinfo {pages}
  {110}},\ \Eprint {https://arxiv.org/abs/1810.10592} {arXiv:1810.10592
  [hep-th]} \BibitemShut {NoStop}%
\bibitem [{\citenamefont {Uhlemann}(2020)}]{Uhlemann:2020bek}%
  \BibitemOpen
  \bibfield  {author} {\bibinfo {author} {\bibfnamefont {C.~F.}\ \bibnamefont
  {Uhlemann}},\ }\bibfield  {title} {\bibinfo {title} {{Wilson loops in 5d long
  quiver gauge theories}},\ }\href@noop {} {\  (\bibinfo {year} {2020})},\
  \Eprint {https://arxiv.org/abs/2006.01142} {arXiv:2006.01142 [hep-th]}
  \BibitemShut {NoStop}%
\bibitem [{\citenamefont {Zagier}(2007)}]{Zagier2007}%
  \BibitemOpen
  \bibfield  {author} {\bibinfo {author} {\bibfnamefont {D.}~\bibnamefont
  {Zagier}},\ }\bibinfo {title} {The dilogarithm function},\ in\ \href
  {https://doi.org/{10.1007/978-3-540-30308-4\_1}} {\emph {\bibinfo {booktitle}
  {Frontiers in Number Theory, Physics, and Geometry II: On Conformal Field
  Theories, Discrete Groups and Renormalization}}},\ \bibinfo {editor} {edited
  by\ \bibinfo {editor} {\bibfnamefont {P.}~\bibnamefont {Cartier}}, \bibinfo
  {editor} {\bibfnamefont {P.}~\bibnamefont {Moussa}}, \bibinfo {editor}
  {\bibfnamefont {B.}~\bibnamefont {Julia}},\ and\ \bibinfo {editor}
  {\bibfnamefont {P.}~\bibnamefont {Vanhove}}}\ (\bibinfo  {publisher}
  {Springer Berlin Heidelberg},\ \bibinfo {address} {Berlin, Heidelberg},\
  \bibinfo {year} {2007})\ pp.\ \bibinfo {pages} {3--65}\BibitemShut {NoStop}%
\bibitem [{\citenamefont {Intriligator}\ \emph {et~al.}(1997)\citenamefont
  {Intriligator}, \citenamefont {Morrison},\ and\ \citenamefont
  {Seiberg}}]{Intriligator:1997pq}%
  \BibitemOpen
  \bibfield  {author} {\bibinfo {author} {\bibfnamefont {K.~A.}\ \bibnamefont
  {Intriligator}}, \bibinfo {author} {\bibfnamefont {D.~R.}\ \bibnamefont
  {Morrison}},\ and\ \bibinfo {author} {\bibfnamefont {N.}~\bibnamefont
  {Seiberg}},\ }\bibfield  {title} {\bibinfo {title} {{Five-dimensional
  supersymmetric gauge theories and degenerations of Calabi-Yau spaces}},\
  }\href {https://doi.org/10.1016/S0550-3213(97)00279-4} {\bibfield  {journal}
  {\bibinfo  {journal} {Nucl. Phys. B}\ }\textbf {\bibinfo {volume} {497}},\
  \bibinfo {pages} {56} (\bibinfo {year} {1997})},\ \Eprint
  {https://arxiv.org/abs/hep-th/9702198} {arXiv:hep-th/9702198} \BibitemShut
  {NoStop}%
\bibitem [{\citenamefont {Bergman}\ and\ \citenamefont
  {Zafrir}(2015{\natexlab{a}})}]{Bergman:2015dpa}%
  \BibitemOpen
  \bibfield  {author} {\bibinfo {author} {\bibfnamefont {O.}~\bibnamefont
  {Bergman}}\ and\ \bibinfo {author} {\bibfnamefont {G.}~\bibnamefont
  {Zafrir}},\ }\bibfield  {title} {\bibinfo {title} {{5d fixed points from
  brane webs and O7-planes}},\ }\href {https://doi.org/10.1007/JHEP12(2015)163}
  {\bibfield  {journal} {\bibinfo  {journal} {JHEP}\ }\textbf {\bibinfo
  {volume} {12}},\ \bibinfo {pages} {163}},\ \Eprint
  {https://arxiv.org/abs/1507.03860} {arXiv:1507.03860 [hep-th]} \BibitemShut
  {NoStop}%
\bibitem [{\citenamefont {Fluder}\ \emph {et~al.}(2019)\citenamefont {Fluder},
  \citenamefont {Hosseini},\ and\ \citenamefont {Uhlemann}}]{Fluder:2019szh}%
  \BibitemOpen
  \bibfield  {author} {\bibinfo {author} {\bibfnamefont {M.}~\bibnamefont
  {Fluder}}, \bibinfo {author} {\bibfnamefont {S.~M.}\ \bibnamefont
  {Hosseini}},\ and\ \bibinfo {author} {\bibfnamefont {C.~F.}\ \bibnamefont
  {Uhlemann}},\ }\bibfield  {title} {\bibinfo {title} {{Black hole microstate
  counting in Type IIB from 5d SCFTs}},\ }\href
  {https://doi.org/10.1007/JHEP05(2019)134} {\bibfield  {journal} {\bibinfo
  {journal} {JHEP}\ }\textbf {\bibinfo {volume} {05}},\ \bibinfo {pages}
  {134}},\ \Eprint {https://arxiv.org/abs/1902.05074} {arXiv:1902.05074
  [hep-th]} \BibitemShut {NoStop}%
\bibitem [{\citenamefont {Benini}\ \emph {et~al.}(2010)\citenamefont {Benini},
  \citenamefont {Tachikawa},\ and\ \citenamefont {Xie}}]{Benini:2010uu}%
  \BibitemOpen
  \bibfield  {author} {\bibinfo {author} {\bibfnamefont {F.}~\bibnamefont
  {Benini}}, \bibinfo {author} {\bibfnamefont {Y.}~\bibnamefont {Tachikawa}},\
  and\ \bibinfo {author} {\bibfnamefont {D.}~\bibnamefont {Xie}},\ }\bibfield
  {title} {\bibinfo {title} {{Mirrors of 3d Sicilian theories}},\ }\href
  {https://doi.org/10.1007/JHEP09(2010)063} {\bibfield  {journal} {\bibinfo
  {journal} {JHEP}\ }\textbf {\bibinfo {volume} {09}},\ \bibinfo {pages}
  {063}},\ \Eprint {https://arxiv.org/abs/1007.0992} {arXiv:1007.0992 [hep-th]}
  \BibitemShut {NoStop}%
\bibitem [{\citenamefont {Aharony}\ and\ \citenamefont
  {Hanany}(1997)}]{Aharony:1997ju}%
  \BibitemOpen
  \bibfield  {author} {\bibinfo {author} {\bibfnamefont {O.}~\bibnamefont
  {Aharony}}\ and\ \bibinfo {author} {\bibfnamefont {A.}~\bibnamefont
  {Hanany}},\ }\bibfield  {title} {\bibinfo {title} {{Branes, superpotentials
  and superconformal fixed points}},\ }\href
  {https://doi.org/10.1016/S0550-3213(97)00472-0} {\bibfield  {journal}
  {\bibinfo  {journal} {Nucl. Phys. B}\ }\textbf {\bibinfo {volume} {504}},\
  \bibinfo {pages} {239} (\bibinfo {year} {1997})},\ \Eprint
  {https://arxiv.org/abs/hep-th/9704170} {arXiv:hep-th/9704170} \BibitemShut
  {NoStop}%
\bibitem [{\citenamefont {Aharony}\ \emph {et~al.}(1998)\citenamefont
  {Aharony}, \citenamefont {Hanany},\ and\ \citenamefont
  {Kol}}]{Aharony:1997bh}%
  \BibitemOpen
  \bibfield  {author} {\bibinfo {author} {\bibfnamefont {O.}~\bibnamefont
  {Aharony}}, \bibinfo {author} {\bibfnamefont {A.}~\bibnamefont {Hanany}},\
  and\ \bibinfo {author} {\bibfnamefont {B.}~\bibnamefont {Kol}},\ }\bibfield
  {title} {\bibinfo {title} {{Webs of (p,q) five-branes, five-dimensional field
  theories and grid diagrams}},\ }\href
  {https://doi.org/10.1088/1126-6708/1998/01/002} {\bibfield  {journal}
  {\bibinfo  {journal} {JHEP}\ }\textbf {\bibinfo {volume} {01}},\ \bibinfo
  {pages} {002}},\ \Eprint {https://arxiv.org/abs/hep-th/9710116}
  {arXiv:hep-th/9710116} \BibitemShut {NoStop}%
\bibitem [{\citenamefont {DeWolfe}\ \emph {et~al.}(1999)\citenamefont
  {DeWolfe}, \citenamefont {Hanany}, \citenamefont {Iqbal},\ and\ \citenamefont
  {Katz}}]{DeWolfe:1999hj}%
  \BibitemOpen
  \bibfield  {author} {\bibinfo {author} {\bibfnamefont {O.}~\bibnamefont
  {DeWolfe}}, \bibinfo {author} {\bibfnamefont {A.}~\bibnamefont {Hanany}},
  \bibinfo {author} {\bibfnamefont {A.}~\bibnamefont {Iqbal}},\ and\ \bibinfo
  {author} {\bibfnamefont {E.}~\bibnamefont {Katz}},\ }\bibfield  {title}
  {\bibinfo {title} {{Five-branes, seven-branes and five-dimensional E(n) field
  theories}},\ }\href {https://doi.org/10.1088/1126-6708/1999/03/006}
  {\bibfield  {journal} {\bibinfo  {journal} {JHEP}\ }\textbf {\bibinfo
  {volume} {03}},\ \bibinfo {pages} {006}},\ \Eprint
  {https://arxiv.org/abs/hep-th/9902179} {arXiv:hep-th/9902179} \BibitemShut
  {NoStop}%
\bibitem [{\citenamefont {Benini}\ \emph {et~al.}(2009)\citenamefont {Benini},
  \citenamefont {Benvenuti},\ and\ \citenamefont {Tachikawa}}]{Benini:2009gi}%
  \BibitemOpen
  \bibfield  {author} {\bibinfo {author} {\bibfnamefont {F.}~\bibnamefont
  {Benini}}, \bibinfo {author} {\bibfnamefont {S.}~\bibnamefont {Benvenuti}},\
  and\ \bibinfo {author} {\bibfnamefont {Y.}~\bibnamefont {Tachikawa}},\
  }\bibfield  {title} {\bibinfo {title} {{Webs of five-branes and N=2
  superconformal field theories}},\ }\href
  {https://doi.org/10.1088/1126-6708/2009/09/052} {\bibfield  {journal}
  {\bibinfo  {journal} {JHEP}\ }\textbf {\bibinfo {volume} {09}},\ \bibinfo
  {pages} {052}},\ \Eprint {https://arxiv.org/abs/0906.0359} {arXiv:0906.0359
  [hep-th]} \BibitemShut {NoStop}%
\bibitem [{\citenamefont {Bergman}\ and\ \citenamefont
  {Zafrir}(2015{\natexlab{b}})}]{Bergman:2014kza}%
  \BibitemOpen
  \bibfield  {author} {\bibinfo {author} {\bibfnamefont {O.}~\bibnamefont
  {Bergman}}\ and\ \bibinfo {author} {\bibfnamefont {G.}~\bibnamefont
  {Zafrir}},\ }\bibfield  {title} {\bibinfo {title} {{Lifting 4d dualities to
  5d}},\ }\href {https://doi.org/10.1007/JHEP04(2015)141} {\bibfield  {journal}
  {\bibinfo  {journal} {JHEP}\ }\textbf {\bibinfo {volume} {04}},\ \bibinfo
  {pages} {141}},\ \Eprint {https://arxiv.org/abs/1410.2806} {arXiv:1410.2806
  [hep-th]} \BibitemShut {NoStop}%
\bibitem [{\citenamefont {D'Hoker}\ \emph {et~al.}(2016)\citenamefont
  {D'Hoker}, \citenamefont {Gutperle}, \citenamefont {Karch},\ and\
  \citenamefont {Uhlemann}}]{DHoker:2016ujz}%
  \BibitemOpen
  \bibfield  {author} {\bibinfo {author} {\bibfnamefont {E.}~\bibnamefont
  {D'Hoker}}, \bibinfo {author} {\bibfnamefont {M.}~\bibnamefont {Gutperle}},
  \bibinfo {author} {\bibfnamefont {A.}~\bibnamefont {Karch}},\ and\ \bibinfo
  {author} {\bibfnamefont {C.~F.}\ \bibnamefont {Uhlemann}},\ }\bibfield
  {title} {\bibinfo {title} {{Warped $AdS_6\times S^2$ in Type IIB supergravity
  I: Local solutions}},\ }\href {https://doi.org/10.1007/JHEP08(2016)046}
  {\bibfield  {journal} {\bibinfo  {journal} {JHEP}\ }\textbf {\bibinfo
  {volume} {08}},\ \bibinfo {pages} {046}},\ \Eprint
  {https://arxiv.org/abs/1606.01254} {arXiv:1606.01254 [hep-th]} \BibitemShut
  {NoStop}%
\bibitem [{\citenamefont {D'Hoker}\ \emph
  {et~al.}(2017{\natexlab{a}})\citenamefont {D'Hoker}, \citenamefont
  {Gutperle},\ and\ \citenamefont {Uhlemann}}]{DHoker:2016ysh}%
  \BibitemOpen
  \bibfield  {author} {\bibinfo {author} {\bibfnamefont {E.}~\bibnamefont
  {D'Hoker}}, \bibinfo {author} {\bibfnamefont {M.}~\bibnamefont {Gutperle}},\
  and\ \bibinfo {author} {\bibfnamefont {C.~F.}\ \bibnamefont {Uhlemann}},\
  }\bibfield  {title} {\bibinfo {title} {{Holographic duals for
  five-dimensional superconformal quantum field theories}},\ }\href
  {https://doi.org/10.1103/PhysRevLett.118.101601} {\bibfield  {journal}
  {\bibinfo  {journal} {Phys. Rev. Lett.}\ }\textbf {\bibinfo {volume} {118}},\
  \bibinfo {pages} {101601} (\bibinfo {year} {2017}{\natexlab{a}})},\ \Eprint
  {https://arxiv.org/abs/1611.09411} {arXiv:1611.09411 [hep-th]} \BibitemShut
  {NoStop}%
\bibitem [{\citenamefont {D'Hoker}\ \emph
  {et~al.}(2017{\natexlab{b}})\citenamefont {D'Hoker}, \citenamefont
  {Gutperle},\ and\ \citenamefont {Uhlemann}}]{DHoker:2017mds}%
  \BibitemOpen
  \bibfield  {author} {\bibinfo {author} {\bibfnamefont {E.}~\bibnamefont
  {D'Hoker}}, \bibinfo {author} {\bibfnamefont {M.}~\bibnamefont {Gutperle}},\
  and\ \bibinfo {author} {\bibfnamefont {C.~F.}\ \bibnamefont {Uhlemann}},\
  }\bibfield  {title} {\bibinfo {title} {{Warped $AdS_6\times S^2$ in Type IIB
  supergravity II: Global solutions and five-brane webs}},\ }\href
  {https://doi.org/10.1007/JHEP05(2017)131} {\bibfield  {journal} {\bibinfo
  {journal} {JHEP}\ }\textbf {\bibinfo {volume} {05}},\ \bibinfo {pages}
  {131}},\ \Eprint {https://arxiv.org/abs/1703.08186} {arXiv:1703.08186
  [hep-th]} \BibitemShut {NoStop}%
\bibitem [{\citenamefont {D'Hoker}\ \emph
  {et~al.}(2017{\natexlab{c}})\citenamefont {D'Hoker}, \citenamefont
  {Gutperle},\ and\ \citenamefont {Uhlemann}}]{DHoker:2017zwj}%
  \BibitemOpen
  \bibfield  {author} {\bibinfo {author} {\bibfnamefont {E.}~\bibnamefont
  {D'Hoker}}, \bibinfo {author} {\bibfnamefont {M.}~\bibnamefont {Gutperle}},\
  and\ \bibinfo {author} {\bibfnamefont {C.~F.}\ \bibnamefont {Uhlemann}},\
  }\bibfield  {title} {\bibinfo {title} {{Warped $AdS_6\times S^2$ in Type IIB
  supergravity III: Global solutions with seven-branes}},\ }\href
  {https://doi.org/10.1007/JHEP11(2017)200} {\bibfield  {journal} {\bibinfo
  {journal} {JHEP}\ }\textbf {\bibinfo {volume} {11}},\ \bibinfo {pages}
  {200}},\ \Eprint {https://arxiv.org/abs/1706.00433} {arXiv:1706.00433
  [hep-th]} \BibitemShut {NoStop}%
\bibitem [{\citenamefont {Hosseini}\ and\ \citenamefont
  {Zaffaroni}(2020)}]{Hosseini:2020mut}%
  \BibitemOpen
  \bibfield  {author} {\bibinfo {author} {\bibfnamefont {S.~M.}\ \bibnamefont
  {Hosseini}}\ and\ \bibinfo {author} {\bibfnamefont {A.}~\bibnamefont
  {Zaffaroni}},\ }\bibfield  {title} {\bibinfo {title} {{Universal AdS black
  holes in theories with sixteen supercharges and their microstates}},\
  }\href@noop {} {\  (\bibinfo {year} {2020})},\ \Eprint
  {https://arxiv.org/abs/2011.01249} {arXiv:2011.01249 [hep-th]} \BibitemShut
  {NoStop}%
\bibitem [{\citenamefont {Closset}\ \emph {et~al.}(2013)\citenamefont
  {Closset}, \citenamefont {Dumitrescu}, \citenamefont {Festuccia},\ and\
  \citenamefont {Komargodski}}]{Closset:2012ru}%
  \BibitemOpen
  \bibfield  {author} {\bibinfo {author} {\bibfnamefont {C.}~\bibnamefont
  {Closset}}, \bibinfo {author} {\bibfnamefont {T.~T.}\ \bibnamefont
  {Dumitrescu}}, \bibinfo {author} {\bibfnamefont {G.}~\bibnamefont
  {Festuccia}},\ and\ \bibinfo {author} {\bibfnamefont {Z.}~\bibnamefont
  {Komargodski}},\ }\bibfield  {title} {\bibinfo {title} {{Supersymmetric Field
  Theories on Three-Manifolds}},\ }\href
  {https://doi.org/10.1007/JHEP05(2013)017} {\bibfield  {journal} {\bibinfo
  {journal} {JHEP}\ }\textbf {\bibinfo {volume} {05}},\ \bibinfo {pages}
  {017}},\ \Eprint {https://arxiv.org/abs/1212.3388} {arXiv:1212.3388 [hep-th]}
  \BibitemShut {NoStop}%
\bibitem [{\citenamefont {Assel}\ \emph
  {et~al.}(2012{\natexlab{b}})\citenamefont {Assel}, \citenamefont {Bachas},
  \citenamefont {Estes},\ and\ \citenamefont {Gomis}}]{Assel:2012cj}%
  \BibitemOpen
  \bibfield  {author} {\bibinfo {author} {\bibfnamefont {B.}~\bibnamefont
  {Assel}}, \bibinfo {author} {\bibfnamefont {C.}~\bibnamefont {Bachas}},
  \bibinfo {author} {\bibfnamefont {J.}~\bibnamefont {Estes}},\ and\ \bibinfo
  {author} {\bibfnamefont {J.}~\bibnamefont {Gomis}},\ }\bibfield  {title}
  {\bibinfo {title} {{IIB Duals of D=3 N=4 Circular Quivers}},\ }\href
  {https://doi.org/10.1007/JHEP12(2012)044} {\bibfield  {journal} {\bibinfo
  {journal} {JHEP}\ }\textbf {\bibinfo {volume} {12}},\ \bibinfo {pages}
  {044}},\ \Eprint {https://arxiv.org/abs/1210.2590} {arXiv:1210.2590 [hep-th]}
  \BibitemShut {NoStop}%
\bibitem [{\citenamefont {Assel}(2013)}]{Assel:2013lpa}%
  \BibitemOpen
  \bibfield  {author} {\bibinfo {author} {\bibfnamefont {B.}~\bibnamefont
  {Assel}},\ }\emph {\bibinfo {title} {{Holographic Duality for
  three-dimensional Super-conformal Field Theories}}},\ \href@noop {} {Ph.D.
  thesis},\ \bibinfo  {school} {Ecole Normale Superieure} (\bibinfo {year}
  {2013}),\ \Eprint {https://arxiv.org/abs/1307.4244} {arXiv:1307.4244
  [hep-th]} \BibitemShut {NoStop}%
\bibitem [{\citenamefont {Assel}\ and\ \citenamefont
  {Tomasiello}(2018)}]{Assel:2018vtq}%
  \BibitemOpen
  \bibfield  {author} {\bibinfo {author} {\bibfnamefont {B.}~\bibnamefont
  {Assel}}\ and\ \bibinfo {author} {\bibfnamefont {A.}~\bibnamefont
  {Tomasiello}},\ }\bibfield  {title} {\bibinfo {title} {{Holographic duals of
  3d S-fold CFTs}},\ }\href {https://doi.org/10.1007/JHEP06(2018)019}
  {\bibfield  {journal} {\bibinfo  {journal} {JHEP}\ }\textbf {\bibinfo
  {volume} {06}},\ \bibinfo {pages} {019}},\ \Eprint
  {https://arxiv.org/abs/1804.06419} {arXiv:1804.06419 [hep-th]} \BibitemShut
  {NoStop}%
\bibitem [{\citenamefont {Lozano}\ \emph
  {et~al.}(2020{\natexlab{a}})\citenamefont {Lozano}, \citenamefont {Nunez},
  \citenamefont {Ramirez},\ and\ \citenamefont {Speziali}}]{Lozano:2020txg}%
  \BibitemOpen
  \bibfield  {author} {\bibinfo {author} {\bibfnamefont {Y.}~\bibnamefont
  {Lozano}}, \bibinfo {author} {\bibfnamefont {C.}~\bibnamefont {Nunez}},
  \bibinfo {author} {\bibfnamefont {A.}~\bibnamefont {Ramirez}},\ and\ \bibinfo
  {author} {\bibfnamefont {S.}~\bibnamefont {Speziali}},\ }\bibfield  {title}
  {\bibinfo {title} {{New AdS$_2$ backgrounds and ${\cal N}=4$ Conformal
  Quantum Mechanics}},\ }\href@noop {} {\  (\bibinfo {year}
  {2020}{\natexlab{a}})},\ \Eprint {https://arxiv.org/abs/2011.00005}
  {arXiv:2011.00005 [hep-th]} \BibitemShut {NoStop}%
\bibitem [{\citenamefont {Lozano}\ \emph
  {et~al.}(2020{\natexlab{b}})\citenamefont {Lozano}, \citenamefont
  {Macpherson}, \citenamefont {Nunez},\ and\ \citenamefont
  {Ramirez}}]{Lozano:2019emq}%
  \BibitemOpen
  \bibfield  {author} {\bibinfo {author} {\bibfnamefont {Y.}~\bibnamefont
  {Lozano}}, \bibinfo {author} {\bibfnamefont {N.~T.}\ \bibnamefont
  {Macpherson}}, \bibinfo {author} {\bibfnamefont {C.}~\bibnamefont {Nunez}},\
  and\ \bibinfo {author} {\bibfnamefont {A.}~\bibnamefont {Ramirez}},\
  }\bibfield  {title} {\bibinfo {title} {{AdS$_3$ solutions in Massive IIA with
  small $\mathcal{N}=(4,0)$ supersymmetry}},\ }\href
  {https://doi.org/10.1007/JHEP01(2020)129} {\bibfield  {journal} {\bibinfo
  {journal} {JHEP}\ }\textbf {\bibinfo {volume} {01}},\ \bibinfo {pages}
  {129}},\ \Eprint {https://arxiv.org/abs/1908.09851} {arXiv:1908.09851
  [hep-th]} \BibitemShut {NoStop}%
\bibitem [{\citenamefont {Lozano}\ \emph
  {et~al.}(2020{\natexlab{c}})\citenamefont {Lozano}, \citenamefont
  {Macpherson}, \citenamefont {Nunez},\ and\ \citenamefont
  {Ramirez}}]{Lozano:2019jza}%
  \BibitemOpen
  \bibfield  {author} {\bibinfo {author} {\bibfnamefont {Y.}~\bibnamefont
  {Lozano}}, \bibinfo {author} {\bibfnamefont {N.~T.}\ \bibnamefont
  {Macpherson}}, \bibinfo {author} {\bibfnamefont {C.}~\bibnamefont {Nunez}},\
  and\ \bibinfo {author} {\bibfnamefont {A.}~\bibnamefont {Ramirez}},\
  }\bibfield  {title} {\bibinfo {title} {{1/4 BPS solutions and the
  AdS$_3$/CFT$_2$ correspondence}},\ }\href
  {https://doi.org/10.1103/PhysRevD.101.026014} {\bibfield  {journal} {\bibinfo
   {journal} {Phys. Rev. D}\ }\textbf {\bibinfo {volume} {101}},\ \bibinfo
  {pages} {026014} (\bibinfo {year} {2020}{\natexlab{c}})},\ \Eprint
  {https://arxiv.org/abs/1909.09636} {arXiv:1909.09636 [hep-th]} \BibitemShut
  {NoStop}%
\bibitem [{\citenamefont {Lozano}\ \emph
  {et~al.}(2020{\natexlab{d}})\citenamefont {Lozano}, \citenamefont
  {Macpherson}, \citenamefont {Nunez},\ and\ \citenamefont
  {Ramirez}}]{Lozano:2019zvg}%
  \BibitemOpen
  \bibfield  {author} {\bibinfo {author} {\bibfnamefont {Y.}~\bibnamefont
  {Lozano}}, \bibinfo {author} {\bibfnamefont {N.~T.}\ \bibnamefont
  {Macpherson}}, \bibinfo {author} {\bibfnamefont {C.}~\bibnamefont {Nunez}},\
  and\ \bibinfo {author} {\bibfnamefont {A.}~\bibnamefont {Ramirez}},\
  }\bibfield  {title} {\bibinfo {title} {{Two dimensional ${\cal N}=(0,4)$
  quivers dual to AdS$_3$ solutions in massive IIA}},\ }\href
  {https://doi.org/10.1007/JHEP01(2020)140} {\bibfield  {journal} {\bibinfo
  {journal} {JHEP}\ }\textbf {\bibinfo {volume} {01}},\ \bibinfo {pages}
  {140}},\ \Eprint {https://arxiv.org/abs/1909.10510} {arXiv:1909.10510
  [hep-th]} \BibitemShut {NoStop}%
\bibitem [{\citenamefont {Aharony}\ \emph {et~al.}(2012)\citenamefont
  {Aharony}, \citenamefont {Berdichevsky},\ and\ \citenamefont
  {Berkooz}}]{Aharony:2012tz}%
  \BibitemOpen
  \bibfield  {author} {\bibinfo {author} {\bibfnamefont {O.}~\bibnamefont
  {Aharony}}, \bibinfo {author} {\bibfnamefont {L.}~\bibnamefont
  {Berdichevsky}},\ and\ \bibinfo {author} {\bibfnamefont {M.}~\bibnamefont
  {Berkooz}},\ }\bibfield  {title} {\bibinfo {title} {{4d N=2 superconformal
  linear quivers with type IIA duals}},\ }\href
  {https://doi.org/10.1007/JHEP08(2012)131} {\bibfield  {journal} {\bibinfo
  {journal} {JHEP}\ }\textbf {\bibinfo {volume} {08}},\ \bibinfo {pages}
  {131}},\ \Eprint {https://arxiv.org/abs/1206.5916} {arXiv:1206.5916 [hep-th]}
  \BibitemShut {NoStop}%
\bibitem [{\citenamefont {N\'u\~nez}\ \emph {et~al.}(2019)\citenamefont
  {N\'u\~nez}, \citenamefont {Roychowdhury}, \citenamefont {Speziali},\ and\
  \citenamefont {Zacar\'\i{}as}}]{Nunez:2019gbg}%
  \BibitemOpen
  \bibfield  {author} {\bibinfo {author} {\bibfnamefont {C.}~\bibnamefont
  {N\'u\~nez}}, \bibinfo {author} {\bibfnamefont {D.}~\bibnamefont
  {Roychowdhury}}, \bibinfo {author} {\bibfnamefont {S.}~\bibnamefont
  {Speziali}},\ and\ \bibinfo {author} {\bibfnamefont {S.}~\bibnamefont
  {Zacar\'\i{}as}},\ }\bibfield  {title} {\bibinfo {title} {{Holographic
  aspects of four dimensional ${\cal N }=2$ SCFTs and their marginal
  deformations}},\ }\href {https://doi.org/10.1016/j.nuclphysb.2019.114617}
  {\bibfield  {journal} {\bibinfo  {journal} {Nucl. Phys. B}\ }\textbf
  {\bibinfo {volume} {943}},\ \bibinfo {pages} {114617} (\bibinfo {year}
  {2019})},\ \Eprint {https://arxiv.org/abs/1901.02888} {arXiv:1901.02888
  [hep-th]} \BibitemShut {NoStop}%
\bibitem [{\citenamefont {Gaiotto}\ and\ \citenamefont
  {Tomasiello}(2014)}]{Gaiotto:2014lca}%
  \BibitemOpen
  \bibfield  {author} {\bibinfo {author} {\bibfnamefont {D.}~\bibnamefont
  {Gaiotto}}\ and\ \bibinfo {author} {\bibfnamefont {A.}~\bibnamefont
  {Tomasiello}},\ }\bibfield  {title} {\bibinfo {title} {{Holography for (1,0)
  theories in six dimensions}},\ }\href
  {https://doi.org/10.1007/JHEP12(2014)003} {\bibfield  {journal} {\bibinfo
  {journal} {JHEP}\ }\textbf {\bibinfo {volume} {12}},\ \bibinfo {pages}
  {003}},\ \Eprint {https://arxiv.org/abs/1404.0711} {arXiv:1404.0711 [hep-th]}
  \BibitemShut {NoStop}%
\bibitem [{\citenamefont {Apruzzi}\ \emph {et~al.}(2015)\citenamefont
  {Apruzzi}, \citenamefont {Fazzi}, \citenamefont {Passias}, \citenamefont
  {Rota},\ and\ \citenamefont {Tomasiello}}]{Apruzzi:2015wna}%
  \BibitemOpen
  \bibfield  {author} {\bibinfo {author} {\bibfnamefont {F.}~\bibnamefont
  {Apruzzi}}, \bibinfo {author} {\bibfnamefont {M.}~\bibnamefont {Fazzi}},
  \bibinfo {author} {\bibfnamefont {A.}~\bibnamefont {Passias}}, \bibinfo
  {author} {\bibfnamefont {A.}~\bibnamefont {Rota}},\ and\ \bibinfo {author}
  {\bibfnamefont {A.}~\bibnamefont {Tomasiello}},\ }\bibfield  {title}
  {\bibinfo {title} {{Six-Dimensional Superconformal Theories and their
  Compactifications from Type IIA Supergravity}},\ }\href
  {https://doi.org/10.1103/PhysRevLett.115.061601} {\bibfield  {journal}
  {\bibinfo  {journal} {Phys. Rev. Lett.}\ }\textbf {\bibinfo {volume} {115}},\
  \bibinfo {pages} {061601} (\bibinfo {year} {2015})},\ \Eprint
  {https://arxiv.org/abs/1502.06616} {arXiv:1502.06616 [hep-th]} \BibitemShut
  {NoStop}%
\bibitem [{\citenamefont {Bergman}\ \emph {et~al.}(2020)\citenamefont
  {Bergman}, \citenamefont {Fazzi}, \citenamefont {Rodr\'\i{}guez-G\'omez},\
  and\ \citenamefont {Tomasiello}}]{Bergman:2020bvi}%
  \BibitemOpen
  \bibfield  {author} {\bibinfo {author} {\bibfnamefont {O.}~\bibnamefont
  {Bergman}}, \bibinfo {author} {\bibfnamefont {M.}~\bibnamefont {Fazzi}},
  \bibinfo {author} {\bibfnamefont {D.}~\bibnamefont
  {Rodr\'\i{}guez-G\'omez}},\ and\ \bibinfo {author} {\bibfnamefont
  {A.}~\bibnamefont {Tomasiello}},\ }\bibfield  {title} {\bibinfo {title}
  {{Charges and holography in 6d (1,0) theories}},\ }\href
  {https://doi.org/10.1007/JHEP05(2020)138} {\bibfield  {journal} {\bibinfo
  {journal} {JHEP}\ }\textbf {\bibinfo {volume} {05}},\ \bibinfo {pages}
  {138}},\ \Eprint {https://arxiv.org/abs/2002.04036} {arXiv:2002.04036
  [hep-th]} \BibitemShut {NoStop}%
\bibitem [{\citenamefont {Heckman}(2020)}]{Heckman:2020otd}%
  \BibitemOpen
  \bibfield  {author} {\bibinfo {author} {\bibfnamefont {J.~J.}\ \bibnamefont
  {Heckman}},\ }\bibfield  {title} {\bibinfo {title} {{Qubit Construction in 6D
  SCFTs}},\ }\href {https://doi.org/10.1016/j.physletb.2020.135891} {\bibfield
  {journal} {\bibinfo  {journal} {Phys. Lett. B}\ }\textbf {\bibinfo {volume}
  {811}},\ \bibinfo {pages} {135891} (\bibinfo {year} {2020})},\ \Eprint
  {https://arxiv.org/abs/2007.08545} {arXiv:2007.08545 [hep-th]} \BibitemShut
  {NoStop}%
\bibitem [{\citenamefont {Baume}\ \emph {et~al.}(2020)\citenamefont {Baume},
  \citenamefont {Heckman},\ and\ \citenamefont {Lawrie}}]{Baume:2020ure}%
  \BibitemOpen
  \bibfield  {author} {\bibinfo {author} {\bibfnamefont {F.}~\bibnamefont
  {Baume}}, \bibinfo {author} {\bibfnamefont {J.~J.}\ \bibnamefont {Heckman}},\
  and\ \bibinfo {author} {\bibfnamefont {C.}~\bibnamefont {Lawrie}},\
  }\bibfield  {title} {\bibinfo {title} {{6D SCFTs, 4D SCFTs, Conformal Matter,
  and Spin Chains}},\ }\href@noop {} {\  (\bibinfo {year} {2020})},\ \Eprint
  {https://arxiv.org/abs/2007.07262} {arXiv:2007.07262 [hep-th]} \BibitemShut
  {NoStop}%
\bibitem [{\citenamefont {Naka}(2002)}]{Naka:2002jz}%
  \BibitemOpen
  \bibfield  {author} {\bibinfo {author} {\bibfnamefont {M.}~\bibnamefont
  {Naka}},\ }\bibfield  {title} {\bibinfo {title} {{Various wrapped branes from
  gauged supergravities}},\ }\href@noop {} {\  (\bibinfo {year} {2002})},\
  \Eprint {https://arxiv.org/abs/hep-th/0206141} {arXiv:hep-th/0206141}
  \BibitemShut {NoStop}%
\bibitem [{\citenamefont {Hong}\ \emph {et~al.}(2018)\citenamefont {Hong},
  \citenamefont {Liu},\ and\ \citenamefont {Mayerson}}]{Hong:2018amk}%
  \BibitemOpen
  \bibfield  {author} {\bibinfo {author} {\bibfnamefont {J.}~\bibnamefont
  {Hong}}, \bibinfo {author} {\bibfnamefont {J.~T.}\ \bibnamefont {Liu}},\ and\
  \bibinfo {author} {\bibfnamefont {D.~R.}\ \bibnamefont {Mayerson}},\
  }\bibfield  {title} {\bibinfo {title} {{Gauged Six-Dimensional Supergravity
  from Warped IIB Reductions}},\ }\href
  {https://doi.org/10.1007/JHEP09(2018)140} {\bibfield  {journal} {\bibinfo
  {journal} {JHEP}\ }\textbf {\bibinfo {volume} {09}},\ \bibinfo {pages}
  {140}},\ \Eprint {https://arxiv.org/abs/1808.04301} {arXiv:1808.04301
  [hep-th]} \BibitemShut {NoStop}%
\bibitem [{\citenamefont {Malek}\ \emph {et~al.}(2018)\citenamefont {Malek},
  \citenamefont {Samtleben},\ and\ \citenamefont
  {Vall~Camell}}]{Malek:2018zcz}%
  \BibitemOpen
  \bibfield  {author} {\bibinfo {author} {\bibfnamefont {E.}~\bibnamefont
  {Malek}}, \bibinfo {author} {\bibfnamefont {H.}~\bibnamefont {Samtleben}},\
  and\ \bibinfo {author} {\bibfnamefont {V.}~\bibnamefont {Vall~Camell}},\
  }\bibfield  {title} {\bibinfo {title} {{Supersymmetric AdS$_{7}$ and AdS$_6$
  vacua and their minimal consistent truncations from exceptional field
  theory}},\ }\href {https://doi.org/10.1016/j.physletb.2018.09.037} {\bibfield
   {journal} {\bibinfo  {journal} {Phys. Lett. B}\ }\textbf {\bibinfo {volume}
  {786}},\ \bibinfo {pages} {171} (\bibinfo {year} {2018})},\ \Eprint
  {https://arxiv.org/abs/1808.05597} {arXiv:1808.05597 [hep-th]} \BibitemShut
  {NoStop}%
\bibitem [{\citenamefont {Hosseini}\ and\ \citenamefont
  {Hristov}(2020)}]{Hosseini:2020wag}%
  \BibitemOpen
  \bibfield  {author} {\bibinfo {author} {\bibfnamefont {S.~M.}\ \bibnamefont
  {Hosseini}}\ and\ \bibinfo {author} {\bibfnamefont {K.}~\bibnamefont
  {Hristov}},\ }\bibfield  {title} {\bibinfo {title} {{4d F(4) gauged
  supergravity and black holes of class $\mathcal{F}$}},\ }\href@noop {} {\
  (\bibinfo {year} {2020})},\ \Eprint {https://arxiv.org/abs/2011.01943}
  {arXiv:2011.01943 [hep-th]} \BibitemShut {NoStop}%
\end{thebibliography}%
\end{document}